\documentclass[a4paper,11pt]{article}
\pdfoutput=1 % if your are submitting a pdflatex (i.e. if you have
\usepackage{jheppub} % for details on the use of the package, please
\usepackage[T1]{fontenc} % if needed
\usepackage{caption}
\usepackage{subfigure}
\usepackage{mathrsfs}
\usepackage{color}
\title{\boldmath Equilibrium and nonequilibrium quantum correlations between two detectors in curved space time}

\author[a,b]{He Wang,}
\author[c,1]{Jin Wang,\note{Corresponding author.}}

\affiliation[a]{College of Physics, Jilin University,\\Changchun 130021, China}
\affiliation[b]{State Key Laboratory of Electroanalytical Chemistry, Changchun Institute of Applied Chemistry,\\Changchun 130021, China}
\affiliation[c]{Department of Chemistry and of Physics and Astronomy, Stony Brook University, Stony Brook,\\NY 11794-3400, USA}

\emailAdd{jin.wang.1@stonybrook.edu}

\abstract{We investigate the equilibrium and nonequilibrium quantum information correlations encoded in two-qubit system (near the horizon of a $Kerr$ black hole). We study the impact of mass and the angular momentum, and further the local curvature or accelerations on the behaviors of the quantum correlations between two qubits. We show the quantum information of two qubits is encoded in the space time structure. In nonequilibrium case, the nonequilibrium can also contribute to the correlations.}

\begin{document}
\maketitle
\flushbottom

\section{Introduction}
\label{sec:Introduction}

Quantum correlations including coherence~\cite{Quantifying Coherence}, entanglement~\cite{Quantum entanglement}~\cite{Entanglement of a Pair of Quantum Bits}, mutual information and the quantum discord~\cite{Quantum Discord} etc have been widely theoretically studied. They provide profound insights on the black hole physics and even cosmology.~\cite{Quantum information transfer and models for black hole mechanics}~\cite{Black holes as mirrors}~\cite{Cosmological quantum entanglement}~\cite{Entanglement in curved spacetimes and cosmology}. They are often needed as the key resources to carry out certain quantum information processing tasks, e.g., quantum teleportation, quantum computation, quantum cryptography,  quantum metrology etc.~\cite{Quantum Computation and Quantum Information}~\cite{Teleporting}~\cite{Quantum Cryptography}~\cite{Teleportation}~\cite{Bell's inequalities versus teleportation} One of the major obstacles to realize the quantum information technologies is the environmental induced decoherence and dissipation effect to the quantum systems, which may give rise to quantum correlations degradation. However, an external environment can also provide indirect interactions between the subsystems through the correlations that exist. A phenomenological and illuminating example is the independent atoms immersed in external quantum fields and weakly coupled to them through $Unruh-DeWitt$ detector interaction~\cite{Cosmological quantum entanglement}~\cite{Entanglement in curved spacetimes and cosmology}~\cite{Notes on black hole evaporation}. The atoms can be usually treated in a non-relativistic approximation, as independent n-level systems (qubits or harmonic oscillator typically), with negligible size, while the environment can be described by a set of quantum fields in a given quantum state, typically either a thermal state or simply the vacuum state. Despite the simplified setting, this can also provide heuristic insight on the relationship between the environment and the quantum correlations. It has been shown that there are certain scenarios where the environment may create quantum correlations rather than destroy them, in both flat space time~\cite{Controlling entanglement generation in external quantum fields}~\cite{Entanglement generation in uniformly accelerating atoms}~\cite{Thermal amplification of field-correlation harvesting}~\cite{Equilibrium and Nonequilibrium Quantum Correlations Between Two Accelerated Detectors}and curved spacetime backgrounds~\cite{Cosmological quantum entanglement}~\cite{Entanglement in curved spacetimes and cosmology}~\cite{Entanglement generation outside a Schwarzschild black hole and the Hawking effect}~\cite{Quantum entanglement generation in de Sitter spacetime}. However, the studies of the quantum information harvesting have so far been focused on special case where two detectors couple to the environment with equal strength and under the same space time background (local curvature or the local accelerations is same, which corresponds to the equilibrium case). How the two detectors at different locations of the curved space time (corresponds to the nonequilibrium case) correlate with each other is still an unresolved problem.

Black hole, which relates gravity, quantum theory and thermodynamics, is fascinating and very challenging subject in physics. The Hawking radiation emerges at the horizon as a pure quantum effect~\cite{Black hole explosions?}, can also be understood by the open quantum system theory. Concretely, consider a single two level atom system near the horizon, treat the massless scalar field in the curved background as the environment and then compute the spontaneous excitation rate of the atom. This reveals that close to the horizon, the ground state detector in the vacuum would spontaneously excite with an excitation rate same as the case when there is a thermal bath around the atom.~\cite{Understanding Hawking radiation in the framework of open quantum systems}~\cite{Researching on Hawking Effect in a Kerr Space Time via Open Quantum System Approach} Therefore, the near-horizon geometry plays the key role to the character of a black hole space time. Moreover it was shown that around the horizon of a Kerr space time, the scalar field theory can be reduced to a 2-dimensional effective field theory.~\cite{Researching on Hawking Effect in a Kerr Space Time via Open Quantum System Approach}~\cite{Hawking radiation from rotating black holes and gravitational anomalies}~\cite{Anomalies} It is then convenient to study the quantum correlations near the black hole with this insight.

In this paper, we study the quantum correlations near the horizon of the Kerr black hole by using the dimensional reduction method. Both equilibrium and nonequilibrium scenarios are discussed. For equilibrium scenario (which means two detectors coupled to the field with the same coupling at the same location near the black horizon). We start with a product state initially, an equilibrium steady state can be found at final time in certain scenario. We focus on the equilibrium steady state and study its quantum correlations. We found that the quantum correlations (including entanglement, coherence, discord and mutual information) in two-qubit system vary non-monotonically with the mass of the black hole and are amplified by the angular momentum. The Von Neumann entropy which measures the entanglement between the system and environment behaves oppositely compared to the quantum correlations with respect to the mass and the angular momentum. Moreover, we found that the curvature can suppress the quantum correlations within the system but enhance the correlation with environment.

For the nonequilibrium case (one of two qubits is weakly coupled to the field that can be viewed equivalently as isolated from the environment.). The system evolves from the maximal entangled state and the information will scramble to environment. The decay rates of the quantum correlations increase at first and then decrease with respect to the mass. The larger angular momentum suppresses the decay rates of the correlations. At a fix time, we found the behavior of the quantum correlations are very similar to the equilibrium case: the quantum correlations in the two-qubit system vary non-monotonically with the mass of black hole and are amplified by the angular momentum, while the curvature suppresses the quantum correlations within the system. We quantify the EPR (entropy production rate) of the system, and found that it decreases in time. The EPR decreases at first and then increases to a constant as the black hole mass becomes larger and increases when the angular momentum becomes larger. Besides, the pace time curvature is found to suppress the information scrambling. The local curvature enhances the decay rate of the quantum correlations and EPR, but reduces the decay rate of the Von Neumann entropy.

Another nonequilibrium scenario is also investigated. There are two types of massless scalar fields equivalent to two different independent bathes coupled to two interacting qubits respectively. The nonequilibrium is measured by the difference in the radius separating of two qubits. We investigate the quantum correlations of the nonequilibrium steady state. On the whole, the correlations can survive and  be maintained at a steady value when $\Delta r$ is large. The entanglement, discord and the mutual information behave non-monotonically in certain parameters. This shows that the quantum correlations can be amplified by the nonequilibrium. The coherence monotonically decreases to a constant. The Von Neumann entropy decreases to a constant which indicates that the nonequilibrium is harmful to produce the correlation between the system and the environment. The flux increases to a constant as the difference in $\Delta r$ or nonequilibrium increases. The EPR as a measure of thermodynamics cost increases as the $\Delta r$ increases.

The organization of our paper is as follows. In section~\ref{sec:Master Equation for open quantum system}, we will describe the simplest model, which can be used for the later generalization. Then we review the basic formulations, including the master equation describing the system of the detector in the $Born-Markov$ approximation. In section~\ref{sec:Measures of quantum correlations}, we introduce certain quantum correlations we are interested in. In section~\ref{sec:Massless scalar field quantized near Kerr horizon and the two vacua}, the dimensional reduction technique is used to investigate the massless scalar field in a Kerr space time, and two types of vacua are discussed. In section~\ref{sec:Equilibrium quantum correlations in curved space time}, we discuss the quantum correlations of the equilibrium steady state for the two atom detectors near Kerr black hole. Importantly, we study the nonequilibrium case. In section~\ref{sec:Nonequilibrium transient quantum correlations in curved space time}, we consider the quantum correlations in the curved space time in a specific nonequilibrium transient scenario. In section~\ref{sec:Nonequilibrium steady quantum correlations}, we study the nonequilibrium quantum correlations in curved space time at the steady state. At last, we will draw a conclusion in section~\ref{sec:Conclusion}

\section{Master Equation for open quantum system}
\label{sec:Master Equation for open quantum system}

Our main objective in this section is to formulate the time evolution of an open quantum system and to obtain the $GSKL$ master equation which properly describes the non-unitary behaviors and can be obtained by
partial trace over the environmental baths i.e. the massless probe scalar field placed on the Kerr black hole space time background. Generally, the open quantum set up can be described by the following Hamiltonian

\begin{equation}
\label{eq:1}
H_{total}=H_{0}+H_{1}=H_{sys}+H_{field}+H_{I}.
\end{equation}

Here $H_{sys}$ is the Hamiltonian of the atom or the detector. For the single two-level atom internal dynamics will be driven by a $2\times2$ hamiltonian matrix. In a given basis can be assumed to have the form: $\frac{\omega}{2}\sigma_{z}$, where $\sigma_{z}$ is the $Pauli$ matrice, while $\omega$ represents the gap between the two energy eigenvalues. Then, the atom Hamiltonian becomes $H_{sys}=\frac{\omega}{2}\sigma_{z}$. We assume that the Hamiltonian describing the interaction between the atom and the scalar field can be take in the form of the $Unruh-DeWitt$ detector interaction: $ H_{I}=\mu(\sigma_{+}+\sigma_{-})\phi(x(\tau))$, in which $\mu$ is the coupling constant. Also, we set $\sigma_{+}(\sigma_{-})$ as the atomic rasing (lowering) operator, and $\phi(x)$ corresponds to the scalar field operator in Kerr spacetime. The time evolution of the total system in the proper time $\tau$ is governed by the $Von Neumann$ equation

\begin{equation}
\label{eq:2}
\partial_{\tau}\rho_{total}=-i[H_{total},\rho_{total}].
\end{equation}

For convenience, one usually performs a unitary transformation to transform the above $Liouville-Von Neumann$ equation into the interaction picture

\begin{equation}
\label{eq:3}
\partial_{\tau}\rho_{total}^{I}=-i[H_{I}^{I},\rho_{total}^{I}].
\end{equation}

The upper index $I$ represents the operator in interaction picture, the unitary transformation reads $\rho_{total}^{I}(\tau)=e^{iH_{0}\tau}\rho_{total}e^{-iH_{0}\tau}$ and $H_{I}^{I}(\tau)=e^{iH_{0}\tau}H_{I}e^{-iH_{0}\tau}$ for $\rho_{total}$ and $H_{I}$ respectively. Integrate the above equation Eqn.\eqref{eq:3}, we get

\begin{equation}
\label{eq:4}
\rho_{total}^{I}(\tau)=\rho_{total}^{I}(0)-i\int_{0}^{\tau}ds[H_{I}^{I}(s),\rho_{total}^{I}(s)].
\end{equation}

Inserting Eqn.\eqref{eq:4} back to Eqn.\eqref{eq:3} and tracing out the field (or environmental) degrees of freedoms, we arrive at

\begin{equation}
\label{eq:5}
\frac{d\rho_{sys}^{I}(\tau)}{d\tau}=-\int_{0}^{\tau}Tr_{field}{[H_{I}^{I}(\tau),[H_{I}^{I}(s),\rho_{total}^{I}(s)]]}.
\end{equation}

where we have taken $Tr_{bath}[H_{I}^{I}(\tau),\rho_{total}^{I}(0)]=0$, meaning that initially the interaction does not create any dynamics in the bath. Eqn.\eqref{eq:5} still contains the density matrix of the total system $\rho_{total}^{I}(\tau)$ on its right-hand side. In order to eliminate $\rho_{total}^{I}(\tau)$ from the equation of motion, on can perform a first approximation, known as the $Born$ approximation: the coupling between the system and the bath is weak such that the influence of the bath is small. Thus one can consider the bath as almost unchanged and then the state of the total system at time $\tau$ may be approximately characterized by a tensor product

\begin{equation}
\label{eq:6}
\rho_{total}^{I}(\tau)=\rho_{sys}^{I}(\tau)\bigotimes\rho_{field}.
\end{equation}

Inserting the tensor product into the exact equation of motion Eqn.\eqref{eq:5}, one obtains a closed integral-differential equation for the reduced density matrix
\begin{equation}
\label{eq:7}
\frac{d\rho_{sys}^{I}(\tau)}{d\tau}=-\int_{0}^{\tau}Tr_{field}{[H_{I}^{I}(\tau),[H_{I}^{I}(s),\rho_{sys}^{I}(s)\bigotimes\rho_{field}]]}.
\end{equation}

In order to simplify the above equation further one can perform the $Markov$ approximation, in which the integrand $\rho_{sys}^{I}(s)$ is firstly replaced by $\rho_{sys}^{I}(\tau)$. In this way
one can obtain an equation of motion for the reduced system's density matrix in which the time development of the state of the system at time $\tau$ only depends on the present state.

\begin{equation}
\label{eq:8}
\frac{d\rho_{sys}^{I}}{d\tau}=-\int_{0}^{\tau}Tr_{field}{[H_{I}^{I}(\tau),[H_{I}^{I}(s),\rho_{sys}^{I}(\tau)\bigotimes\rho_{field}]]}.
\end{equation}

The $Markov$ approximation grouped together with the Born approximation is often regarded as the $Born-Markov$ approximation. However, under this approximation alone the resulting master equation does not guarantee to generate a quantum dynamical semigroup. One therefore performs a further secular approximation which involves an averaging over and discard the rapidly oscillating terms in the master
equation~\cite{The Theory of Open Quantum Systems}. With the aid of all these approximations, one can go back to the Schr\"{o}dinger picture where we obtain the following $Markovian$ master equation in $Kossakowski-Lindblad$ form ~\cite{Completely positive dynamical semigroups of N-level systems}:

\begin{equation}\begin{split}
\label{eq:9}
\frac{d\rho_{sys}(\tau)}{d\tau}&=-i[H_{eff},\rho_{sys}(\tau)]+\mathscr{L}[\rho_{sys}(\tau)]\\
&=-i[H_{eff},\rho_{sys}(\tau)]+\sum_{j=1}^{3}[2L_{j}\rho_{sys}L_{j}^{\dag}-L_{j}^{\dag}L_{j}\rho_{sys}-\rho_{sys}L_{j}^{\dag}L_{j}]
\end{split}
\end{equation}

where $H_{eff}$ and $L_{j}$ are given as

\begin{equation}\begin{split}
\label{eq:10}
H_{eff}=\frac{\Omega}{2}\sigma_{z}=\frac{\omega+i(\mathscr{K}(-\omega)-\mathscr{K}(\omega))}{2}\sigma_{z}\\
L_{1}=\sqrt{\frac{\gamma_{-}}{2}}\sigma_{-},L_{2}=\sqrt{\frac{\gamma_{+}}{2}}\sigma_{+},L_{3}=\sqrt{\frac{\gamma_{0}}{2}}\sigma_{z}
\end{split}
\end{equation}

where $\gamma_{\pm}=\mu^{2}\int_{-\infty}^{+\infty}e^{\mp i\omega s}G^{+}(s-i\epsilon)ds$ and $\gamma_{0}=0$, $G^{+}(s-i\epsilon)=\langle0|\phi(x)\phi(x')|0\rangle$ is the $Wightman$ function of the massless scalar
field($s=\tau-\tau'$ here). And $\mathscr{K}(\lambda)=\frac{P}{i\pi}\int\frac{\mathscr{G}(\omega)}{\omega-\lambda} d\omega$ ($P$ denotes principal value) where $\mathscr{G}(\omega)$ is the Fourier transformation of $G^{+}$.

\section{Measures of quantum correlations}
\label{sec:Measures of quantum correlations}

In this section, we introduce certain important measures which are required for the quantification of the quantum correlations. Coherence, being at the heart of interference
phenomena, plays a central role in physics as it enables applications that are impossible within classical mechanics or ray optics and can be measured as~\cite{Quantifying Coherence}
\begin{equation}
\label{eq:11}
\mathscr{C}_{l_{1}}=\sum_{i\neq j}\mid\rho_{ij}\mid
\end{equation}

Quantum entanglement has remained a major resource for accomplishing quantum information processing tasks such as teleportation~\cite{Teleporting}, quantum key distribution~\cite{Quantum Cryptography}, and quantum computing~\cite{Quantum entanglement} etc.

Among many measures of entanglement of a two-qubit system, concurrence is extensively used so far in many contexts. The concurrence of a two-qubit mixed state $\rho$ is defined as~\cite{Entanglement of a Pair of Quantum Bits}
\begin{equation}
\label{eq:12}
\mathscr{C}=Max(0,\lambda_{1}-\lambda_{2}-\lambda_{3}-\lambda_{4})
\end{equation}

where $\lambda_{i}$ represents the square root of the $ith$ eigenvalue, in descending order of the matrix $\rho\widetilde{\rho}$ with $\widetilde{\rho}=(\sigma_{2}\bigotimes\sigma_{2})\rho^{T}(\sigma_{2}\bigotimes\sigma_{2})$, while $T$ denotes transposition.

%$Bell$ inequality was also one of the first tools used to detect entanglement. Originally, $Bell$ inequalities were introduced as an attempt to rule out local hidden variable (LHV) models. One can express the most general form of $Bell-CHSH$ inequality for the mixed state $\rho_{AB}=\frac{1}{4}(I_{a}\bigotimes I_{b}+\sum_{i=1}^{3}(a_{i}\sigma_{i}\bigotimes I_{b}+I_{a}\bigotimes b_{i}\sigma_{i})+\sum_{i,j=1}^{3}C_{ij}\sigma_{i}\bigotimes\sigma_{i})$, where ~\cite{Teleportation}

%\begin{equation}
%\label{eq:13}
%M(\rho)=max(u_{i}+u_{j})
%\end{equation}

%$u_{i}$ is the eigenvalue of the matrix $(CC^{\dagger})$. With the Bell's operator,~\cite{Bell inequality, Bell states and maximally entangled states for n qubits} the maximal violation of the Bell inequalities (MVBI) has the following relation: $\gamma(\rho)=2\sqrt{u_{1}+u_{2}}$, where $u_{1}$ and $u_{2}$ are the first two largest eigenvalues of the matrix $(CC^{\dagger})$.~\cite{Bell's inequality, generalized concurrence and entanglement in qubits.} Notice that, the violation of Bell inequality for a given quantum state indicates that the state is entangled. But at the same time, there are certain entangled states which do not violate Bell¡¯s inequality.

A bipartite quantum state contains both classical and quantum correlations which are quantified jointly by their quantum mutual information, an information-theoretic measure of the total correlation in
a bipartite quantum state. In particular, if $\rho_{AB}$ denotes the density operator of a composite bipartite system AB, and $\rho_{A}$ ($\rho_{B}$) denotes the density operator of part A(B), respectively, then the quantum mutual information is defined as~\cite{Quantum Discord}

\begin{equation}
\label{eq:14}
\emph{I}(\rho_{AB})=S(\rho_{A})+S(\rho_{B})-S(\rho_{AB})
\end{equation}
where $S(\rho)=-tr(\rho\log_{2}\rho)$ is the Von Neumann entropy. The whole system is a pure state at any time because of the unitary evolution, and $S(\rho_{total})=0$. When we trace out the degrees of freedoms of the field, and only consider $S(\rho_{AB})$ which is non-vanishing, this measures the entanglement between the system and the environment.

Quantum discord is a measure of non classical correlation between two subsystems of a quantum system. It includes correlations that are due to quantum physical effects, but do not necessarily involve the
concept of quantum entanglement. In fact it is a different type of quantum correlation than the entanglement because separable mixed states (that is, with no entanglement) can have non-zero quantum discord. Sometimes it is also identified as the measure of quantumness of the correlation functions.  It is defined as~\cite{Quantum Discord}~\cite{Quantum discord for two-qubit X-states}

\begin{equation}
\label{eq:15}
\mathscr{Q}(\rho_{AB})=\emph{I}(\rho_{AB})-\mathscr{CC}(\rho_{AB})
\end{equation}

$\mathscr{CC}(\rho_{AB})$ is the classical correlation which depends on the projection operator and we use the maximum in computing discord. For a general state, quantum discord is hard to compute and only for $X-type$ state there is an exact expression. For any two qubit state the density matrix is given by the following expression:

\begin{equation}
\label{eq:16}
\rho_{AB}=\frac{1}{4}(I_{a}\bigotimes I_{b}+\sum_{i=1}^{3}(a_{i}\sigma_{i}\bigotimes I_{b}+I_{a}\bigotimes b_{i}\sigma_{i})+\sum_{i,j=1}^{3}C_{ij}\sigma_{i}\bigotimes\sigma_{j})
\end{equation}
For the class of a $"X"$ state, the Bloch vector is along the z-axis, the above expression can be simplified as

\begin{equation}
\label{eq:rho2}
\rho_{AB}=\frac{1}{4}(I_{a}\bigotimes I_{b}+(a\sigma_{z}\bigotimes I_{b}+I_{a}\bigotimes b\sigma_{z})+\sum_{i,j=1}^{3}C_{ij}\sigma_{i}\bigotimes\sigma_{j})
\end{equation}

The quantum discord is invariant under the local unitary transformations. It has been shown that the $\rho_{AB}$ can be further simplified as
\begin{equation}
\label{eq:rho3}
\rho_{AB}=\frac{1}{4}(I_{a}\bigotimes I_{b}+(a\sigma_{z}\bigotimes I_{b}+I_{a}\bigotimes b\sigma_{z})+\sum_{i=1}^{3}C_{i}\sigma_{i}\bigotimes\sigma_{i})
\end{equation}
with the local unitary transformations.~\cite{Quantum discord for two-qubit systems}
The $\mathscr{CC}$ can be measured as~\cite{Nonequilibrium effects on quantum correlations: Discord}

\begin{equation}
\label{eq:classical correlation}
\mathscr{CC}=S(\rho_{A})-\min{S_{1},S_{2}}
\end{equation}
where
\begin{equation}\begin{split}
\label{eq:s1}
S_{1}=&-\frac{1+a+b+c_{3}}{4}log_{2}\frac{1+a+b+c_{3}}{2(1+b)}\\
&-\frac{1-a+b-c_{3}}{4}log_{2}\frac{1-a+b-c_{3}}{2(1+b)}\\
&-\frac{1+a-b-c_{3}}{4}log_{2}\frac{1+a-b-c_{3}}{2(1-b)}\\
&-\frac{1-a-b+c_{3}}{4}log_{2}\frac{1-a-b+c_{3}}{2(1-b)}
\end{split}
\end{equation}
and
\begin{equation}
\label{eq:s2}
S_2=1+f(\sqrt{a^{2}+C_{1}^2})
\end{equation}
The $f(t)$ is defined as $f(t)=-\frac{1-t}{2}log_{2}(1-t)-\frac{1+t}{2}log_{2}(1+t)$. Finally, the quantum discord is given as $\mathscr{Q}(\rho_{AB})=\emph{I}(\rho_{AB})-\mathscr{CC}(\rho_{AB})$.

\section{Massless scalar field quantized near Kerr horizon and the two vacua}
\label{sec:Massless scalar field quantized near Kerr horizon and the two vacua}

In order to find out out how the reduced density matrix evolves with proper time from Eqn.\eqref{eq:9}, we will review the scalar wave equation of the Kerr black hole space time by following Liu et al ~\cite{Researching on Hawking Effect in a Kerr Space Time via Open Quantum System Approach}. The metric of Kerr spacetime in $Boyer-Lindquist$ coordinates is given as

\begin{equation}\begin{split}
\label{eq:20}
ds^{2}=-\frac{\Delta}{R^{2}}(dt-a\sin^{2}\theta d\varphi)^2+\frac{\sin^{2}\theta}{R^{2}}[(r^{2}+a^{2})d\varphi-adt]^{2}+\frac{R^{2}}{\Delta}dr^{2}+R^{2}d\theta^{2}
\end{split}
\end{equation}

where $\Delta=(r-r_{+})(r-r_{-})$, $R^{2}=r^{2}+a^{2}\cos^{2}\theta$ and $r_{\pm}=M\pm\sqrt{M^{2}-a^{2}}$. $M$ and $a$ represent the mass and the angular momentum per unit mass of the black
hole, respectively. The event horizon of the Kerr black hole is located at $r=r_{+}$. Then, Liu et al show that the scalar field theory in the background Eqn.\eqref{eq:20} can be reduced to a $2-dimensional$ field theory in the near-horizon region with the dimensional reduction technique.~\cite{Researching on Hawking Effect in a Kerr Space Time via Open Quantum System Approach} This technique firstly has been employed for the Kerr black hole by Murata and Soda~\cite{Hawking radiation from rotating black holes and gravitational anomalies} and developed with a more general technique by Iso et al.~\cite{Anomalies}

First of all, we write further the action of the massless scalar field as
\begin{equation}\begin{split}
\label{eq:21}
S[\phi]=\frac{1}{2}\int dx^{4}\sqrt{-g}g^{\mu\nu}\partial_{\mu}\phi\partial_{\nu}\phi
\end{split}
\end{equation}

By substituting Eqn.\eqref{eq:20} into Eqn.\eqref{eq:21}, and then transform the radial coordinate $r$ into the tortoise
coordinate $r_{*}$ defined by
\begin{equation}\begin{split}
\label{eq:22}
\frac{dr_{*}}{dr}=\frac{1}{F(r)}=\frac{r^{2}+a^{2}}{\Delta}
\end{split}
\end{equation}

Now the action reads
\begin{equation}\begin{split}
\label{eq:23}
S[\phi]=&-\frac{1}{2}\int dr_{*}dtd\theta d\varphi\sin\theta\phi\\
&\times[-((r^{2}+a^{2})-F(r)a^{2}\sin^{2}\theta)\partial_{t}^{2}-2a(1-F(r))\partial_{t}\partial_{\varphi}+(\frac{F(r)}{\sin^{2}\theta}-\frac{a^{2}}{r^{2}+a^{2}})\partial_{\varphi}^{2}\\
&+\partial_{r_{*}}(r^{2}+a^{2})\partial_{r_{*}}+\frac{F(r)}{\sin\theta}\partial_{\theta}\sin\theta\partial_{\theta}]\phi
\end{split}
\end{equation}

We only consider the region near Kerr horizon. Since $F(r_{+})\rightarrow 0$ when $r\rightarrow r_{+}$, we can only consider dominant terms in Eqn.\eqref{eq:23}.

\begin{equation}\begin{split}
\label{eq:24}
S[\phi]=&-\frac{1}{2}\int dr_{*}dtd\theta d\varphi\sin\theta\phi\\
&\times[-(r^{2}+a^{2})\partial_{t}^{2}-2a\partial_{t}\partial_{\varphi}-\frac{a^{2}}{r^{2}+a^{2}}\partial_{\varphi}^{2}+\partial_{r_{*}}(r^{2}+a^{2})\partial_{r_{*}}]\phi
\end{split}
\end{equation}

And then one returns to the $r$ coordinate system and uses a globally corotating coordinate system as

\begin{equation}\begin{split}
\label{eq:25}
\psi&=\varphi-\frac{a}{a^{2}+r^{2}}t\\
\xi&=t
\end{split}
\end{equation}

In the new coordinates, we can rewrite Eqn.\eqref{eq:24} as

\begin{equation}\begin{split}
\label{eq:26}
S[\phi]=&\frac{1}{2}\int dr_{*}dtd\theta d\varphi\sin\theta\phi\\
&\times[(r^{2}+a^{2})\sin\theta\phi(-\frac{1}{F(r)}\partial_{\xi}^{2}+\partial_{r}F(r)\partial_{r})\phi]
\end{split}
\end{equation}

Therefore the angular terms disappear completely. Using the spherical harmonics expansion $\phi=\sum_{l,m}\phi_{lm}(\xi,r)Y_{lm}(\theta,\psi)$, we obtain the effective $2-dimensional$ action

\begin{equation}\begin{split}
\label{eq:27}
S[\phi]=\sum_{l,m}-\frac{1}{2}\int (r^{2}+a^{2})drd\xi \phi_{lm}\times(-\frac{1}{F(r)}\partial_{\xi}^2+\partial_{r}F(r)\partial_{r})\phi_{lm}
\end{split}
\end{equation}

where we have used the orthonormal condition for the spherical harmonics as follows:

\begin{equation}\begin{split}
\label{eq:28}
\int d\psi d\theta \sin\theta Y_{l'm'}^{*}Y_{lm}=\delta_{ll'}\delta_{mm'}
\end{split}
\end{equation}

From the action Eqn.\eqref{eq:27}, it is obvious to find that $\phi$ can be considered as a (1+1)-dimensional massless scalar field in the backgrounds of the dilaton $\Phi$. The effective 2-dimensional metric near the horizon and the dilaton can be written as

\begin{equation}\begin{split}
\label{eq:29}
ds^{2}&=-F(r)d\xi^{2}+\frac{1}{F(r)}dr^{2}\\
\Phi&=r^{2}+a^{2}
\end{split}
\end{equation}

Hence, we have reduced the 4-dimensional field theory to a 2-dimensional case. This is consistent with~\cite{Hawking radiation from rotating black holes and gravitational anomalies}~\cite{Anomalies}. From Eqn.\eqref{eq:29}, we can define two types of vacua: the $Boulware$ Vacuum and the $Unruh$ Vacuum.%The effective scalar curvature is $\frac{4Mr(-3a^{2}+r^{2})}{(a^{2}+r^{2})^{3}}$.

According to Eqn.\eqref{eq:22}, the effective 2-dimensional metric Eqn.\eqref{eq:29} changes to

\begin{equation}\begin{split}
\label{eq:30}
ds^{2}&=F(r)(-d\xi^{2}+dr_{*}^{2})
\end{split}
\end{equation}

We can see Eqn.\eqref{eq:30} is exactly conformal to $Minkowski$ metric form, hence, the scalar field equation reads

\begin{equation}\begin{split}
\label{eq:31}
(\partial_{\xi}^{2}-\partial_{r_{*}}^{2})\phi(\xi,r_{*})=0
\end{split}
\end{equation}

We can derive the standard ingoing and outgoing orthonormal mode solutions of Eqn.\eqref{eq:31}: $\phi(\xi,r_{*})\sim(e^{-i\omega(\xi+r_{*})},e^{-i\omega(\xi-r_{*})})$. The particle can be suitably defined:the modes
are positive frequency modes with respect to the killing vector field $\frac{\partial}{\partial\xi}$ for $\omega>0$. Near the horizon, one only considers the outgoing modes $\phi(\xi,r_{*})=\frac{1}{\sqrt{4\pi\omega}}e^{-i\omega(\xi-r_{*})}$, so the massless scalar field near horizon can be quantized as

\begin{equation}\begin{split}
\label{eq:32}
\phi^{B}=\sum_{\omega}[a_{\omega}^{B}\phi(\xi,r_{*})+a_{\omega}^{B\dag}\phi(\xi,r_{*})]
\end{split}
\end{equation}

where $a_{\omega}^{B}$ and $a_{\omega}^{B\dag}$ are the the annihilation and creation operators acting on the $Boulware$ vacuum state. The $Fock$ vacuum state corresponds to
$a_{\omega}^{B}|0\rangle=0$. The $Wightman$ function of $Boulware$ vacuum state can be showen as

\begin{equation}\begin{split}
\label{eq:33}
G^{B+}(x,x')=-\frac{1}{4\pi^{2}}\frac{1}{(\Delta\xi-i\epsilon)^{2}}
\end{split}
\end{equation}

with the proper $i\epsilon$ prescription. Its Fourier transform with respect to the proper time $\mathscr{G}^{B+}(\omega)=0$. In fact, the $Boulware$ vacuum corresponds to our familiar notion of a vacuum state.

Now one can define the $Unruh$ vacuum state following the method as mentioned above. First of all, we write down the Kerr space time line element according to $Kruskal-like$ coordinates

\begin{equation}\begin{split}
\label{eq:34}
ds^{2}&=C(r)(-dT^{2}+dR^{2})
\end{split}
\end{equation}

where $T=\kappa^{-1}e^{\kappa r_{*}}\sinh\kappa\xi$ , $R=\kappa^{-1}e^{\kappa r_{*}}\cosh\kappa\xi$ and $\kappa=\frac{r_{+}-r_{-}}{2(r_{+}^{2}+a^{2})}$, and $C(r)=e^{-2\kappa r_{*}}F(r)$ is a finite constant near horizon. As seen from Eqn.\eqref{eq:34}, we can see that Eqn.\eqref{eq:34} is exactly conformal to $Minkowski$ metric form, hence, the scalar field equation reads

\begin{equation}\begin{split}
\label{eq:35}
(\partial_{T}^{2}-\partial_{R}^{2})\phi(T,R)=0
\end{split}
\end{equation}

\begin{figure}[htbp]
\centering
\subfigure[]{
\begin{minipage}{7cm}
\centering
\includegraphics[scale=0.5]{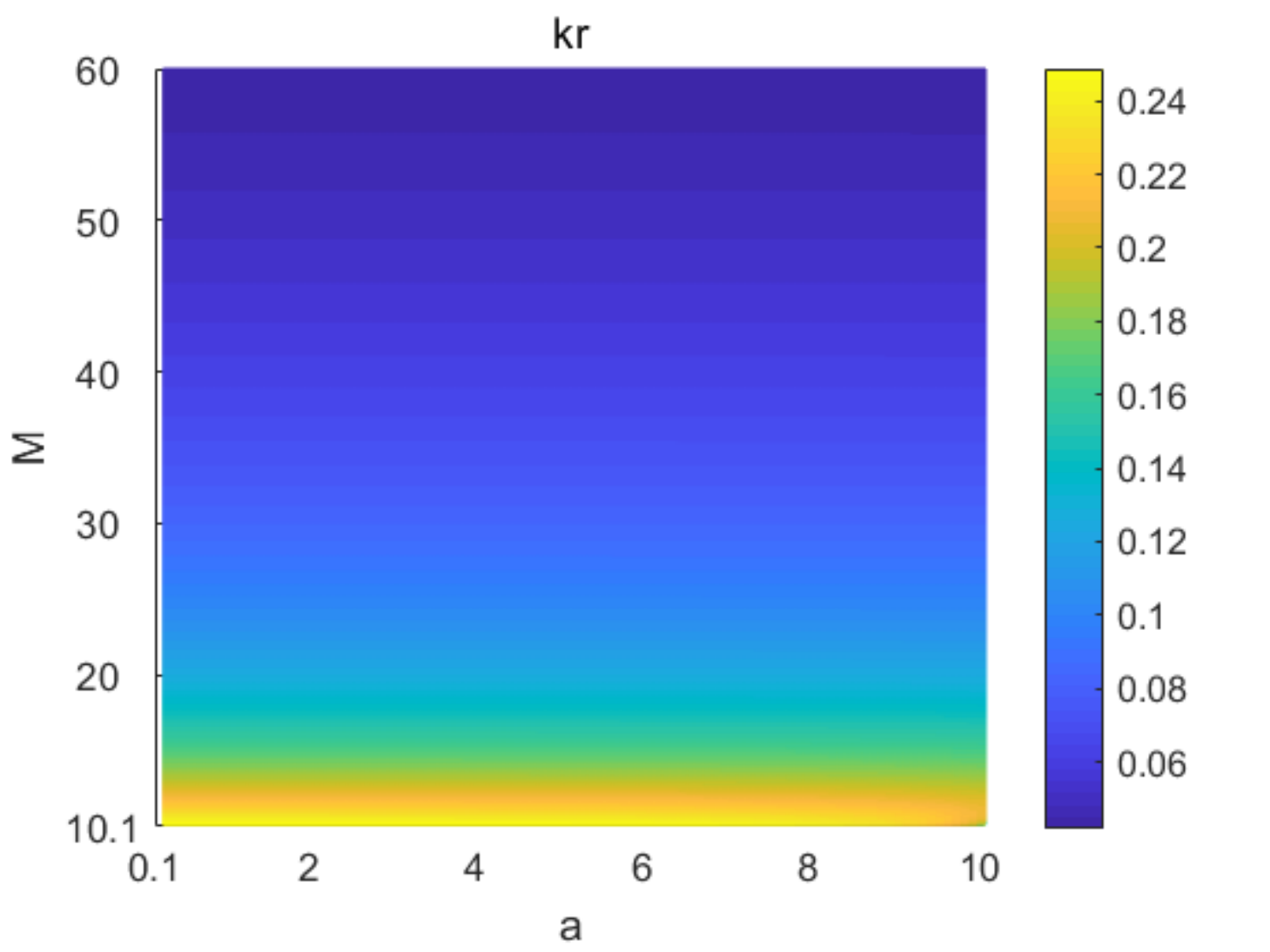}
\end{minipage}
}
\subfigure[]{
\begin{minipage}{7cm}
\centering
\includegraphics[scale=0.5]{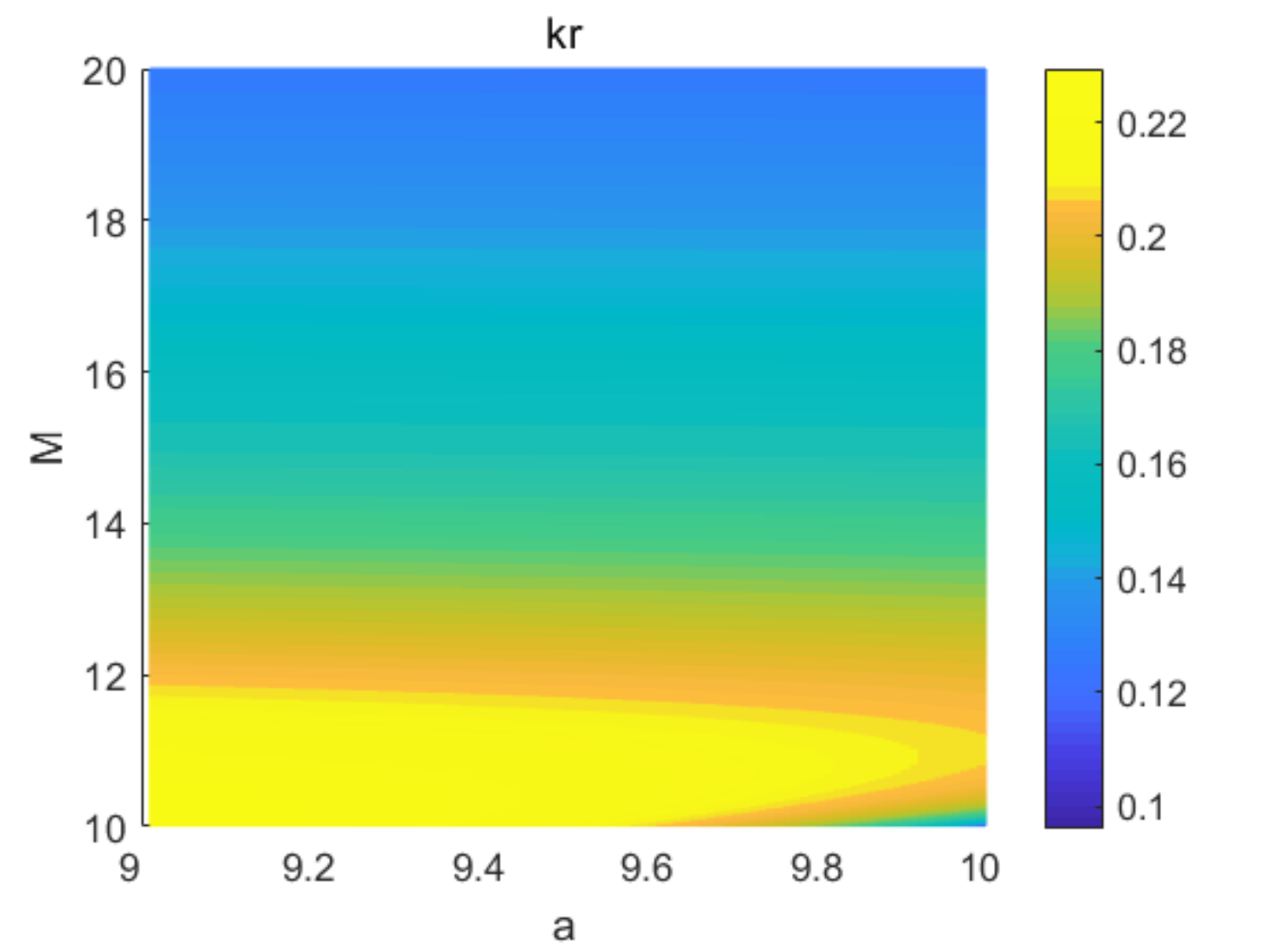}
\end{minipage}
}
\caption{Local acceleration $\kappa_{r}$ vs mass and angular momentum per mass. The system is located at 1.01$r_{+}$}
\label{fig:kr}
\end{figure}

Similar to the previous proceeding, we can derive the outgoing wave equation as $\phi(\xi,r_{*})\sim e^{-i\omega(T-R)}$. The particle can be suitably defined: the modes are positive frequency modes with respect to the killing vector field $\frac{\partial}{\partial T}$ for $\omega>0$. Near the horizon , we only consider the outgoing modes $\phi(T,R)=\frac{1}{\sqrt{4\pi\omega}}e^{-i\omega(T-R)}$, so the massless scalar field near horizon can be quantized as

\begin{equation}\begin{split}
\label{eq:36}
\phi^{U}=\sum_{\omega}[a_{\omega}^{U}\phi(T,R)+a_{\omega}^{U\dag}\phi(T,R)]
\end{split}
\end{equation}

where $a_{\omega}^{U}$ and $a_{\omega}^{U\dag}$ are the the annihilation and creation operators acting on the $Unruch$ vacuum state. The $Fock$ vacuum state corresponds to
$a_{\omega}^{U}|0\rangle=0$. The $Wightman$ function of $Unruch$ vacuum state can be shown as

\begin{equation}\begin{split}
\label{eq:37}
G^{U+}(x,x')&=-\frac{1}{4\pi^{2}}\frac{1}{(\Delta T-i\epsilon)^{2}-\Delta R^{2}}\\
&=-\frac{1}{16\pi^{2}\kappa^{2}\sinh^{2}[\frac{\Delta\xi}{2\kappa}-i\epsilon]}
\end{split}
\end{equation}

with the proper $i\epsilon$ prescription. Its Fourier transform with respect to the proper time is given as $\mathscr{G}^{U+}(\omega)=\frac{\omega}{2\pi}\frac{1}{1-e^{-2\pi\kappa_{r}^{-1}\omega}}$, where $\kappa_{r}=\frac{\kappa}{\sqrt{F(r)}}$. It is found that the detector in the $Unruh$ vacuum can spontaneously get excited with a nonvanishing probability, in the same way as the thermal radiation with Hawking temperature from a Kerr black hole. $Hawking-Unruh$ effect of a Kerr spacetime can be understood as a manifestation of thermalization behavior in an open quantum system.~\cite{Researching on Hawking Effect in a Kerr Space Time via Open Quantum System Approach} We will only consider the $Unruch$ vacuum in the following studies. In fact, the local $\kappa_{r}$ plays an essential role when we study the characteristics of the quantum correlations. This not only reflects the local curvature of the space time, but also embodies thermal nature of the black hole. In Fig.\ref{fig:kr}, at fixed angular momentum, the local acceleration $\kappa_{r}$ decreases to a steady value as the mass increases. The local acceleration only shows non-monotonic behavior when the mass is close to the angular momentum per mass. At fixed mass, the local acceleration keeps a steady value when the angular momentum per mass is less than and away from the mass and only significantly decreases when the angular momentum per mass is close to the mass.

The proper acceleration of the stationary detector near the horizon is divergent, therefore one can define a renormalized value termed as surface gravity. The surface gravity is generally the local proper acceleration multiplied by the gravitational time dilation factor (which goes to zero at the horizon). It corresponds to the Newtonian gravitational value in the non-relativistic limit. For a asymptotic observer, we can use Newtonian gravity to obtain the surface gravity at Schwarzschild black hole horizon $\kappa=\frac{M}{r^2}\mid_{r=r_0}=\frac{1}{4M}$.~\cite{Black Holes: An Introduction} For a $3+1$ dimensional asymptotically flat Kerr black hole with angular speed $\Omega_{+}=\frac{a}{r_{+}^{2}+a^2}$, one can use it to define an effective spring constant $k=M\Omega_{+}^{2}$. The surface gravity of Kerr black hole can be formulated as $\kappa_{Kerr}=\frac{1}{4M}-k$ which decreases when $M$ and angular speed $\Omega_{+}$ increase.~\cite{Are black holes springlike?} One can naively consider that the reduction of the Schwarzschild black hole surface gravity compensates as a centripetal force for the detector co-rotating with the black hole. The behaviors of $\kappa$ with respect to the mass and the angular momentum are very similar to those of $\kappa_{r}$ except that the $\kappa$ decreases monotonically with respect to the mass. In our derivation steps, the effective surface gravity is changed as $\kappa_{r}=\frac{\kappa}{\sqrt{F(r)}}$ due to the use of dimensional reduction. Finally it leads to non-trivial behaviors of $\kappa_{r}$ on the mass and the angular momentum. %The behaviors of differences between $\kappa_{r}$ and $\kappa$ with respect to the mass are also caused by the dimensional reduction.

\section{Equilibrium quantum correlations in curved space time}
\label{sec:Equilibrium quantum correlations in curved space time}

We consider that both two detectors are coupled to the field. Generalizing Eqn.\eqref{eq:9} from one atom to two atoms, generalizing free hamiltonian as $H_{0}=\frac{\omega}{2}\vec{n}\cdot\vec{\sigma}$. The interaction hamiltonian becomes $H_{I}=\sum_{\mu=0}^{3}[\sigma_{\mu}^{\alpha}\Phi_{\mu}(x_{\alpha})]$ where $\Phi_{\mu}(x)=\sum_{i=1}^{N}\chi_{\mu}^{i}\phi^{-}(x)+\chi_{\mu}^{i*}\phi^{+}(x)$.  $\phi^{\pm}(x)$ are positive and negative energy field operators of the massless scalar field, and $\chi$ are the corresponding complex coefficients. After assuming $\sum_{i=1}^{N}\chi_{\mu}^{i}\chi_{\nu}^{i*}=\delta_{\mu\nu}$ the master equation reads~\cite{Controlling entanglement generation in external quantum fields}~\cite{Entanglement generation in uniformly accelerating atoms}

\begin{equation}\begin{split}
\label{eq:38}
\frac{d\rho_{sys}(\tau)}{d\tau}&=-i[H_{eff},\rho_{sys}(\tau)]+\mathscr{L}[\rho_{sys}(\tau)]\\
H_{eff}&=\frac{\omega}{2}\sum_{i=1}^{3}\sum_{\alpha=1}^{2}n_{i}^{\alpha}\sigma_{i}^{\alpha}-\frac{i}{2}\sum_{\alpha,\beta=1}^{2}\sum_{i,j=1}^{3}H_{ij}^{\alpha\beta}\sigma_{i}^{\alpha}\sigma_{j}^{\beta}\\
\mathscr{L}[\rho_{sys}]&=\frac{1}{2}\sum_{\alpha,\beta=1}^{2}\sum_{i,j=1}^{3}\mathscr{C}_{ij}^{\alpha\beta}[\sigma_{j}^{\beta}\rho_{sys}\sigma_{i}^{\alpha}-\{\sigma_{i}^{\alpha}\sigma_{j}^{\beta},\rho_{sys}\}]
\end{split}
\end{equation}

where $\sigma_{i}^{1}=\sigma_{i}\otimes\sigma_{0}$ and $\sigma_{i}^{2}=\sigma_{0}\otimes\sigma_{i}$. The $GSKL$ matrix $\mathscr{C}_{ij}^{\alpha\beta}$ is given by the following expression

\begin{equation}\begin{split}
\label{eq:39}
\mathscr{C}_{ij}^{\alpha\beta}=A^{\alpha\beta}\delta_{ij}-iB^{\alpha\beta}\epsilon_{ijk}n_{k}+C^{\alpha\beta}n_{i}n_{j}
\end{split}
\end{equation}

where the $A^{\alpha\beta}$ and $B^{\alpha\beta}$ and $C^{\alpha\beta}$ for the two atomic system are defined as:

\begin{equation}\begin{split}
\label{eq:40}
A^{\alpha\beta}&=\frac{\mu^{2}}{4}(G^{\alpha\beta}(\omega)+G^{\alpha\beta}(-\omega))\\
B^{\alpha\beta}&=\frac{\mu^{2}}{4}(G^{\alpha\beta}(\omega)-G^{\alpha\beta}(-\omega))
\end{split}
\end{equation}

and $C^{\alpha\beta}$ is given as $G(0)-A^{\alpha\beta}$.
Similarly, the coefficients of $H_{ij}^{\alpha\beta}$ can be obtained by replacing $G^{\alpha\beta}(\omega)$ with $K^{\alpha\beta}(\omega)$ in the above equations where $K^{\alpha\beta}(\lambda)=\frac{P}{i\pi}\int^{\infty}_{-\infty}\frac{G^{\alpha\beta}(\omega)}{\omega-\lambda} d\omega$.~\cite{Controlling entanglement generation in external quantum fields} In the following we set $\mu_{\alpha}=\mu_{\beta}=0.01$.

These results for the Hamiltonian contributions require some further comments. The $K^{11}$ can be splitted (similar results hold also for $K^{12}$):

\begin{equation}\begin{split}
\label{eq:42}
K^{11}(\lambda)=\frac{1}{2\pi^{2}i}[P\int_{0}^{\infty}d\omega\frac{\omega}{\omega-\lambda}+P\int_{0}^{\infty}d\omega\frac{\omega}{1-e^{2\pi\kappa_{r}^{-1}\omega}}(\frac{1}{\omega+\lambda}-\frac{1}{\omega-\lambda})]
\end{split}
\end{equation}

 to a flat and a curvature-dependent piece. Although we do not calculate the above function concretely, the curvature dependent second term is a finite, odd function of $\lambda$, vanishing as $\kappa_{r}$  becomes less. The first contribution in Eqn.\eqref{eq:42} is however divergent. Despite some cancellations that occur in $H_{eff}$, the effective Hamiltonian turns out in general to be infinite, and its definition requires the introduction of a suitable cutoff and a renormalization procedure. The appearance of divergences comes from the non-relativistic treatment of the two-level atoms, while any reasonable calculation of energy shifts would have required the quantum field theory approaches. In our quantum mechanical setting, the procedure needed to make $H_{eff}$ well defined is therefore clear: perform a suitable curvature independent subtraction, so that $H_{eff}$ reproduces the correct quantum field theory result. However, since we are interested in analyzing the effects due to the curvature, we do not need to do this explicitly. In the following we only consider the correlations induced by the curvature effect,~\cite{Controlling entanglement generation in external quantum fields}~\cite{Entanglement generation in uniformly accelerating atoms}~\cite{Entanglement generation outside a Schwarzschild black hole and the Hawking effect}, by disregarding the Hamiltonian contribution in Eqn.\eqref{eq:9} and only concentrate on the study of the effects induced by the dissipative part.
We consider a situation that there is no real distance between two qubits, which means the system in an equilibrium common environment. The presence of an equilibrium state $\rho^{\infty}$ can be in general determined by setting $\mathscr{L}[\rho_{sys}(\tau)]=0$. Consider a general density matrix of the two-atom system in the form of $\rho(\tau)=\frac{1}{4}[\textbf{1}\otimes\textbf{1}+\rho_{0i}(\tau)\sigma_{0}\otimes\sigma_{i}+\rho_{i0}(\tau)\sigma_{i}\otimes\sigma_{0}+\rho_{ij}(\tau)\sigma_{i}\otimes\sigma_{j}]$, inserting it into $\mathscr{L}[\rho_{sys}(\tau)]=0$, one derives the following result~\cite{Controlling entanglement generation in external quantum fields}:

\begin{equation}\begin{split}
\label{eq:43}
\rho^{\infty}_{0i}&=\rho^{\infty}_{i0}=-\frac{R}{3+R^2}(\tau_{*}+3)n_{i}\\
\rho^{\infty}_{ij}&=\frac{1}{3+R^2}[(\tau_{*}-R^2)\delta_{ij}+R^{2}(\tau_{*}+3)n_{i}n_{j}]
\end{split}
\end{equation}

where $R = B/A$, $\tau_{*}$ is the trace of the density matrix $\tau_{*}=\sum_{i=1}^{3}\rho_{ii}$, which is actually a constant of motion, and the positivity of $\rho(0)$ requires that $-3 \leq \tau_{*} \leq1$. At the initial state, consider the direct product of two pure states: $\rho(0)=\rho_{\vec{a}}\otimes\rho_{\vec{b}}$, where $\rho_{\vec{a}}=\frac{1}{2}(\textbf{1}+\vec{a}\cdot\vec{\sigma})$, $\rho_{\vec{b}}=\frac{1}{2}(\textbf{1}+\vec{b}\cdot\vec{\sigma})$, and $\vec{a}$ and $\vec{b}$ are two unit vectors. In this case, one easily finds that $\tau=\vec{a}\cdot\vec{b}$. In this paper, we set $\vec{a}=(0,0,1)$ and $\vec{b}=(0,0,-1)$, the hamiltonian for the single atom is $H_{0}=\frac{\omega}{2}\sigma_{z}$.

The system finally reaches the equilibrium steady state since the two subsystems are at the same temperature. More concretely speaking, the space time curvature leads to a Hawking temperature due to the $Unruch$ effect, i.e., there is a common thermal bath around the two-qubits system which is near the black hole horizon and the temperature depends on the curvature.

\begin{figure}[htbp]
\centering
\subfigure[]{
\begin{minipage}{7cm}
\centering
\includegraphics[scale=0.5]{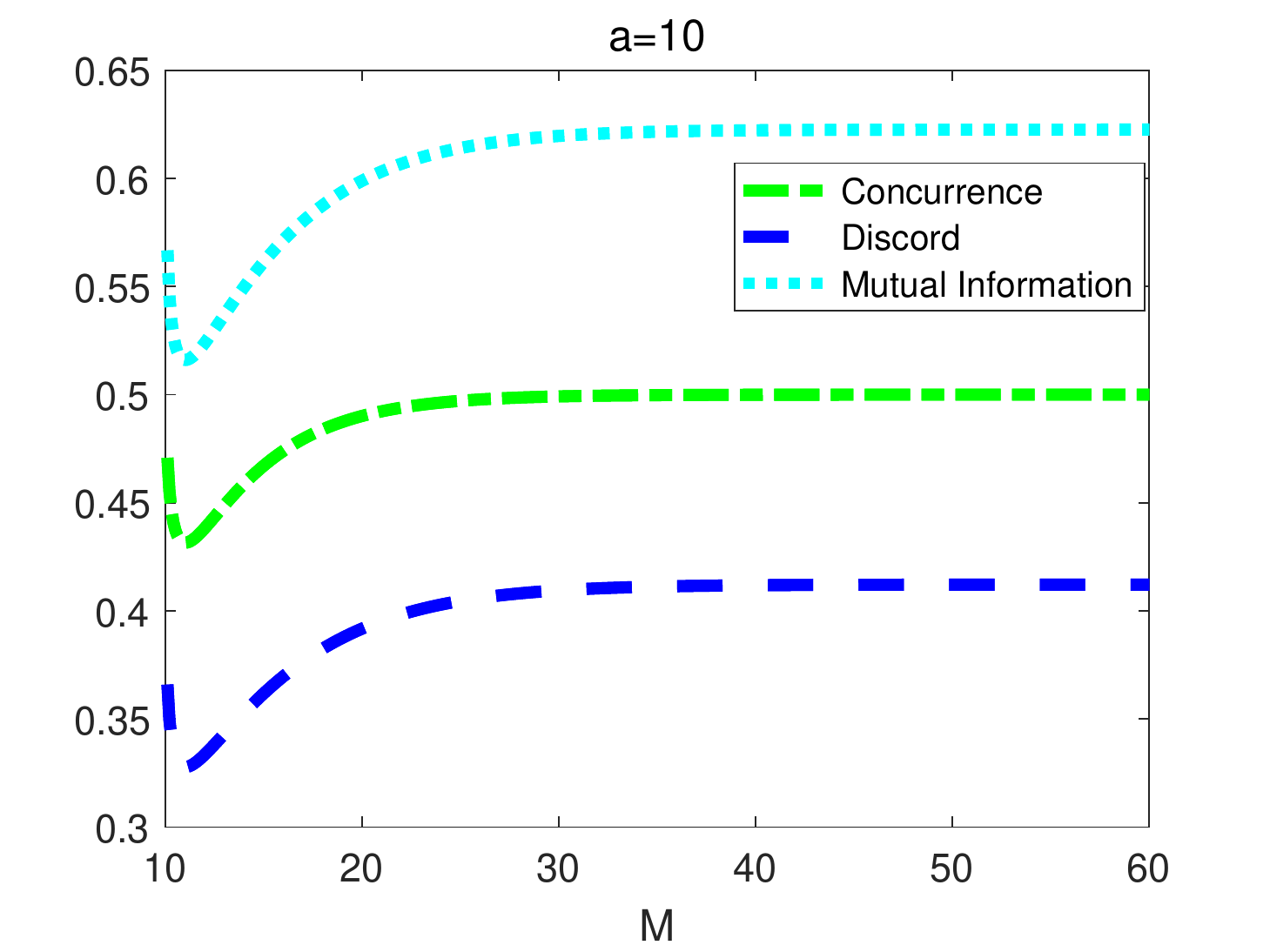}
\end{minipage}
}
\subfigure[]{
\begin{minipage}{7cm}\centering
\includegraphics[scale=0.5]{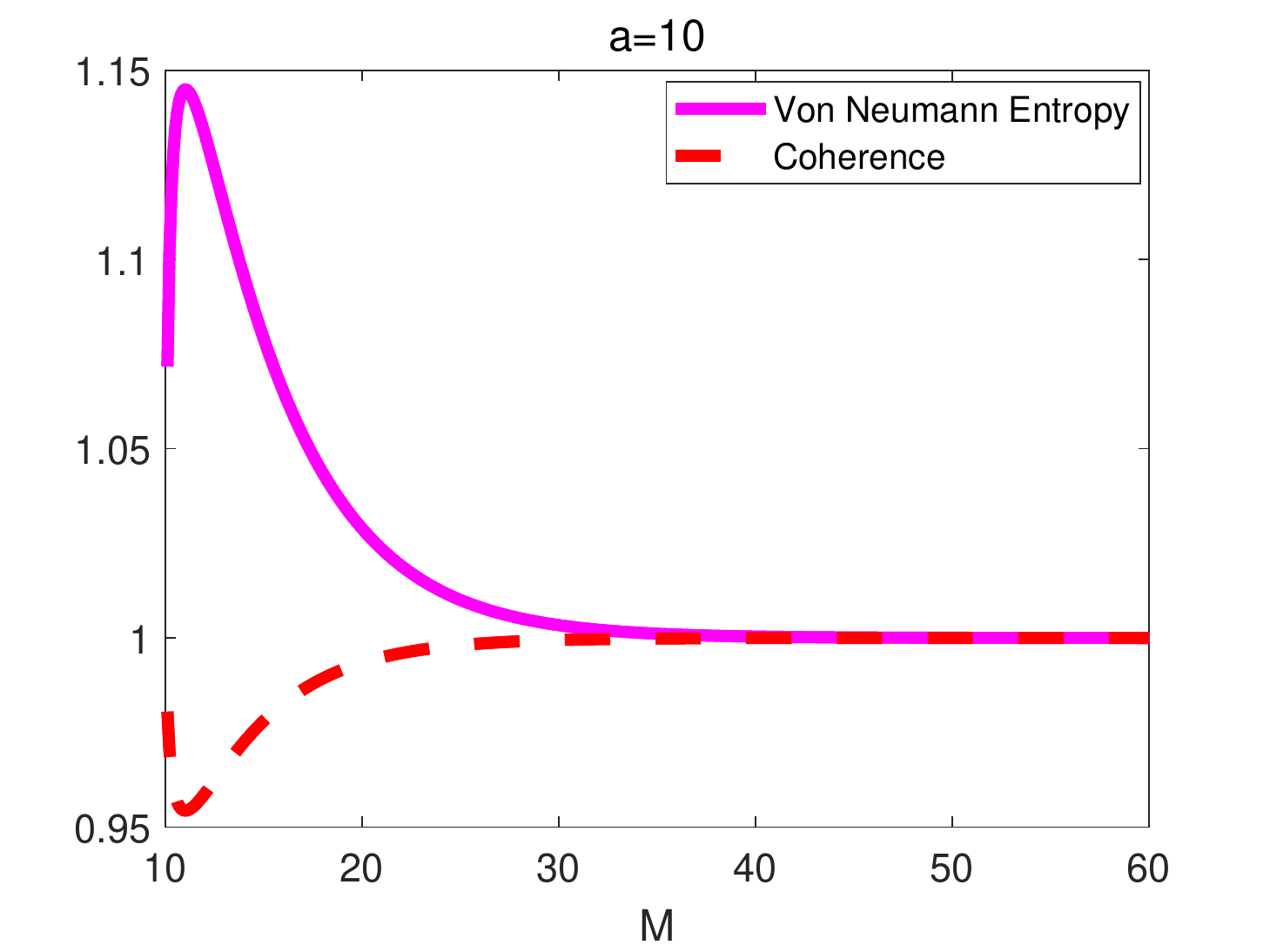}
\end{minipage}
}
\subfigure[]{
\begin{minipage}{7cm}\centering
\includegraphics[scale=0.5]{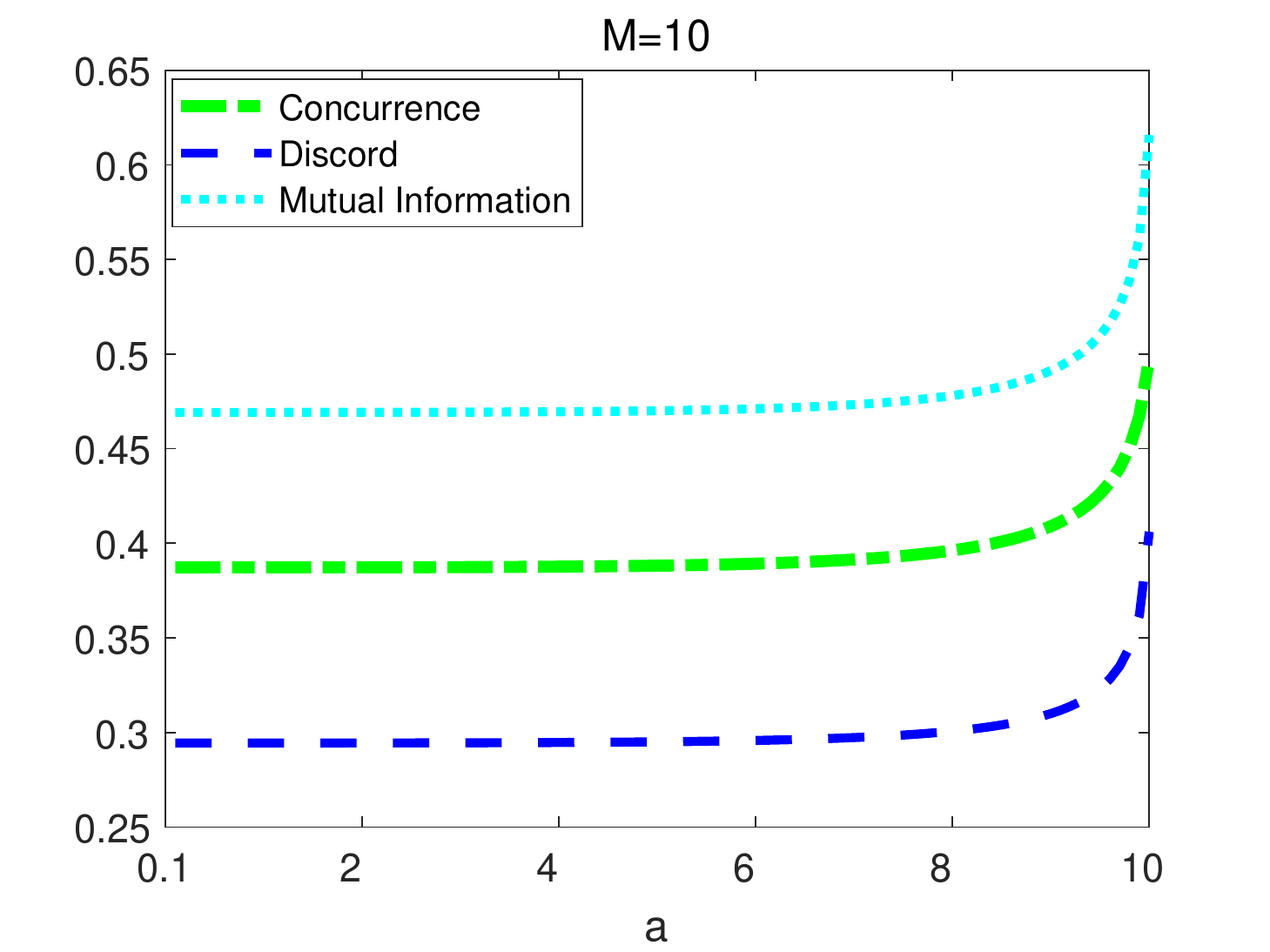}
\end{minipage}
}
\subfigure[]{
\begin{minipage}{7cm}\centering
\includegraphics[scale=0.5]{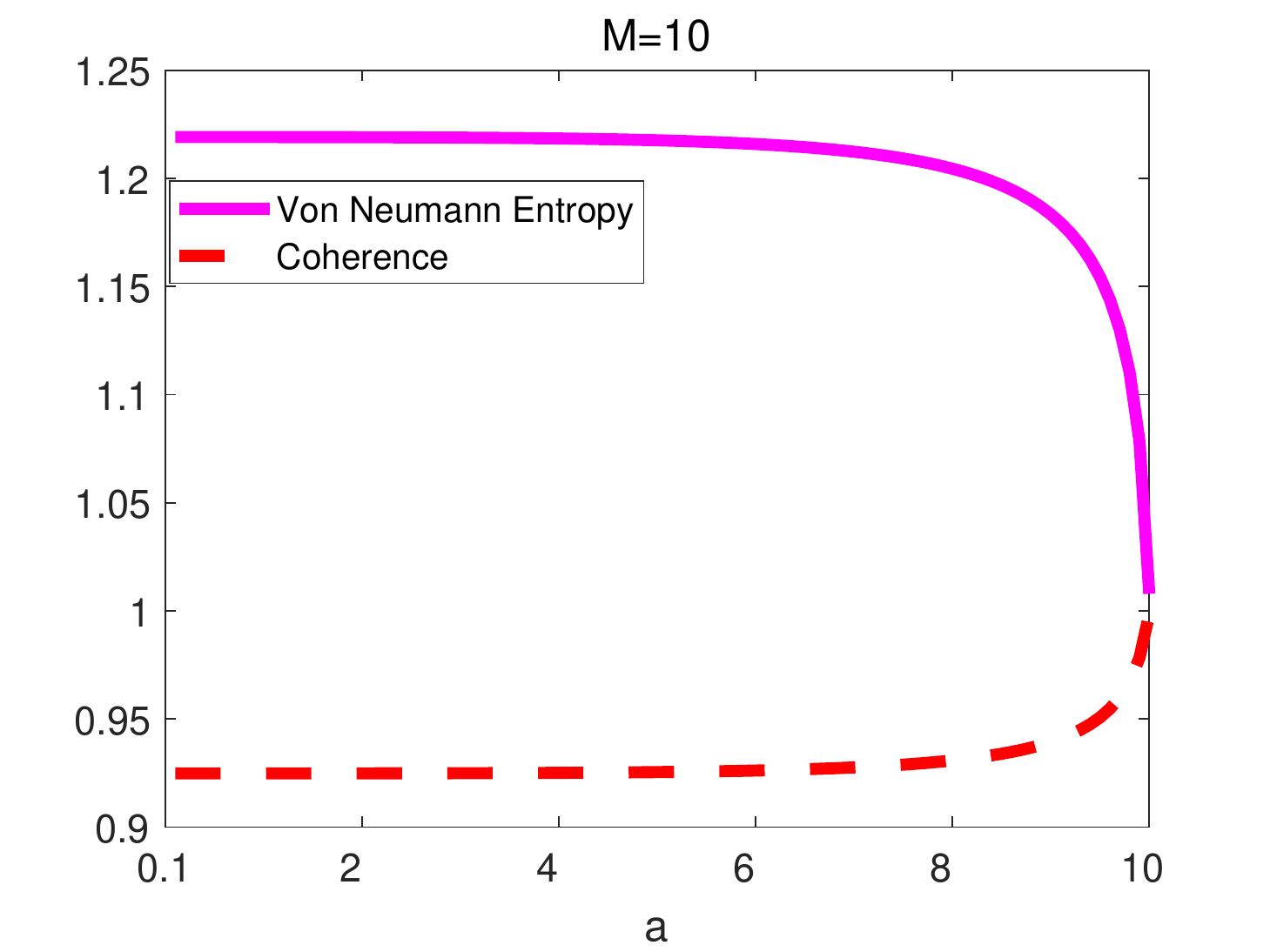}
\end{minipage}
}
\caption{Quantum correlations at equilibrium state vs mass or angular momentum per mass. The system is located at 1.01$r_{+}$, the angular momentum per mass $a$ is set up 10 and the mass is changed from 10 in Fig.\ref{fig:1}(a)(b). The system is located at 1.01$r_{+}$, the mass $a$ is set up as 10.01 and the angular momentum per mass is changed from 0.1 in Fig.\ref{fig:1}(c)(d). $\tau=-1$. The eigenfrequencies of the qubits are $\omega_{1}=\omega_{2}=0.1$}
\label{fig:1}
\end{figure}

On the above basis, we study the quantum correlations of two-qubit system near the $Kerr$ black hole with a global corotating coordinate and as shown in Fig.\ref{fig:1}.

In our setting, the initial state is a separable state, i.e. there is no any quantum correlation initially. After evolution, it has been shown that the system reaches a steady state. More remarkably, the system harvests the quantum correlation from the $Unruch$ vacuum in Fig.\ref{fig:1}, consistent with the suggestions made in Ref~\cite{Thermal amplification of field-correlation harvesting}~\cite{Equilibrium and Nonequilibrium Quantum Correlations Between Two Accelerated Detectors}~\cite{Entanglement generation outside a Schwarzschild black hole and the Hawking effect}. The initial state of the system is a pure state where the Von Neumann entropy vanishes and the final state is a mixed state which the Von Neumann entropy is nonvanishing. Apparently, the system has experienced a nonequilibrium process and finally forms a steady state with quantum correlations under the parameters of the black hole. We focus on this steady state. The quantum correlation is derived from nonunitary evolution which is caused by the interaction between the field and the system. In the Fig.\ref{fig:1}(a)(b), the concurrence, mutual information, discord and coherence all decrease first and then increase to a constant as the mass of black hole increases while the angular momentum per mass is unchanged. On the contrary, the Von Neumann entropy of the system which measures the entanglement between the system and the environment increases initially and then decreases to a constant. For the black hole with larger mass, the quantum correlations are neither more sensitive to the change of the angular momentum per mass, nor to the mass. The angular momentum per mass has significant effect on the quantum correlations only when it is comparable to the mass. This is also demonstrated in Fig.\ref{fig:1}(c)(d). The quantum correlations are boosted by the angular momentum per mass except that the Von Neumann entropy decreases.

\begin{figure}[htbp]
\centering
\subfigure[]{
\begin{minipage}{7cm}
\centering
\includegraphics[scale=0.5]{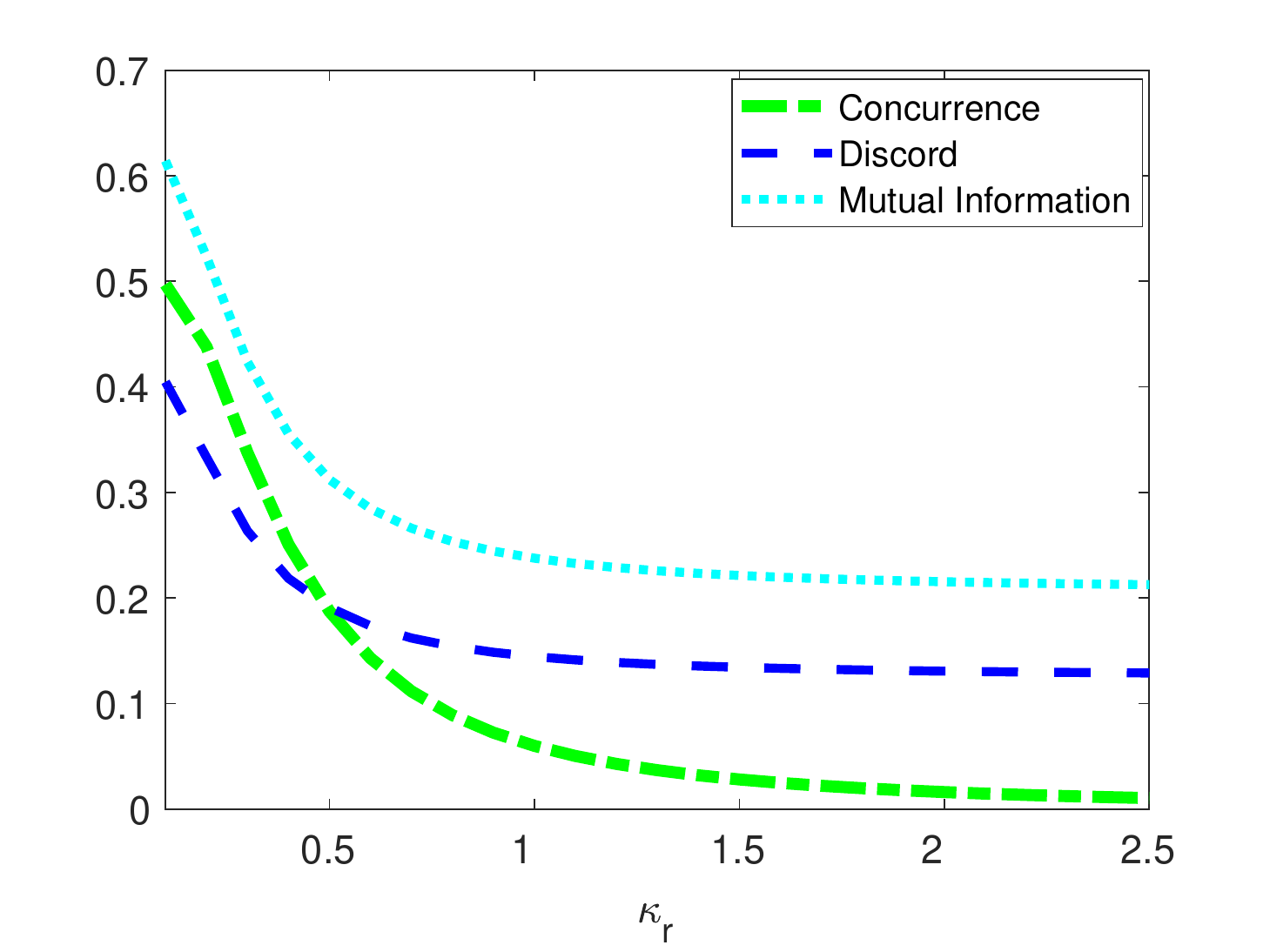}
\end{minipage}
}
\subfigure[]{
\begin{minipage}{7cm}\centering
\includegraphics[scale=0.5]{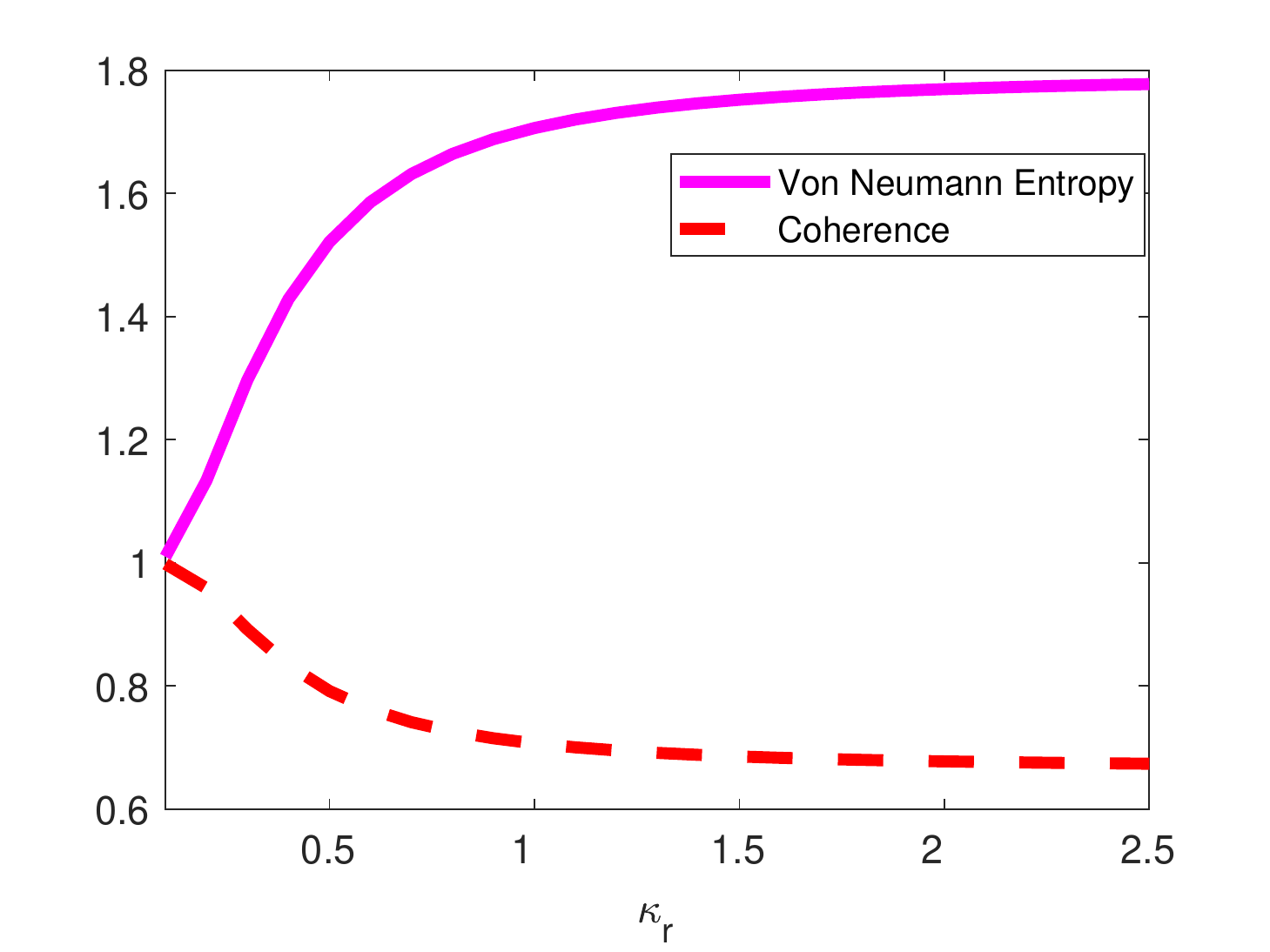}
\end{minipage}
}
\caption{Quantum correlations at equilibrium state vs $\kappa_{r}$. Other parameters are the same as those of the Fig.\ref{fig:1}}
\label{fig:2}
\end{figure}

The above non-trivial result comes from the fact that the dependence of the quantum correlations on the local acceleration $\kappa_{r}$. The $\kappa_{r}$ is directly related to the curvature of space time. All the quantum information between the two qubits are reduced by larger curvature. On the contrary the Von Neumann entropy which measures the entanglement between the system and the environment increases as shown in Fig.\ref{fig:2}. At fixed angular momentum, the $\kappa_{r}$ decreases to a steady value as the mass increases, and only shows non-monotonously behavior when the mass is close to the angular momentum per mass. At fixed mass, it keeps a steady value when the angular momentum per mass is away and less than mass and only significantly decreases when the angular momentum per mass is close to the mass. Thus the behavior of the $\kappa_{r}$ with respect to the mass and angular momentum determines the behavior of the correlations. Moreover, the $\kappa_{r}$ appears to be inversely related to the quantum correlations in the system and similarly to the system-environment entanglement. The larger $\kappa_{r}$ correspond to the higher space time curvature and higher effective temperature. This makes the system more classical and weakens the quantum nature. On the contrary, the higher temperature leads to strengthening of the interaction between the system and the field. Therefore this makes more easily for them to correlate. Here we see the impact of the space time structure on the quantum correlations.

\section{Nonequilibrium transient quantum correlations in curved space time}
\label{sec:Nonequilibrium transient quantum correlations in curved space time}

\begin{figure}
\centering
\includegraphics[scale=0.45]{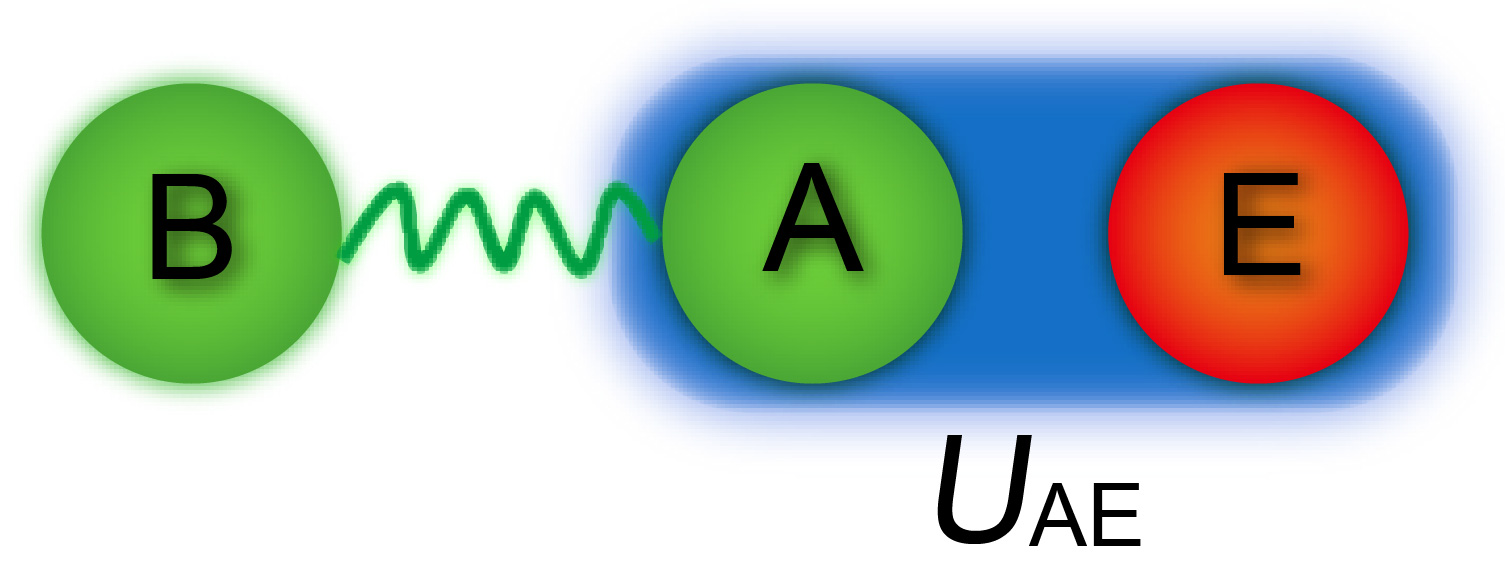}
\caption{schematic diagram}
\label{fig:3}
\end{figure}

To study the quantum correlation in the Kerr black hole space time, we introduce an auxiliary system (the same two-level atom) which is isolated from the environment, meaning the coupling between the atom and environment is extremely weak.~\cite{Loss of Spin Entanglement For Accelerated Electrons in Electric and Magnetic Fields} ~\cite{Dynamics and quantum entanglement of two-level atoms in de Sitter spacetime} The schematic diagram is shown in Fig.\ref{fig:3}. The qubit coupled to the field is called A while the qubit free of the field coupling is called B. The field is respresented by E. Initially, A and B is maximally entangled, only the A interacts with E. After initial time, the initial quantum correlations between A and B is expected to transferred to the correlations between A and E. We can expand any general density matrix for the bipartite two-level atom system as follows

\begin{equation}\begin{split}
\label{eq:44}
\rho=\sum_{i,j=0}^{3}\rho_{ij}\sigma_{i}\bigotimes\sigma_{j}
\end{split}
\end{equation}

where $\{\sigma_{i}\bigotimes\sigma_{j}|i,j \in 0,...,3\}$ forms sixteen linearly independent complete vector basis and we choose

\begin{equation}
\label{eq:45}
\sigma_{1}=\left(
  \begin{array}{cc}
    0 & 1\\
    1 & 0 \\
  \end{array}
\right)
\sigma_{2}=\left(
  \begin{array}{cc}
    0 & -i\\
    i & 0 \\
  \end{array}
\right)
\sigma_{3}=\left(
  \begin{array}{cc}
    1 & 0\\
    0 & -1\\
  \end{array}
\right)
\end{equation}

and $\sigma_{0}=\textbf{1}$. A good property about this choice of basis is that the expansion coefficients $\rho_{ij}$ are real and satisfy $\rho^{\dag}=\rho$ and $Tr\rho=1$. Furthermore the expansion coefficients can be computed directly using $\rho_{ij}=Tr\{\rho\sigma_{i}\bigotimes\sigma_{j}\}$.

Substituting Eqn.\eqref{eq:44} into Eqn.\eqref{eq:9}, we derive

\begin{equation}\begin{split}
\label{eq:46}
\sum_{i,j=0}^{3}\frac{d\rho_{ij}(\tau)}{d\tau}\sigma_{i}\bigotimes\sigma_{j}&=-i[H_{eff},\sum_{i,j=0}^{3}\rho_{ij}(\tau)\sigma_{i}\bigotimes\sigma_{j}]\\
&+\sum_{i.j=0}^{3}\sum_{m=1}^{3}\rho_{ij}[2L_{m}\sigma_{i}L_{m}^{\dag}-L_{m}^{\dag}L_{m}\sigma_{i}-\sigma_{i}L_{m}^{\dag}L_{m}]\bigotimes\sigma_{j}
\end{split}
\end{equation}

%We can understanding Eqn.\eqref{eq:46} from the follow dynamics viewpoint.%
We comment more about the above equation. The auxiliary atom is isolated from the environment meaning the environment has no interaction or dissipative effect on it. During the evolution, the system and the environment exchange the energy and the information. In the transient process, the interior of system is unbalanced.

From Eqn.\eqref{eq:42}, we can derive the time dependent density matrix elements

\begin{equation}\begin{split}
\label{eq:47}
\rho_{0j}(\tau)&=\rho_{0j}(0),\\
\rho_{1j}(\tau)&=\rho_{1j}(0)e^{-\frac{A\tau}{2}}\cos(\Omega\tau)-\rho_{2j}(0)e^{-\frac{A\tau}{2}}\sin(\Omega\tau),\\
\rho_{2j}(\tau)&=\rho_{1j}(0)e^{-\frac{A\tau}{2}}\sin(\Omega\tau)+\rho_{2j}(0)e^{-\frac{A\tau}{2}}\cos(\Omega\tau),\\
\rho_{3j}(\tau)&=\rho_{1j}(0)e^{-A\tau}-\frac{B}{A}\rho_{0j}(0)(1-e^{-A\tau}).
\end{split}
\end{equation}

where $A=\gamma_{+}+\gamma_{-}$ and $B=\gamma_{+}-\gamma_{-}$. In the following we consider that two atoms initially share a maximally entangled state, i.e, $\rho_{00}=\rho_{11}=-\rho_{22}=\rho_{33}=\frac{1}{4}$, while the rest of $\rho_{ij}$ vanish.

\begin{figure}[htbp]
\centering
\subfigure[]{
\begin{minipage}{7cm}
\centering
\includegraphics[scale=0.5]{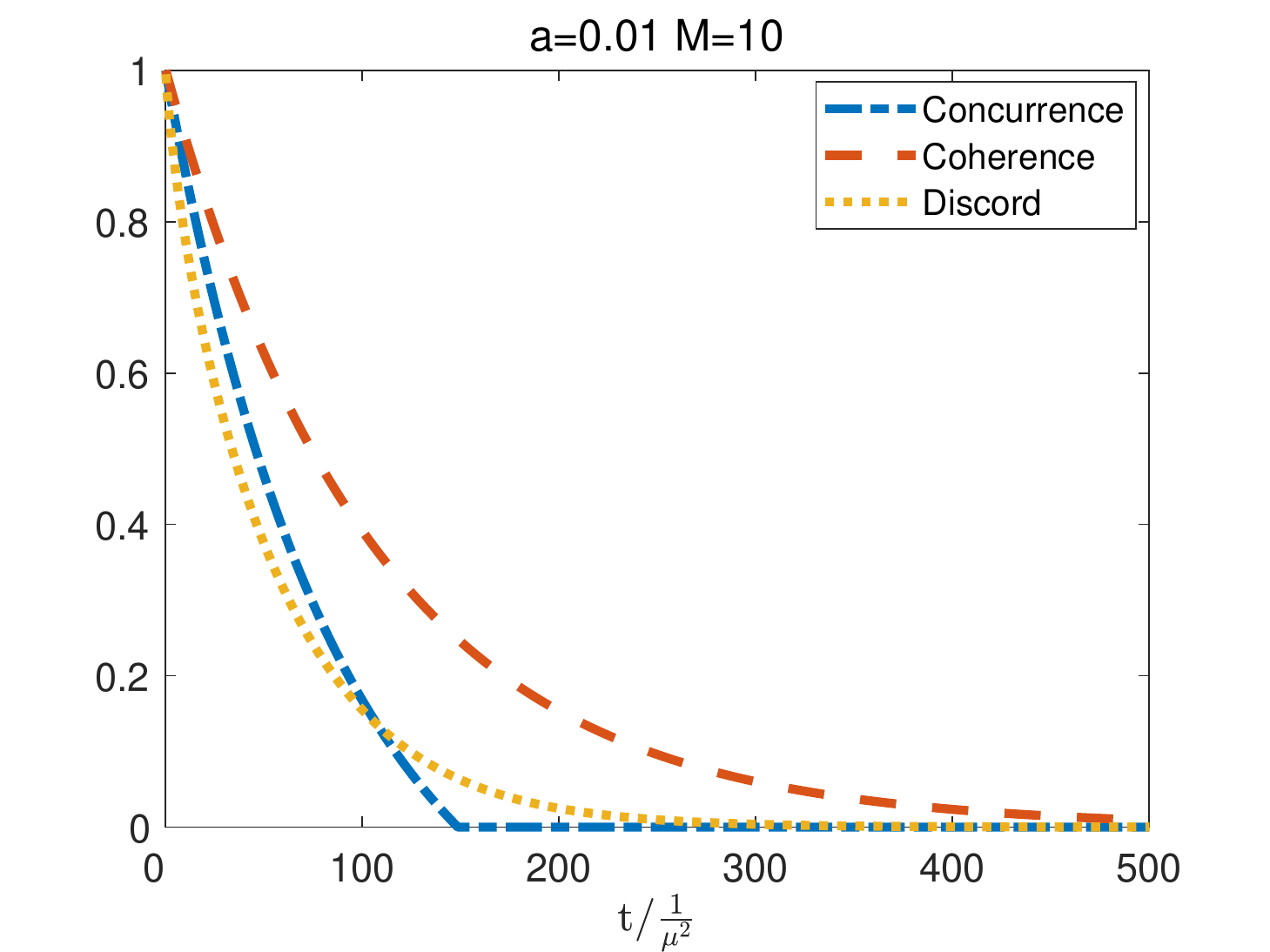}
\end{minipage}
}
\subfigure[]{
\begin{minipage}{7cm}\centering
\includegraphics[scale=0.5]{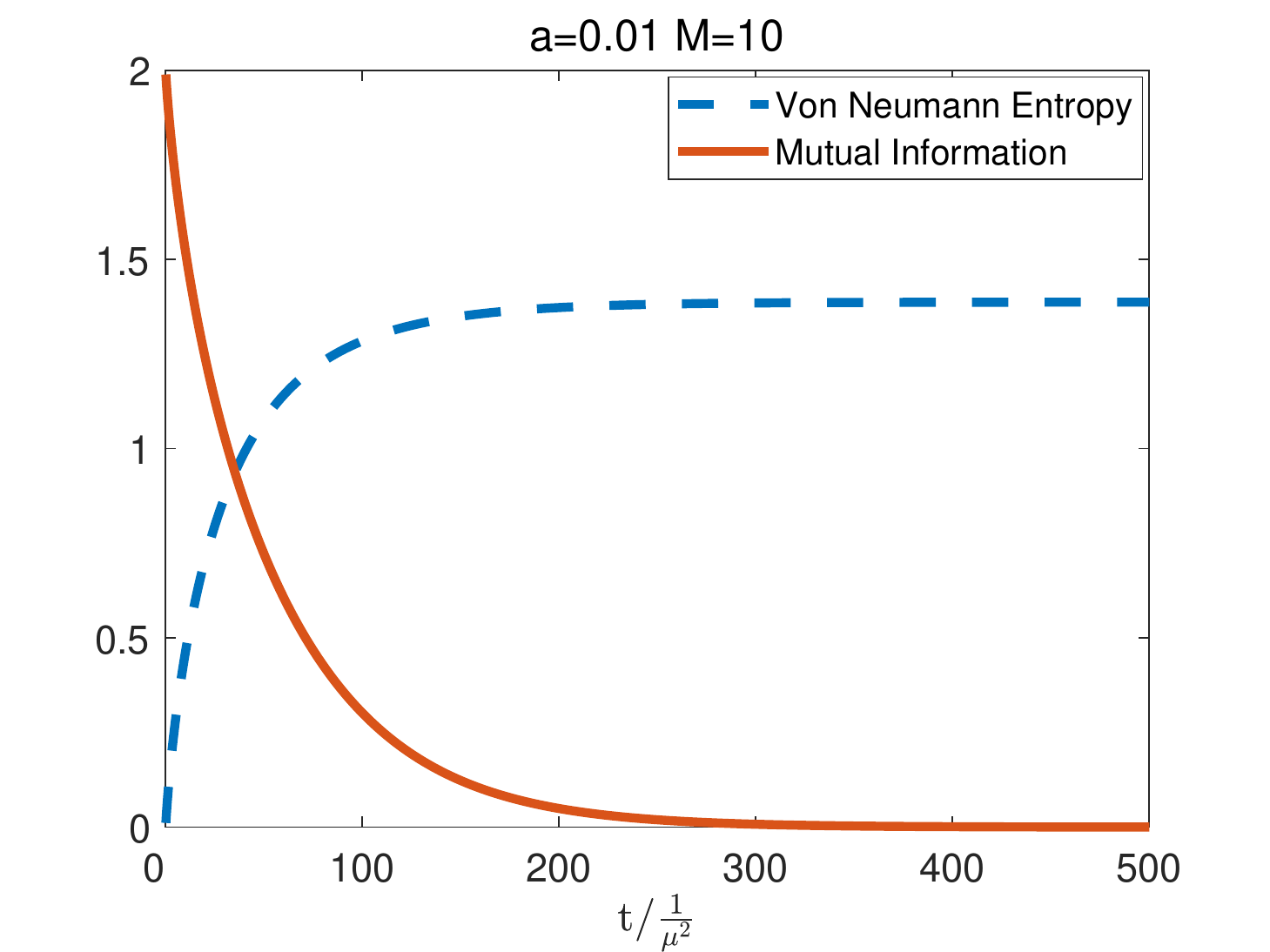}
\end{minipage}
}
\subfigure[]{
\begin{minipage}{7cm}\centering
\includegraphics[scale=0.5]{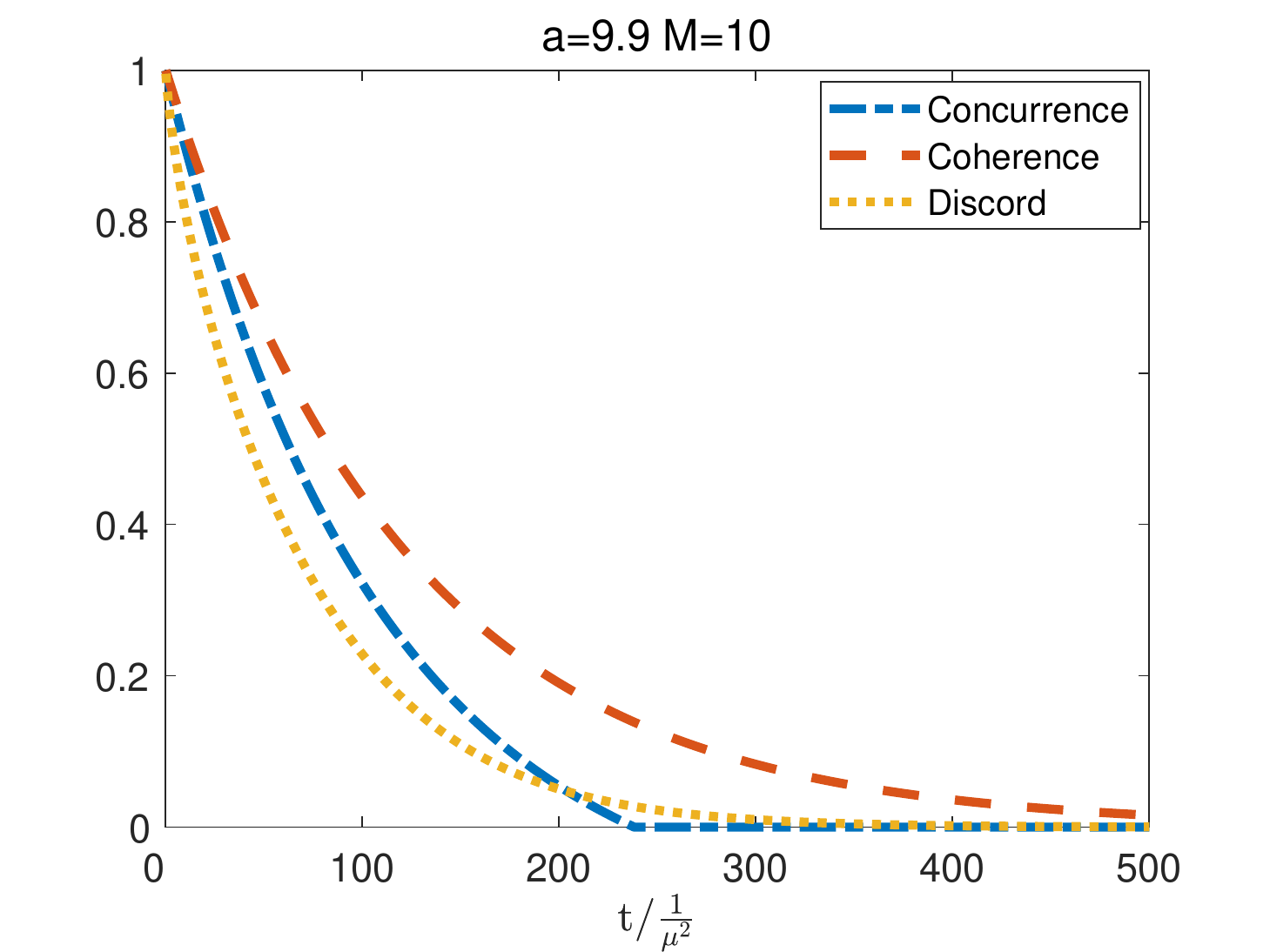}
\end{minipage}
}
\subfigure[]{
\begin{minipage}{7cm}\centering
\includegraphics[scale=0.5]{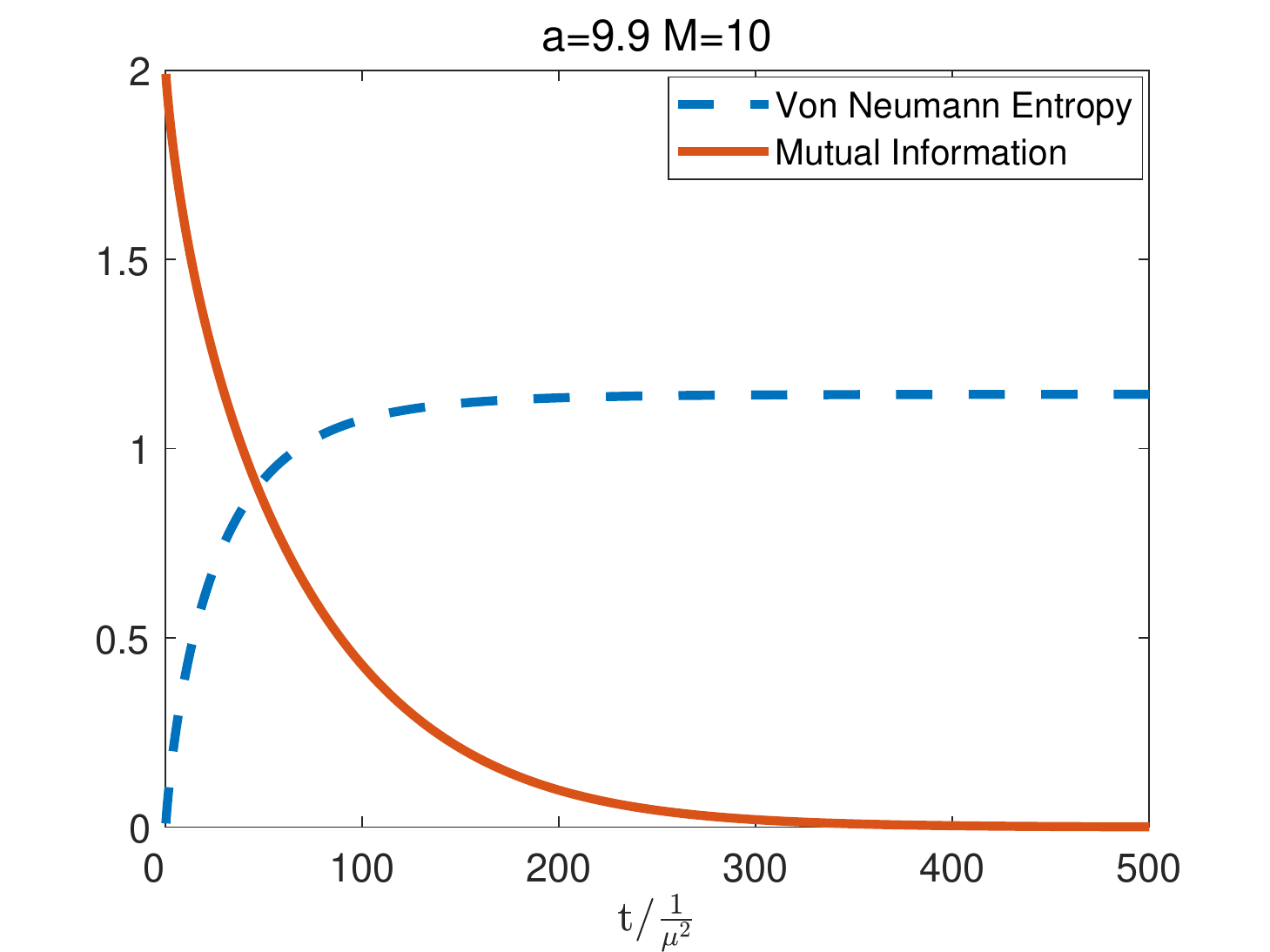}
\end{minipage}
}
\caption{Nonequilibrium quantum correlation evolutions vs time with the different angular momentum per mass in the unit of $\frac{1}{\mu^{2}}$. Other parameters are the same as those of the Fig.\ref{fig:1}}
\label{fig:4}
\end{figure}

\begin{figure}[htbp]
\centering
\subfigure[]{
\begin{minipage}{7cm}
\centering
\includegraphics[scale=0.5]{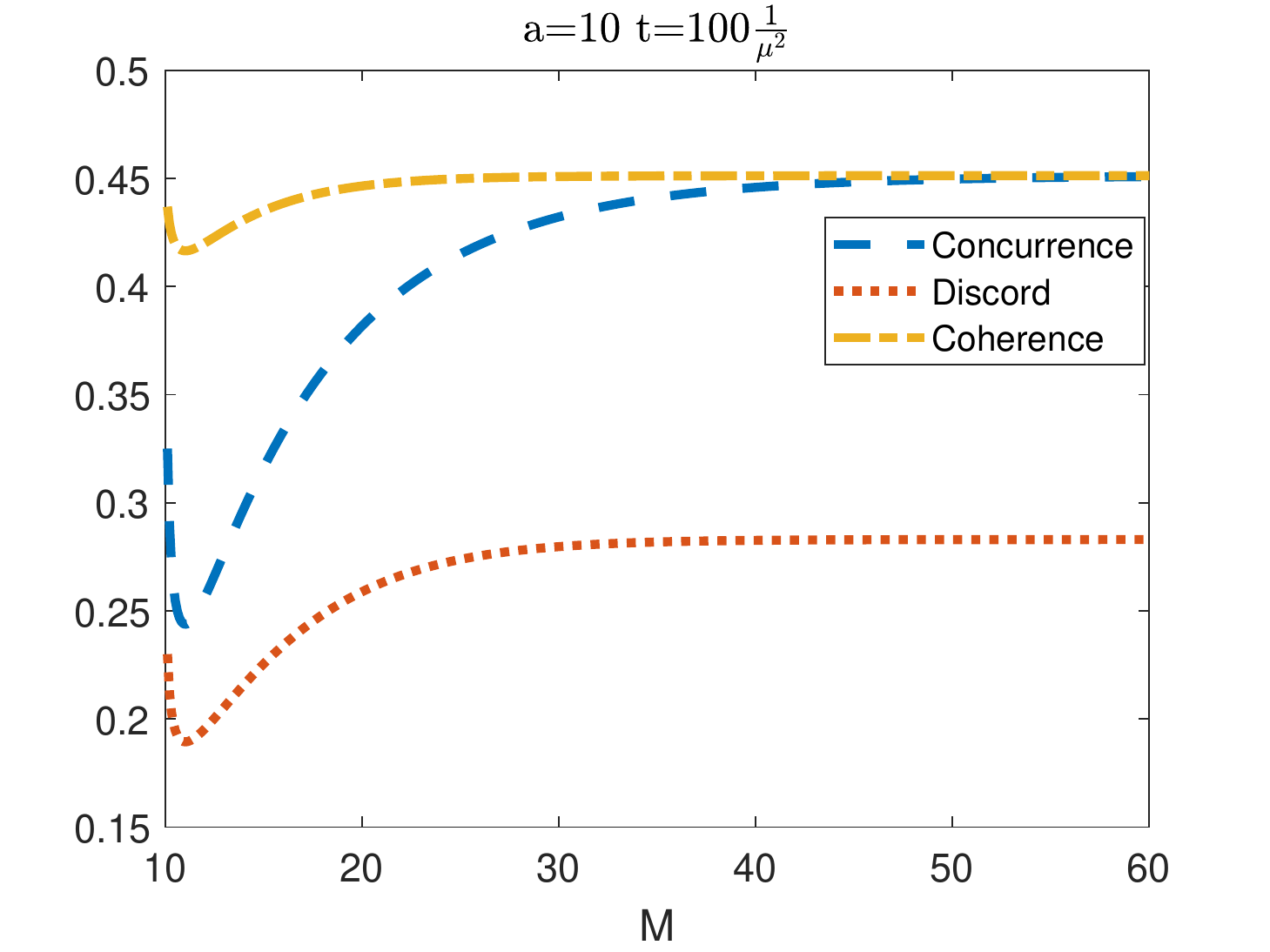}
\end{minipage}
}
\subfigure[]{
\begin{minipage}{7cm}\centering
\includegraphics[scale=0.5]{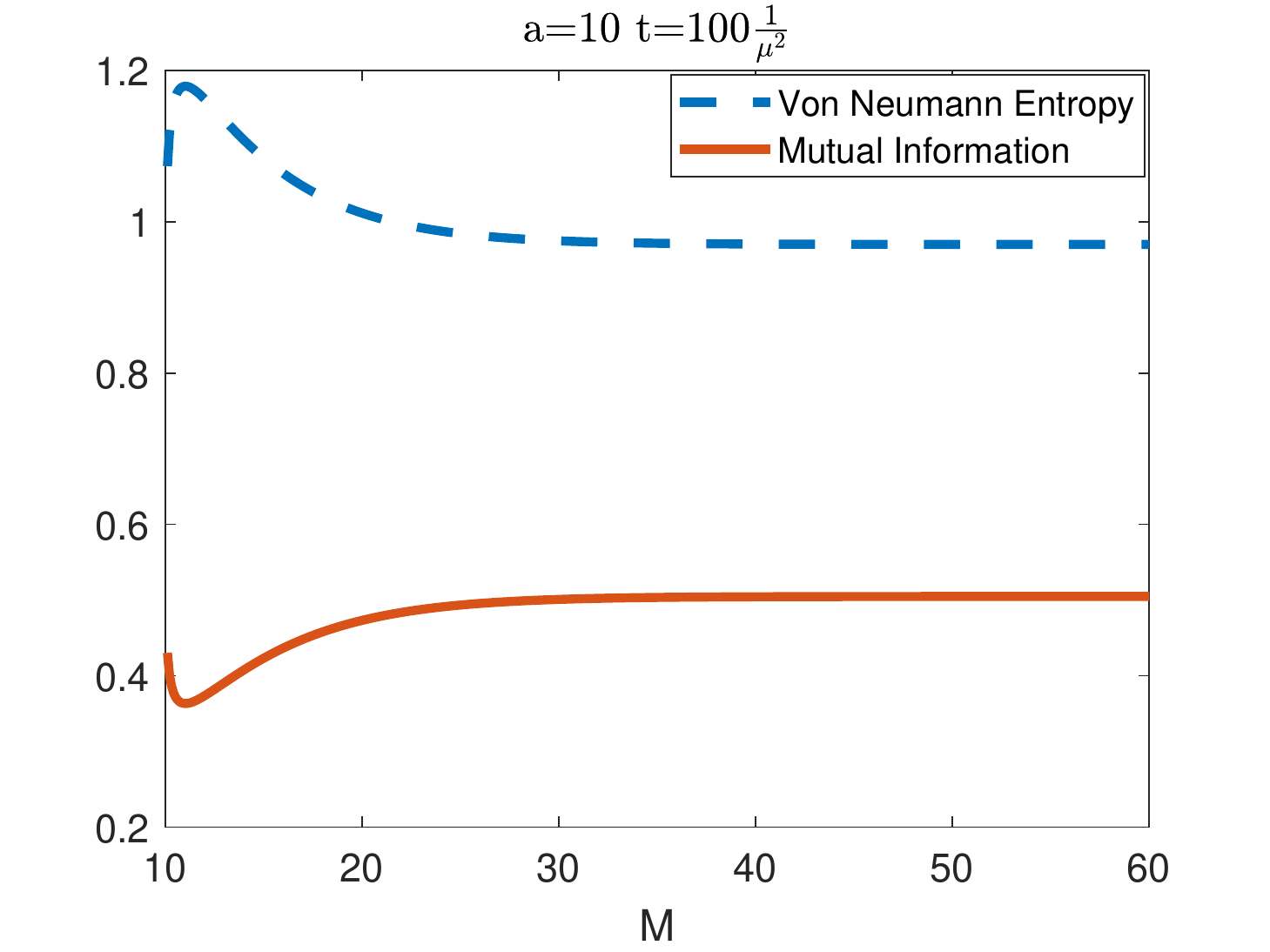}
\end{minipage}
}
\subfigure[]{
\begin{minipage}{7cm}\centering
\includegraphics[scale=0.5]{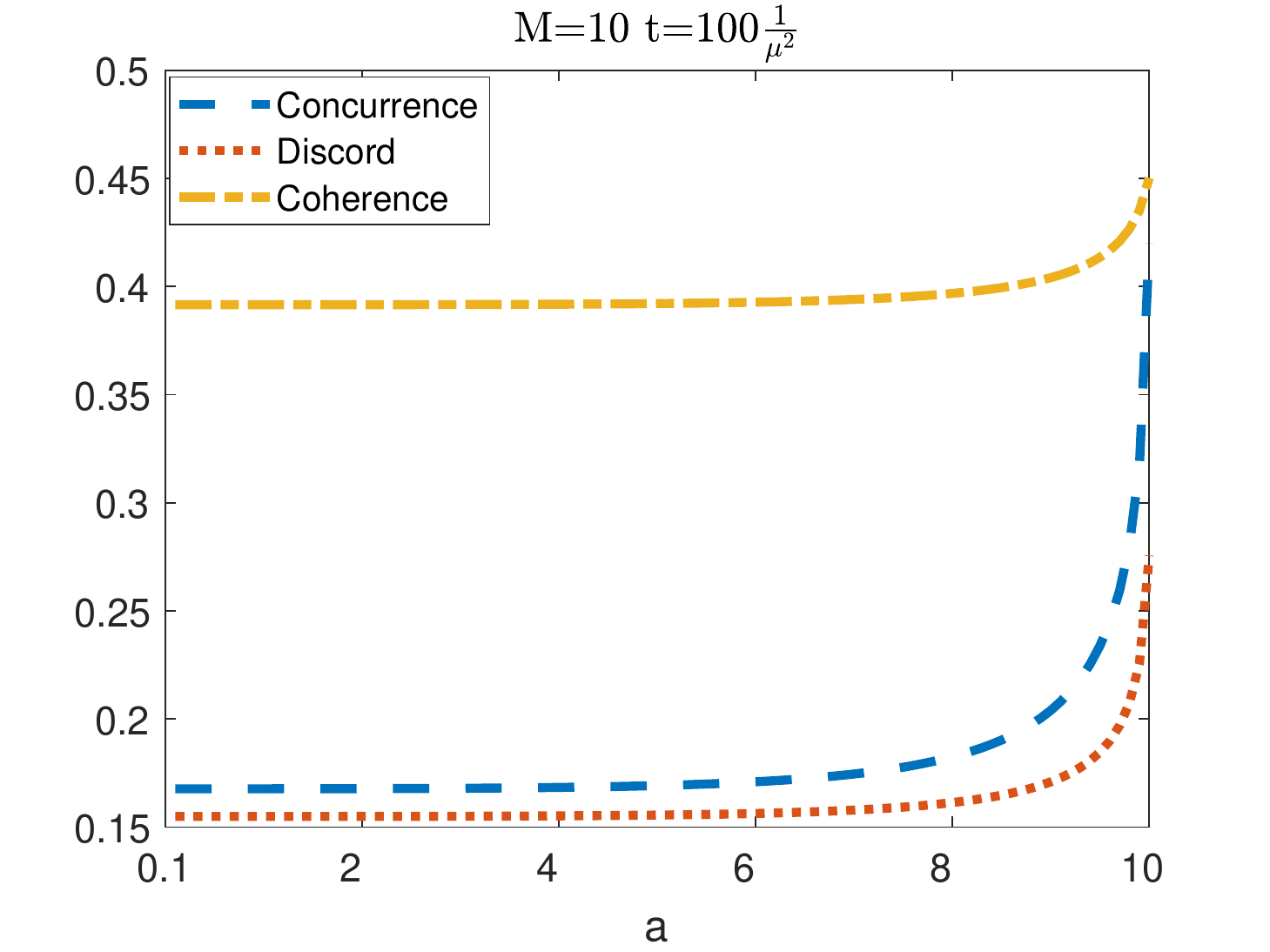}
\end{minipage}
}
\subfigure[]{
\begin{minipage}{7cm}\centering
\includegraphics[scale=0.5]{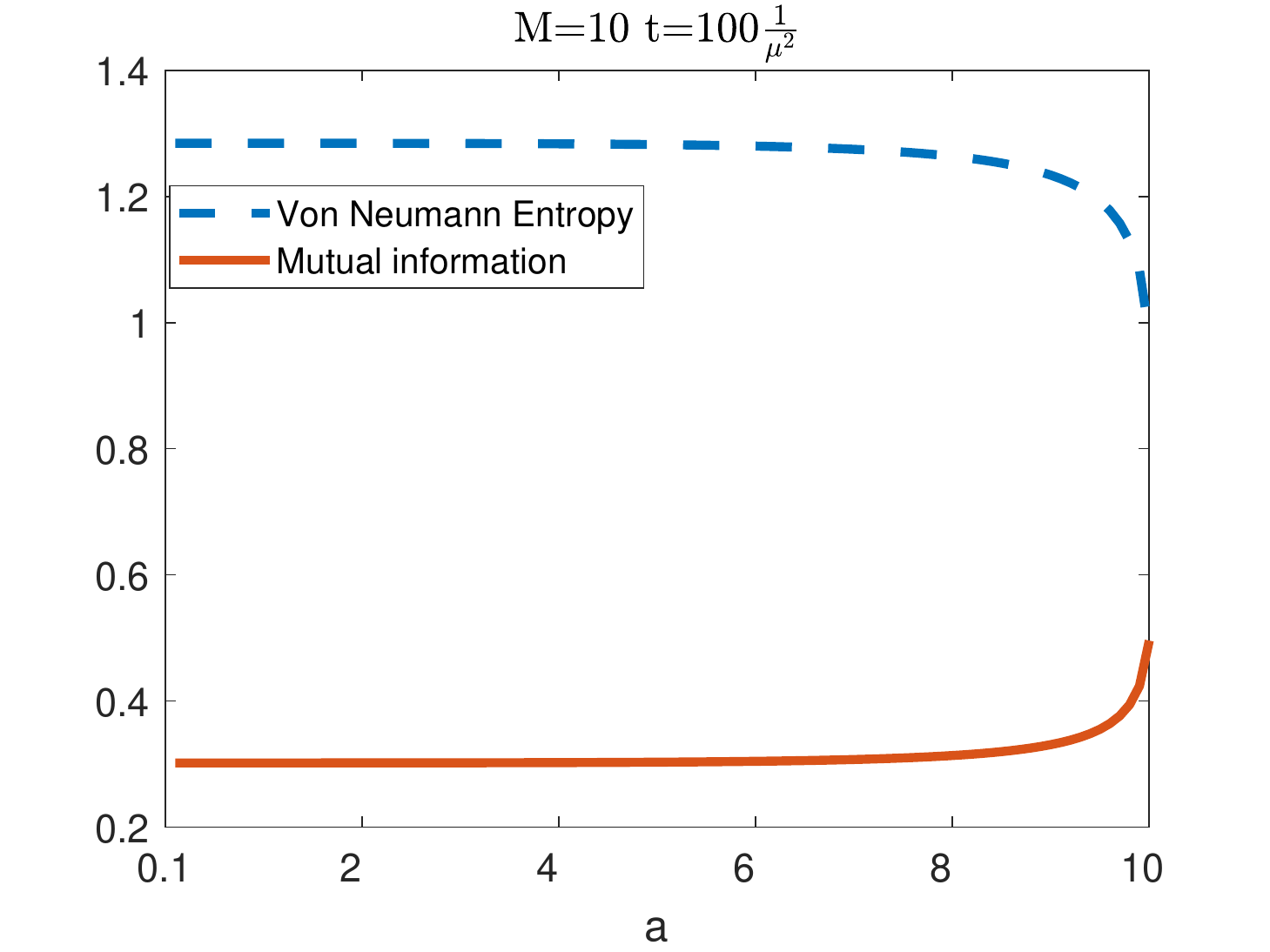}
\end{minipage}
}
\caption{Fig.\ref{fig:5}(a)(b): Quantum correlations at nonequilibrium transient state vs mass when $t=\frac{100}{\mu^{2}}$ and $a=10$. Fig.\ref{fig:5}(c)(d): Quantum correlations at nonequilibrium transient state vs the angular momentum per mass when $t=\frac{100}{\mu^{2}}$ and $m=10$. Other parameters are the same as those of the Fig.\ref{fig:1}}
\label{fig:5}
\end{figure}

\begin{figure}[htbp]
\centering
\subfigure[]{
\begin{minipage}{7cm}
\centering
\includegraphics[scale=0.5]{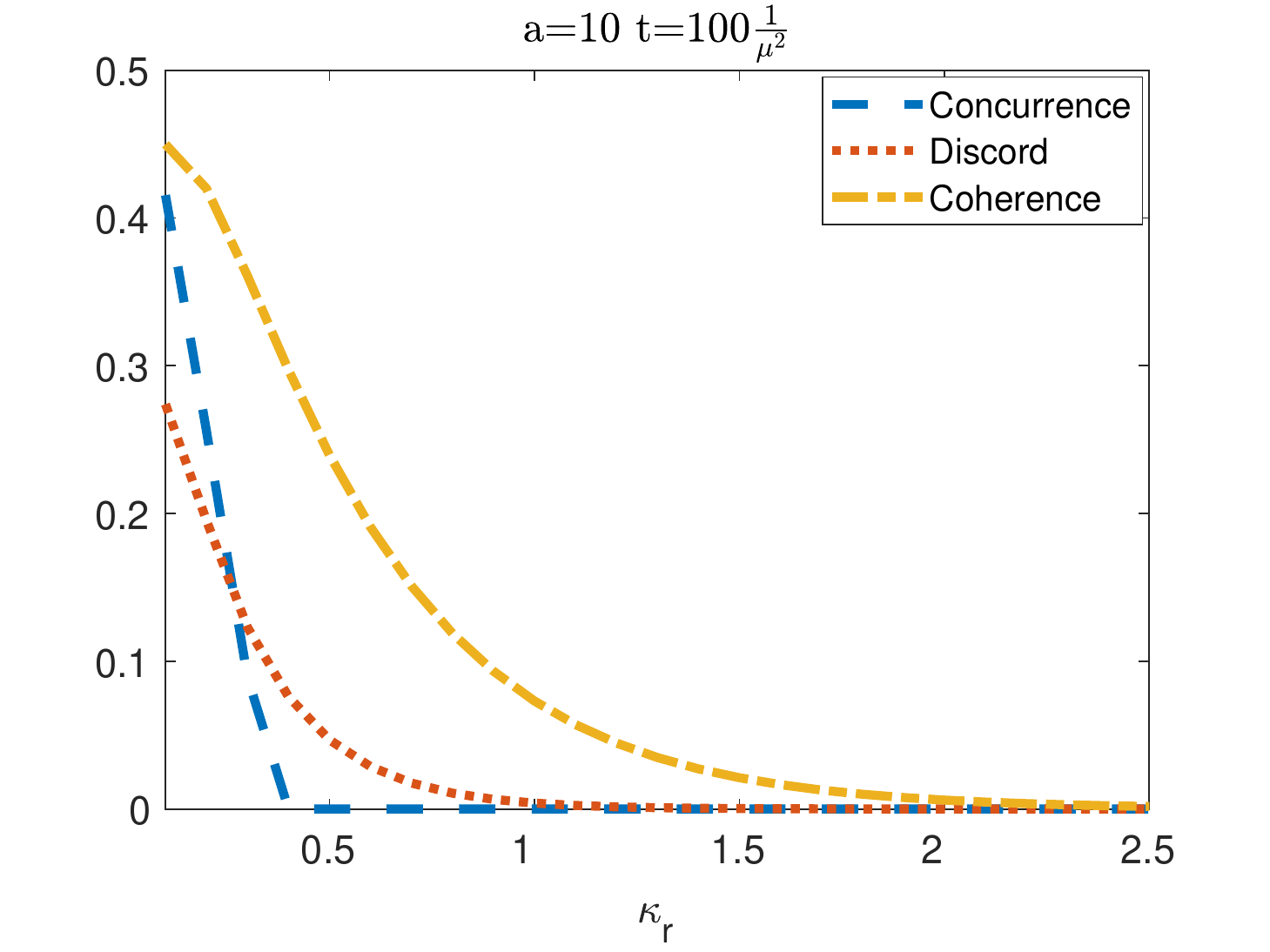}
\end{minipage}
}
\subfigure[]{
\begin{minipage}{7cm}\centering
\includegraphics[scale=0.5]{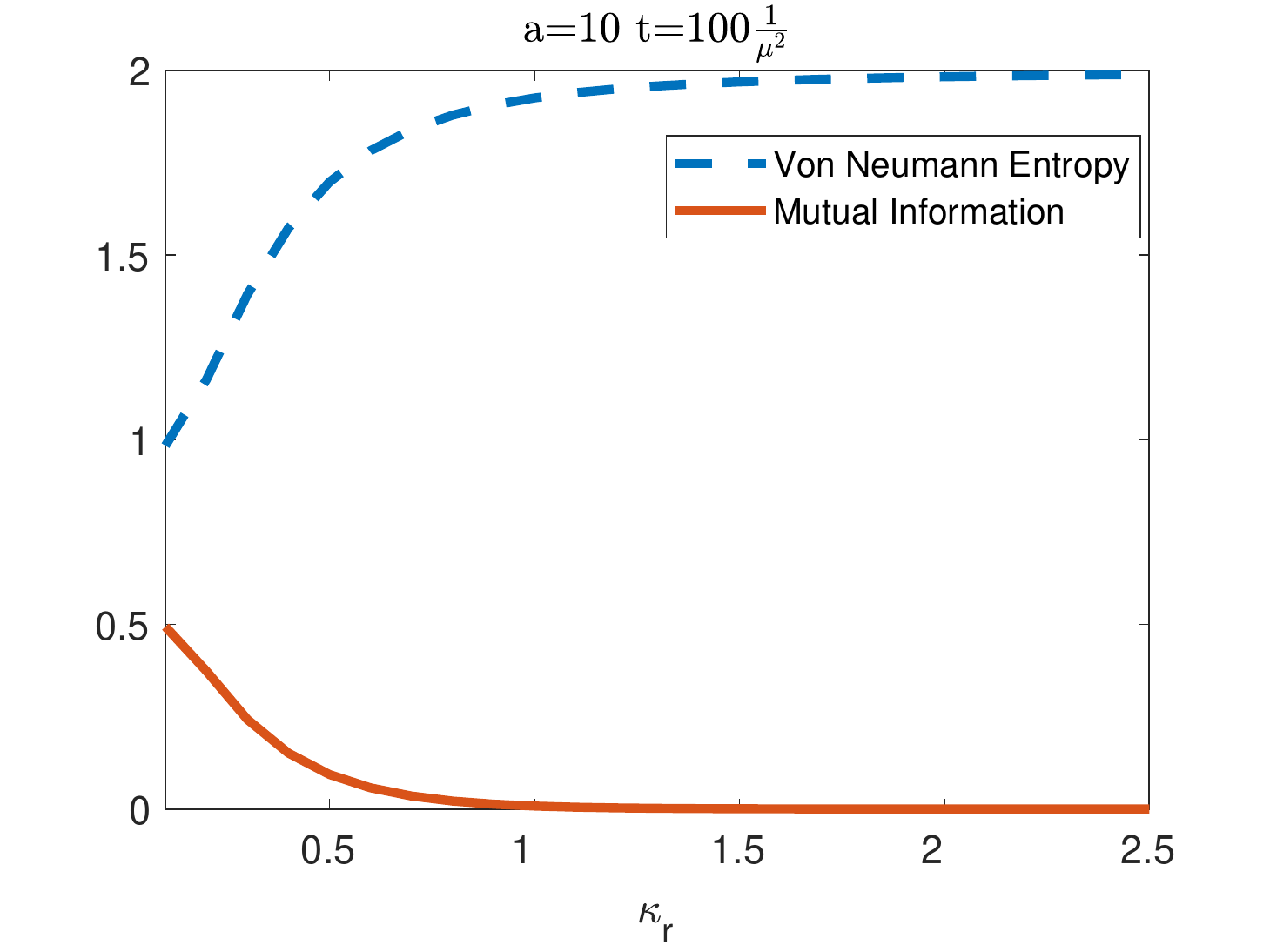}
\end{minipage}
}
\caption{Quantum correlations at nonequilibrium transient state vs $\kappa_{r}$ when $t=\frac{100}{\mu^{2}}$ and $a=10$. Other parameters are the same as those of the Fig.\ref{fig:1}}
\label{fig:6}
\end{figure}

In Fig.\ref{fig:4} we plot various quantum correlations vs time in the unit of $\frac{1}{\mu^{2}}$ at $a=0.01$, $M=10$ and $a=9.9$, $M=10$ respectively. In Fig.\ref{fig:4}, all quantum correlations between two qubits decrease to zero due to the dissipative effect of the environment but the Von Neumann entropy increases to a constant in time. Comparing Fig.\ref{fig:4}(a)(b) and Fig.\ref{fig:4}(c)(d), for a black hole with a larger mass relative to angular momentum, the quantum correlations decrease faster and reach a larger entropy. In the meanwhile, we study the dependence of the quantum correlations on the mass in Fig.\ref{fig:5}(a)(b) and on the angular momentum per mass in Fig.\ref{fig:5}(c)(d). The behaviors of the quantum correlations are very similar to the equilibrium case:the concurrence, the mutual information, the discord and the coherence all decrease first and then increase to a constant as the mass of black hole increases while the angular momentum per mass is unchanged. On the contrary, the entropy increases initially and then decreases to a constant. The angular momentum can amplify quantum correlations, especially when the angular momentum becomes larger. Also the quantum correlations vs the local curvature $\kappa_{r}$ is plotted in Fig.\ref{fig:6}(6a)(6b). All the quantum correlations between the two atoms decrease by larger curvature. On the contrary the Von Neumann entropy increases. These imply that the quantum correlations in this nonequilibrium model is also determined by the local curvature or acceleration $\kappa_{r}$. The larger $\kappa_{r}$ makes the system more classical and this weakens the quantum nature. On the contrary, the larger $\kappa_{r}$ leads to strengthening of the interaction between the system and the field, therefore makes them more easily correlate.

We try to quantify the nonequilibrium by considering entropy production $(EP)$ and entropy production rate $(EPR)$ for our setting. The initial state of the correlated system $AB$ is denoted by $\rho_{AB}^{i}$ and the initial state of the field is denoted by $\rho_{E}^{i}$. In our setup, we assume there is no correlation between the system $AB$ and $E$ initially, so the initial state of total system reads $\rho_{ABE}^{i}=\rho_{AB}^{i}\bigotimes\rho_{E}^{i}$. Although the system $AB$ experiences a nonunitary evolution, the total system is isolated and follows unitary evolution. At the finial state of $ABE$, the density matrix is given by

\begin{equation}\begin{split}
\label{eq:evolution}
\rho_{ABE}^{f}=U_{AE}\rho_{AB}^{i}\bigotimes\rho_{E}^{i}U_{AE}^{\dag}
\end{split}
\end{equation}

and the evolution of $AE$ is given as

\begin{equation}\begin{split}
\label{eq:48}
\rho_{AE}^{f}=U_{AE}\rho_{AE}U_{AE}^{\dag}
\end{split}
\end{equation}

 where $U_{AE}$ is associated to the unitary transformation. The entropy production of the system $A$ about the evolution $U_{AE}$ can be given as~\cite{Entropy production as correlation between system and reservoir}~\cite{Kun Zhang}

\begin{equation}\begin{split}
\label{eq:49}
\Sigma_{A}(t_{i}:t_{f}):=\emph{I}_{A:E}(t_{f})+\emph{S}(\rho_{E}^{f}\mid\mid\rho_{E}^{i})
\end{split}
\end{equation}

where $\emph{S}(\rho\mid\mid\sigma)=Tr\rho\ln\rho-\rho\ln\sigma$ is the relative entropy.

The entropy production of the system $AB$ about the evolution $U_{AE}$ is

\begin{equation}\begin{split}
\label{eq:50}
\Sigma_{AB}(t_{i}:t_{f})=\emph{I}_{AB:E}(t_{f})+\emph{S}(\rho_{E}^{f}\mid\mid\rho_{E}^{i})
\end{split}
\end{equation}

The entropy production can be rewritten as $\Sigma_{AB}(t_{i}:t_{f})$

\begin{equation}\begin{split}
\label{eq:51}
\Sigma_{AB}(t_{i}:t_{f}):=\Delta\emph{I}_{A:B}(t_{i}:t_{f})+\Sigma_{A}(t_{i}:t_{f})
\end{split}
\end{equation}

where $\Delta\emph{I}_{A:B}(t_{i}:t_{f}):=\emph{I}_{A:B}(t_{i})-\emph{I}_{A:B}(t_{f})$, $\emph{I}_{A:B}$ is mutual information between $A$ and $B$. For the detailed derivation, see appendix~\ref{sec:Derivation of the Entropy production}. %For the fix-point evolution $U_{AE}\rho_{A}\bigotimes\rho_{E}U_{AE}^{\dag}=\rho_{A}\bigotimes\rho_{E}$, in general, we have $\Sigma_{AB}(t_{i}:t_{f}):=\Delta\emph{I}_{A:B}(t_{i}:t_{f})>0$.%
In our early setup, we have assumed a Born approximation: the coupling between the system and the bath is weak such that the influence of the bath is small. Thus we can consider the bath is almost unchanged. Then the state of the total system at time $\tau$ may be approximately characterized by a tensor product as Eqn.\eqref{eq:6}, so that $U_{AE}\rho_{AB}(\tau^{i})\bigotimes\rho_{E}U_{AE}^{\dag}\approx\rho_{AB}(\tau^{f})\bigotimes\rho_{E}$. Hence $\Sigma_{AB}(t_{i}:t_{f})\geq\Delta\emph{I}_{A:B}(t_{i}:t_{f})\geq 0$ are due to the positivity of $\emph{I}_{A:B}(t_{i}:t_{f})$ and the quantum relative entropy. Consider the Born approximation, we further derive a lower bound of the entropy production:

%\begin{equation}\begin{split}
%\label{eq:52}
%\Sigma_{AB}(t_{i}:t_{f})&=\emph{I}_{AB:E}(t_{f})+\emph{S}(\rho_{E}^{f}\mid\mid\rho_{E}^{i})\\
%&=\Delta\emph{I}_{A:B}(t_{i}:t_{f})+\emph{I}_{A:E}(t_{f})+\emph{S}(\rho_{E}^{f}\mid\mid\rho_{E}^{i})\geq0
%\end{split}
%\end{equation}%
\begin{equation}\begin{split}
\label{eq:53}
\Delta\emph{I}_{A:B}(t_{i}:t_{f})&\geq\emph{I}_{A:E}(t_{f})\\
&=\emph{S}_{A}^{f}+\emph{S}_{E}^{f}-\emph{S}_{AE}^{f}\approx0\\
\emph{S}(\rho_{E}^{f}\mid\mid\rho_{E}^{i})\approx0
\end{split}
\end{equation}

The first greater than or equal to sign comes from the fact that the scrambling information between $A$ and $B$ is greater than the increased information between $A$ and $E$. The later approximately equal sign is due to the Born approximation.

\begin{figure}[htbp]
\centering
\subfigure[]{
\begin{minipage}{7cm}
\centering
\includegraphics[scale=0.5]{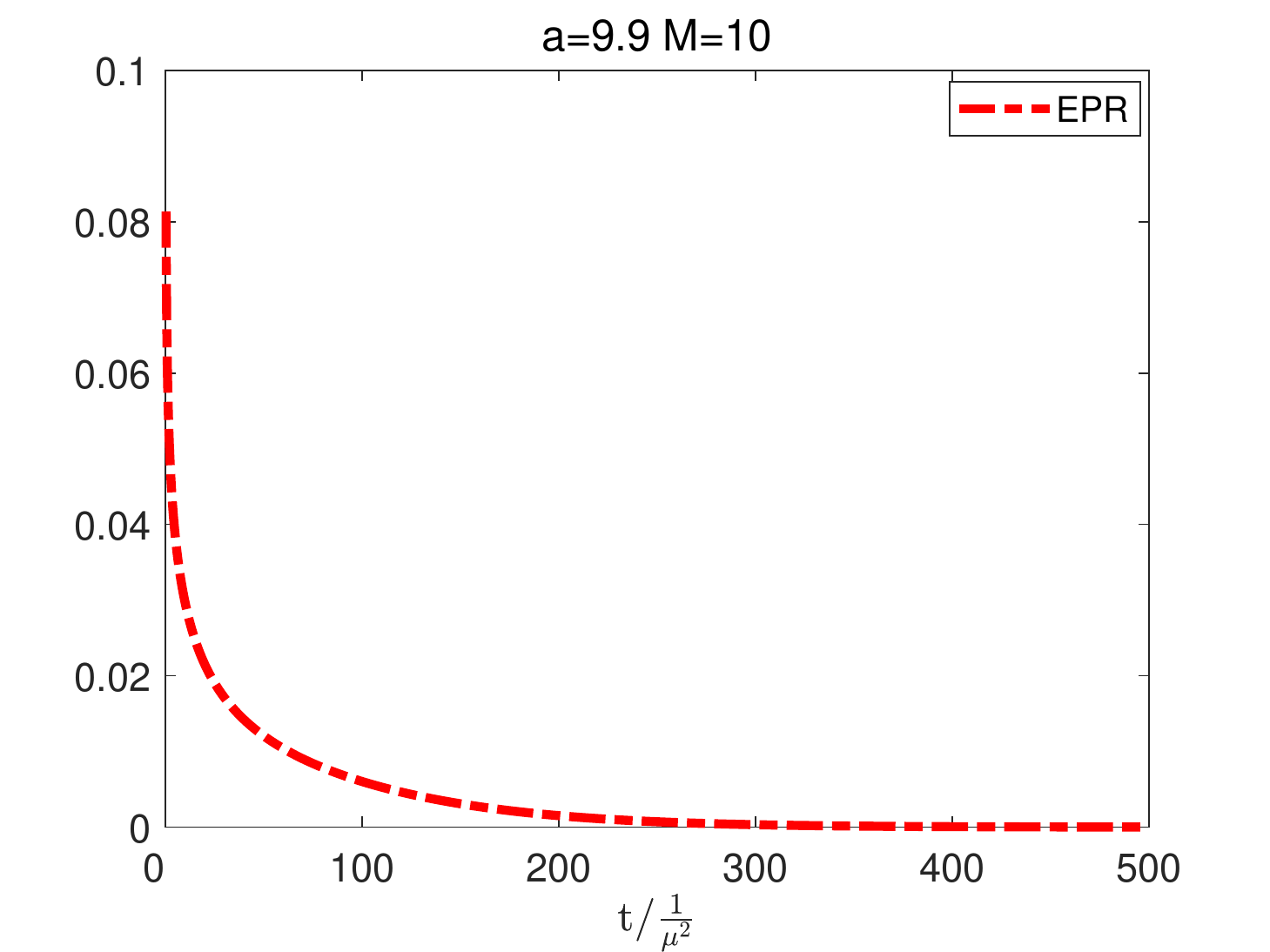}
\end{minipage}
}
\subfigure[]{
\begin{minipage}{7cm}\centering
\includegraphics[scale=0.5]{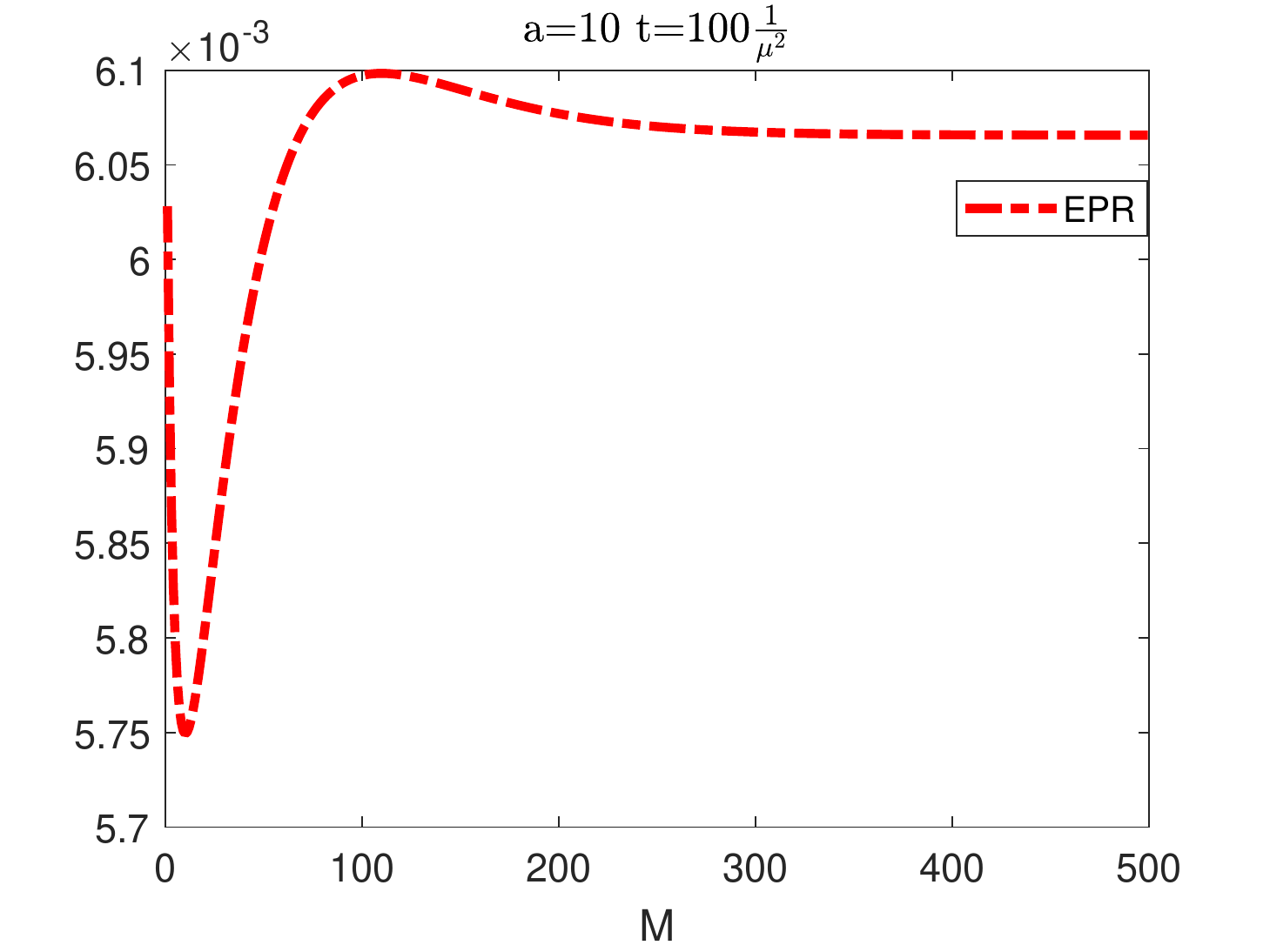}
\end{minipage}
}
\subfigure[]{
\begin{minipage}{7cm}
\centering
\includegraphics[scale=0.5]{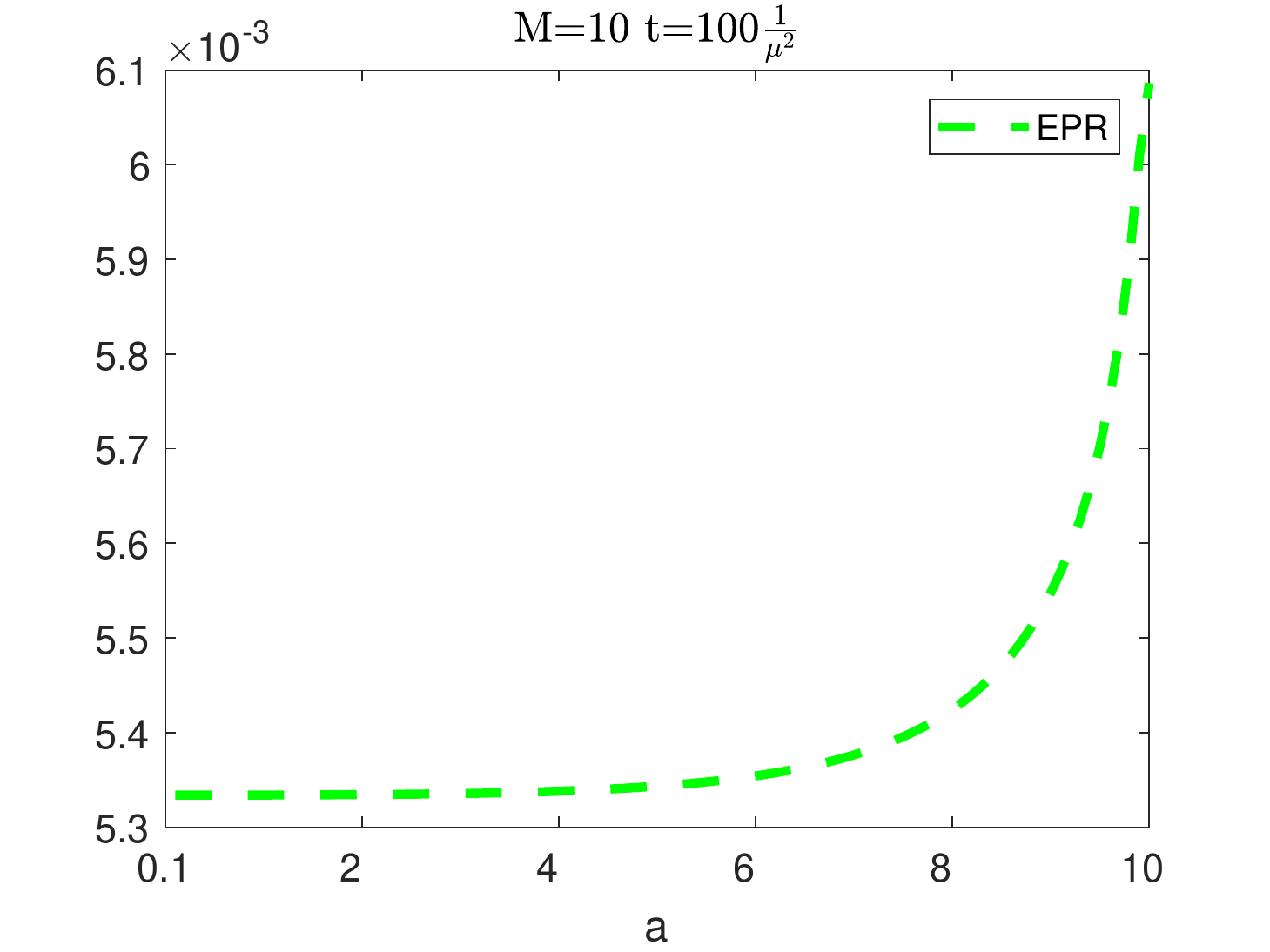}
\end{minipage}
}
\subfigure[]{
\begin{minipage}{7cm}\centering
\includegraphics[scale=0.5]{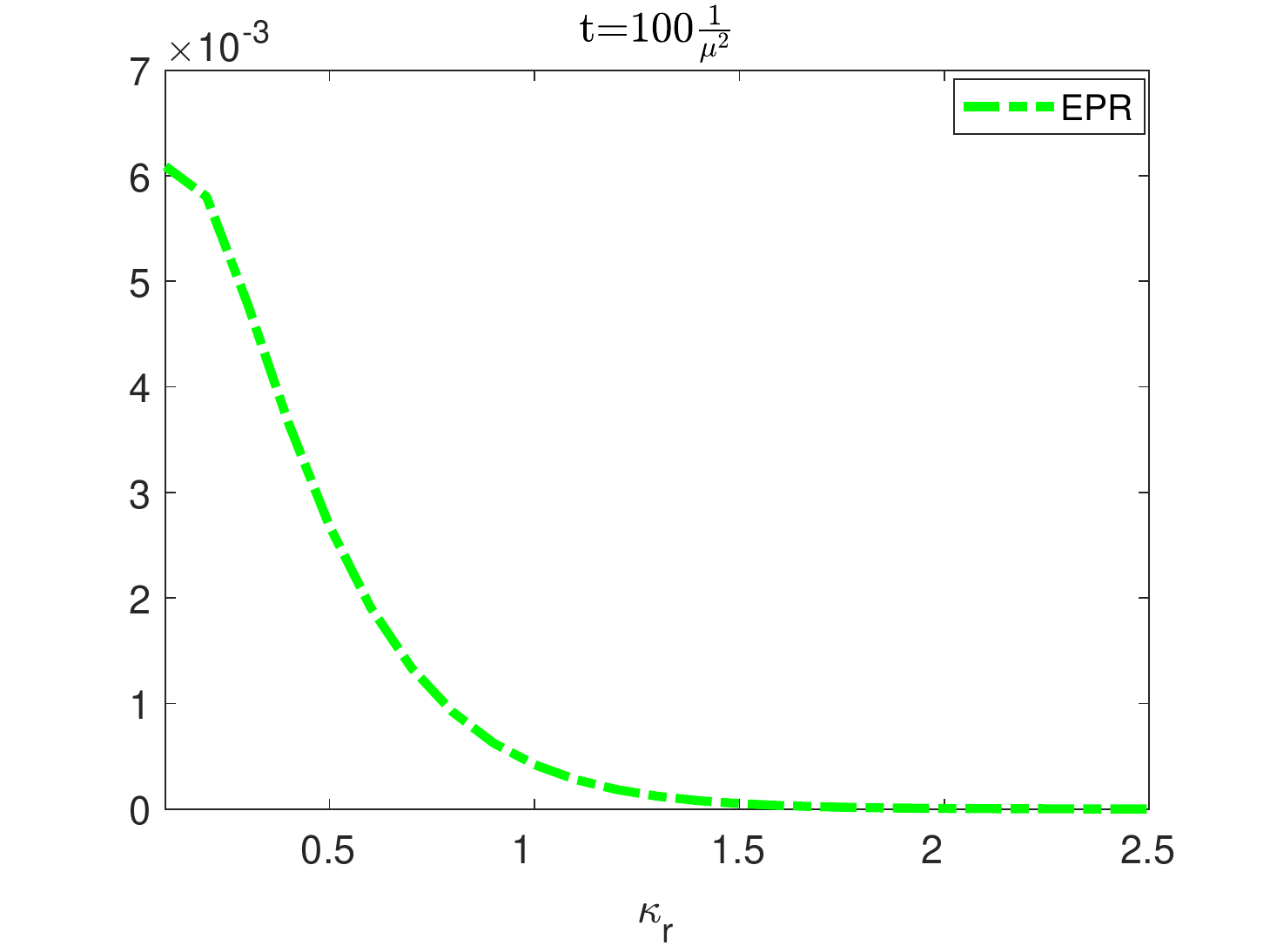}
\end{minipage}
}
\caption{ a) EPR at nonequilibrium state varying with time $t$ at angular momentum $a=9.9$ and the mass $M=10$. b) EPR versus mass when $a=10$ and $t=\frac{100}{\mu^{2}}$. c) EPR versus the angular momentum per mass when $m=10$ and $t=\frac{100}{\mu^{2}}$. d) EPR versus $\kappa_{r}$ when $t=\frac{100}{\mu^{2}}$. Other parameters are the same as the Fig.\ref{fig:1}}
\label{fig:7}
\end{figure}

For a nonequilibrium state the EPR quantifying the dissipative cost is always larger than zero, the $Born-Markov$ approximation is assumed such that the two positive parts are omitted, hence the EPR in our paper is actually a lower bound. The EPR of the system is plotted in Fig.\ref{fig:7}. The EPR decreases in time, and finally vanishes. At fixed time, the EPR varies non-monotonically with respect to the mass. The increase in angular momentum can amplify EPR. The non-trivial behavior of EPR on the mass and the angular momentum also comes from the fact that the mass and the angular momentum are directly related to the local curvature or acceleration $\kappa_{r}$. The EPR varying with $\kappa_{r}$ is plotted in Fig.\ref{fig:7}d. The EPR decreases monotonically with the local curvature. Hence, we see that the increase of the local curvature reduces the dissipative cost of information scrambling.

\begin{figure}[htbp]
\centering
\subfigure[]{
\begin{minipage}{7cm}
\centering
\includegraphics[scale=0.5]{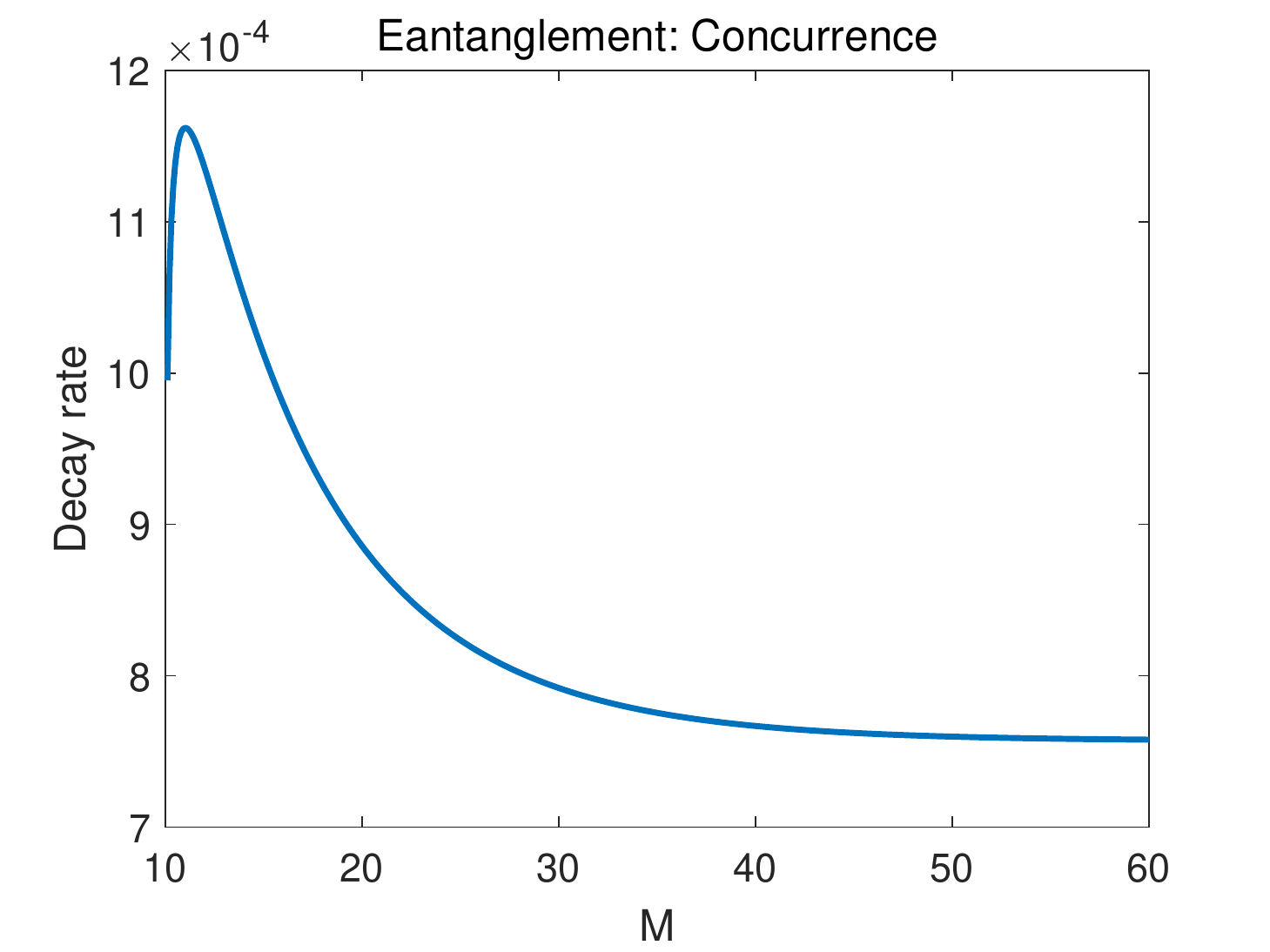}
\end{minipage}
}\subfigure[]{
\begin{minipage}{7cm}\centering
\includegraphics[scale=0.5]{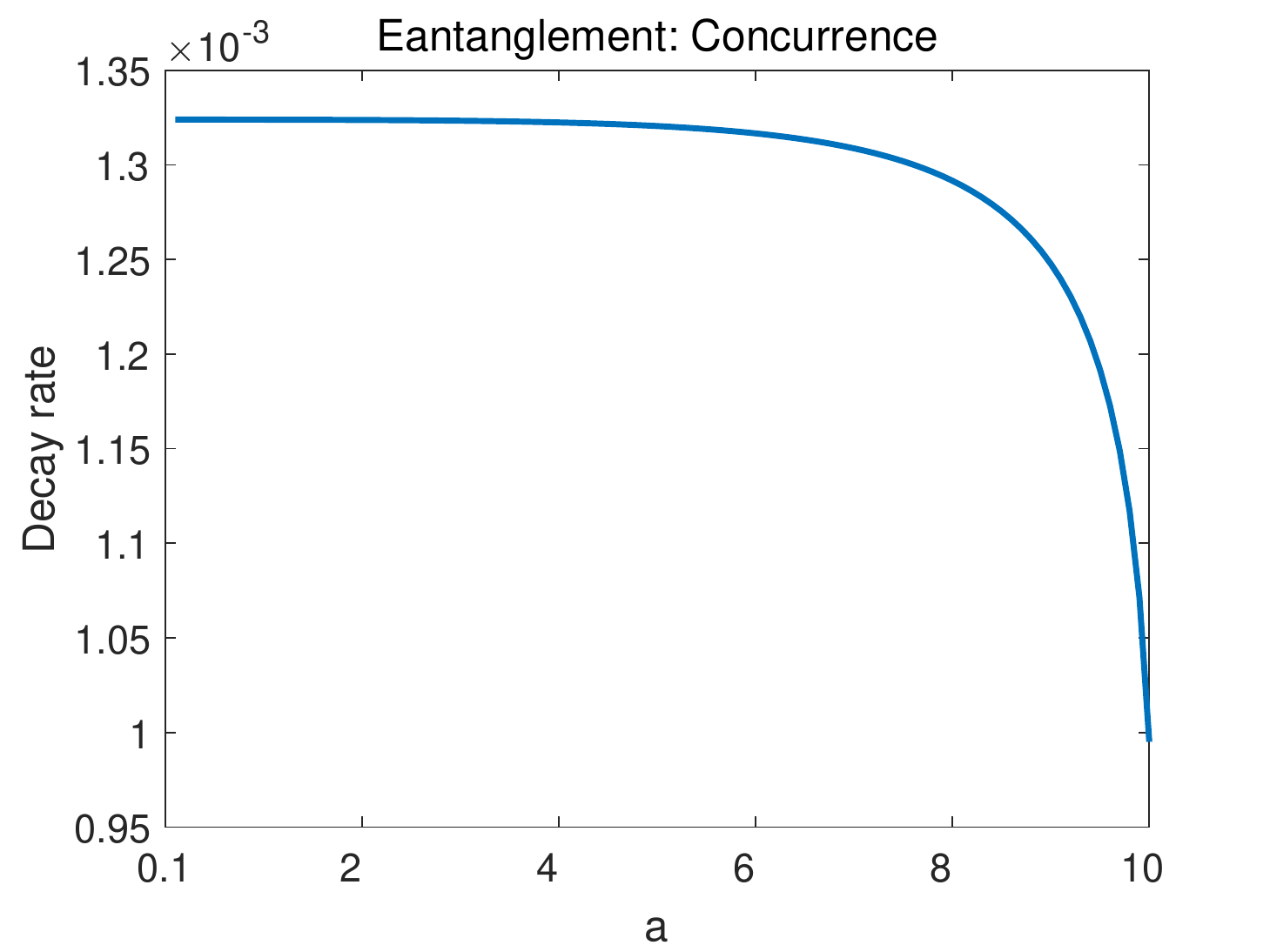}
\end{minipage}
}
\subfigure[]{
\begin{minipage}{7cm}
\centering
\includegraphics[scale=0.5]{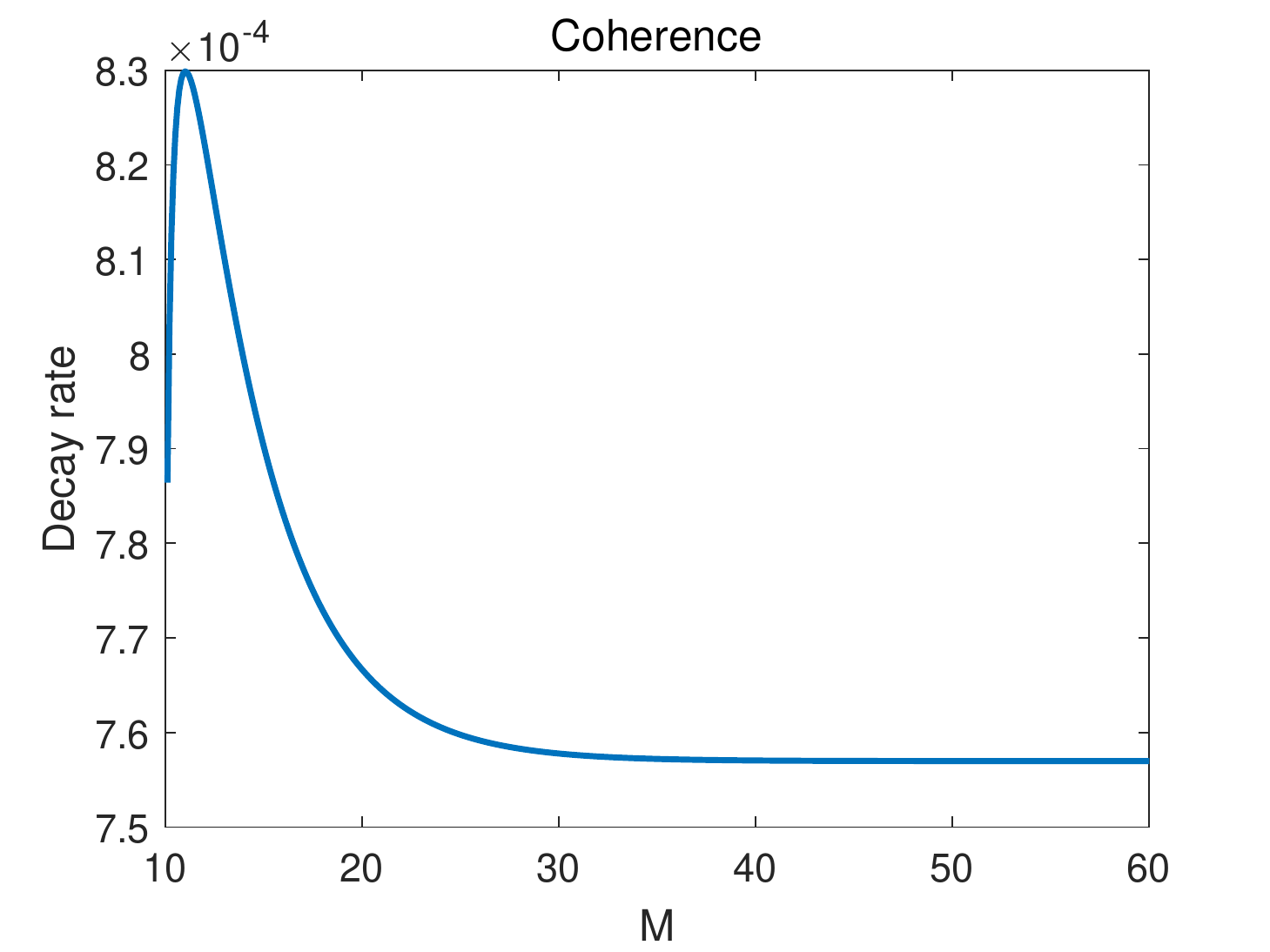}
\end{minipage}
}\subfigure[]{
\begin{minipage}{7cm}\centering
\includegraphics[scale=0.5]{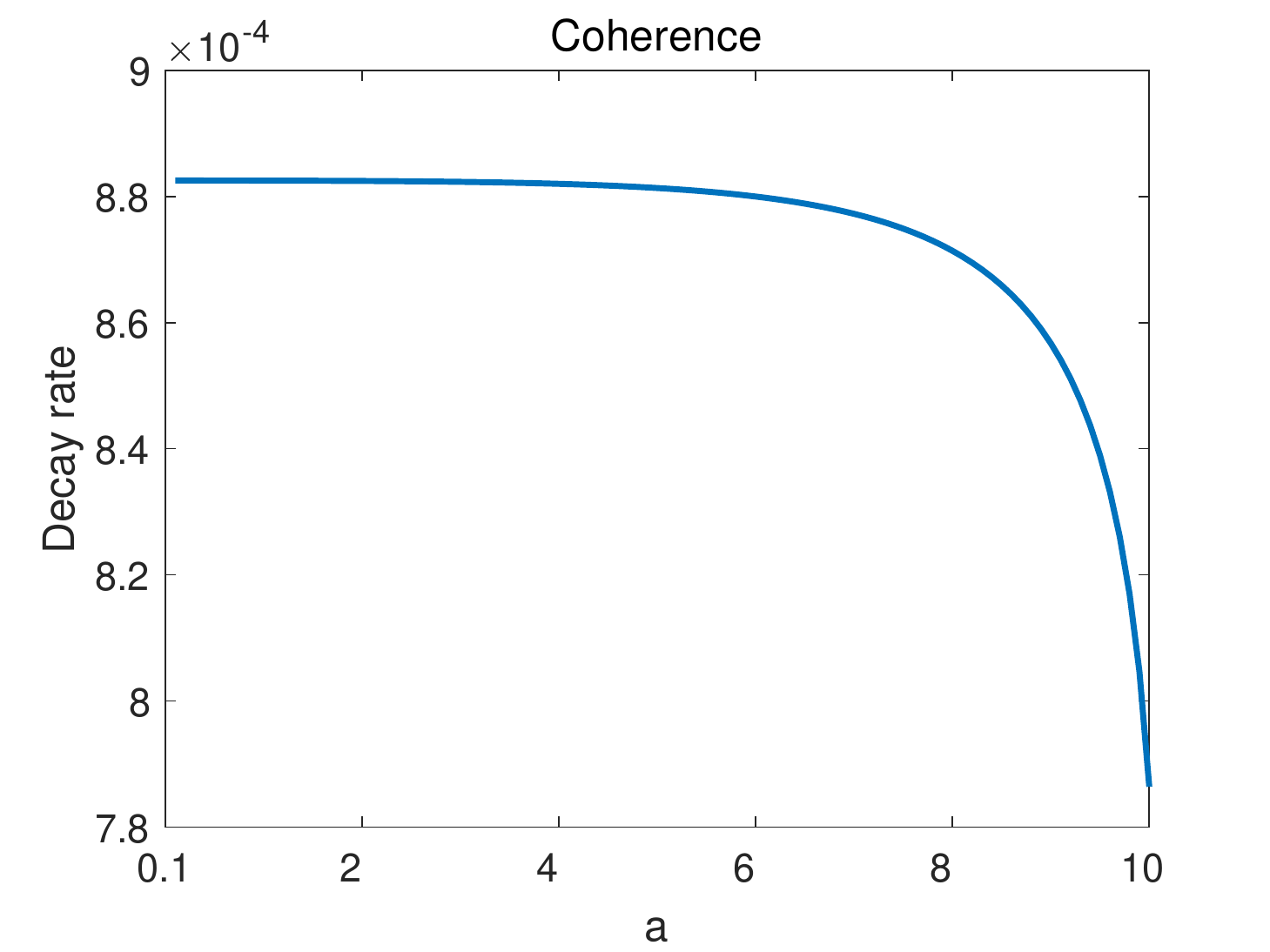}
\end{minipage}
}
\subfigure[]{
\begin{minipage}{7cm}\centering
\includegraphics[scale=0.5]{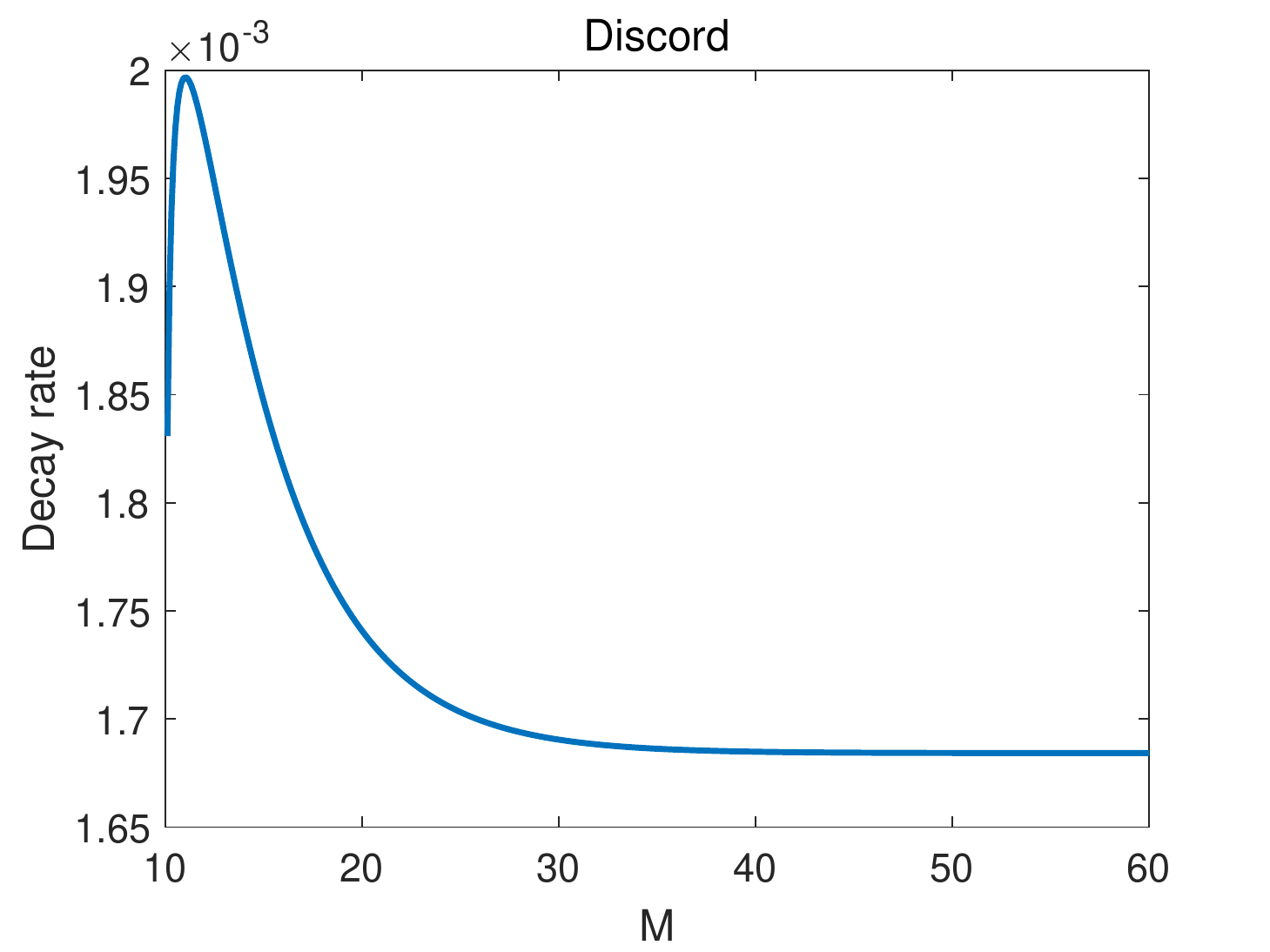}
\end{minipage}
}\subfigure[]{
\begin{minipage}{7cm}\centering
\includegraphics[scale=0.5]{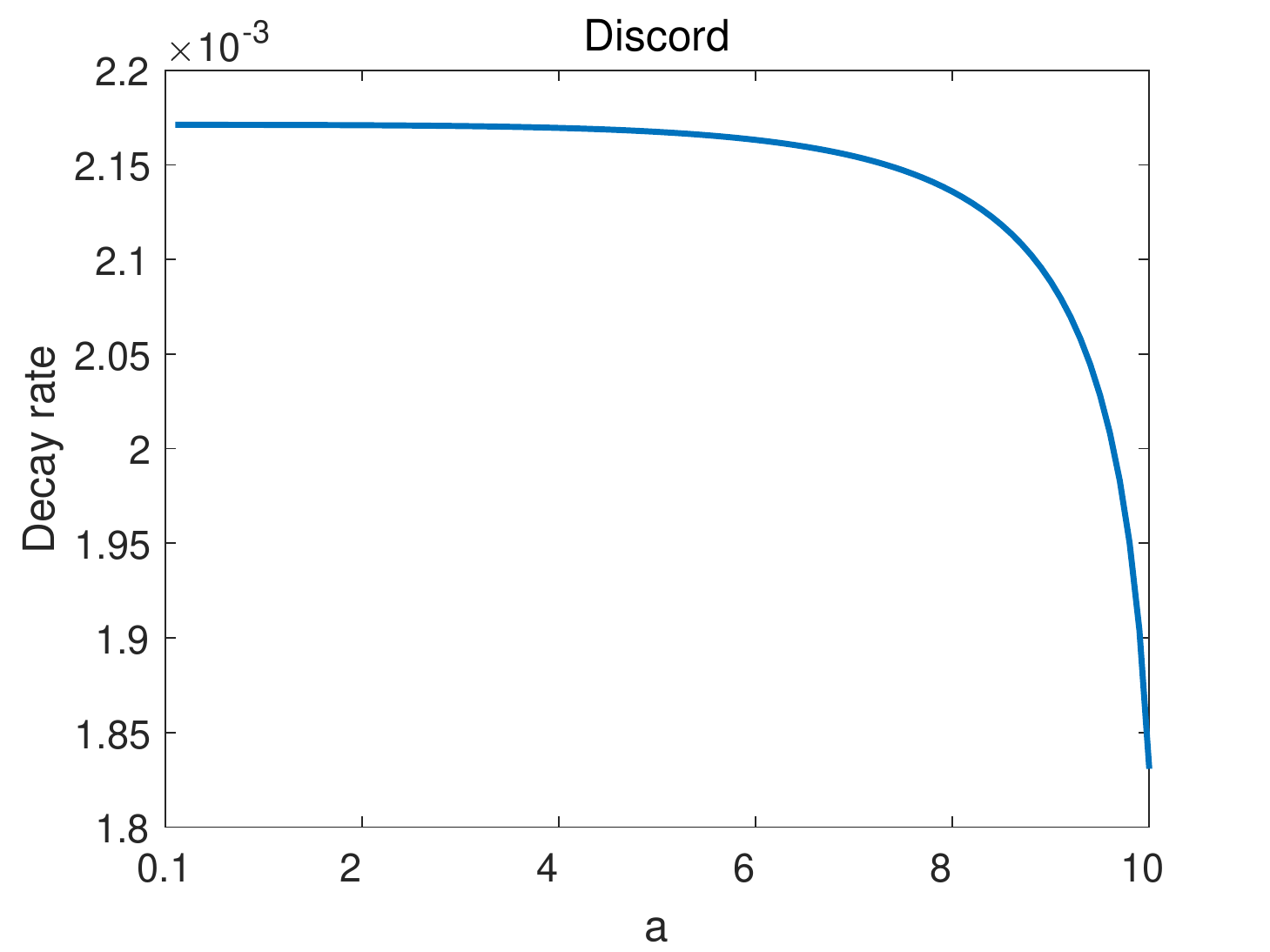}
\end{minipage}
}
\caption{The decay rates of the quantum correlations with respect to the mass when $a=10$, $t=\frac{100}{\mu^{2}}$ or w.r.t. the angular momentum per mass when $m=10$, $t=\frac{100}{\mu^{2}}$. Other parameters are the same as those of the Fig.\ref{fig:1}}
\label{fig:8.1}
\end{figure}

\begin{figure}[htbp]
\centering
\subfigure[]{
\begin{minipage}{7cm}\centering
\includegraphics[scale=0.5]{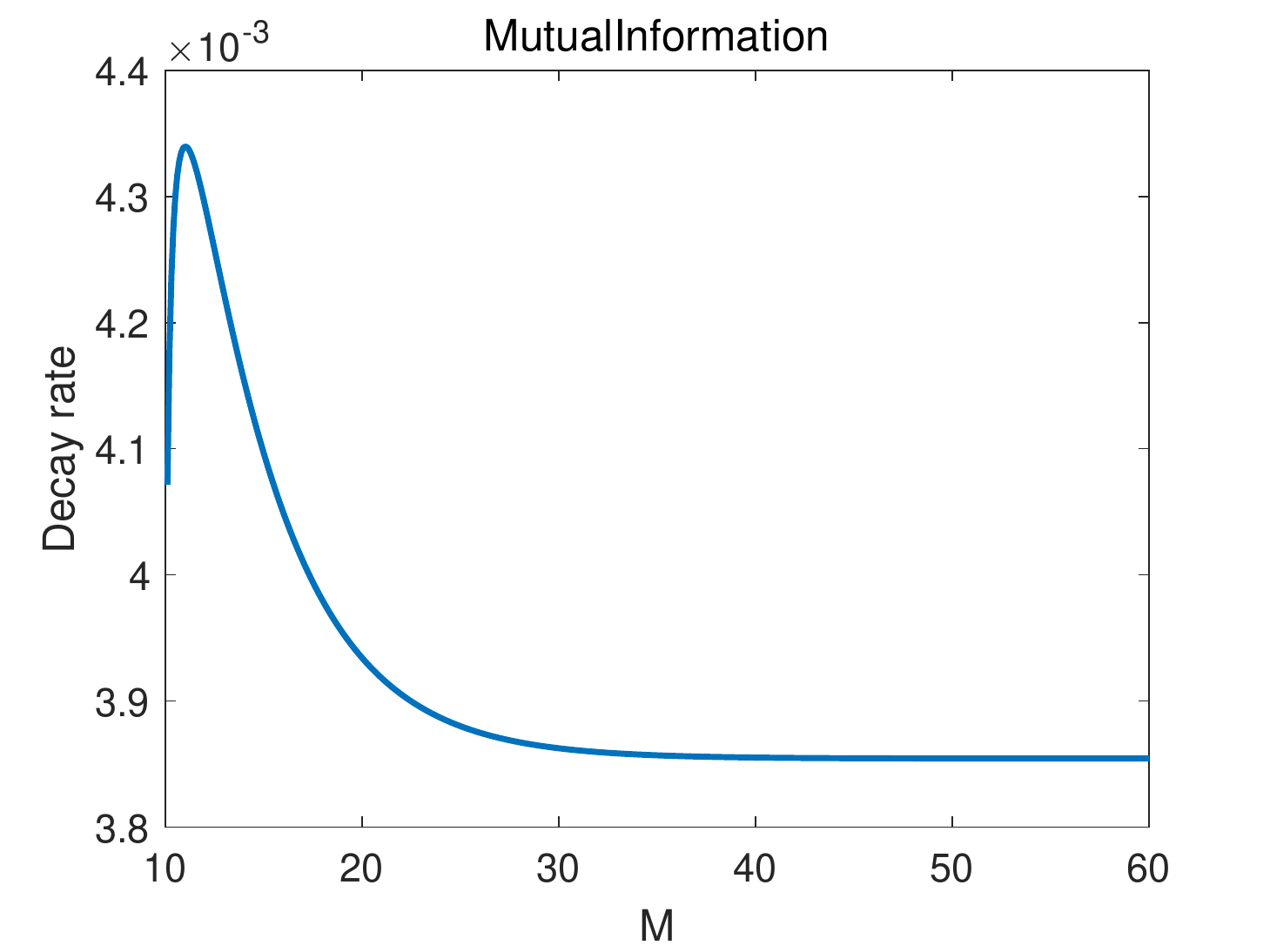}
\end{minipage}
}\subfigure[]{
\begin{minipage}{7cm}\centering
\includegraphics[scale=0.5]{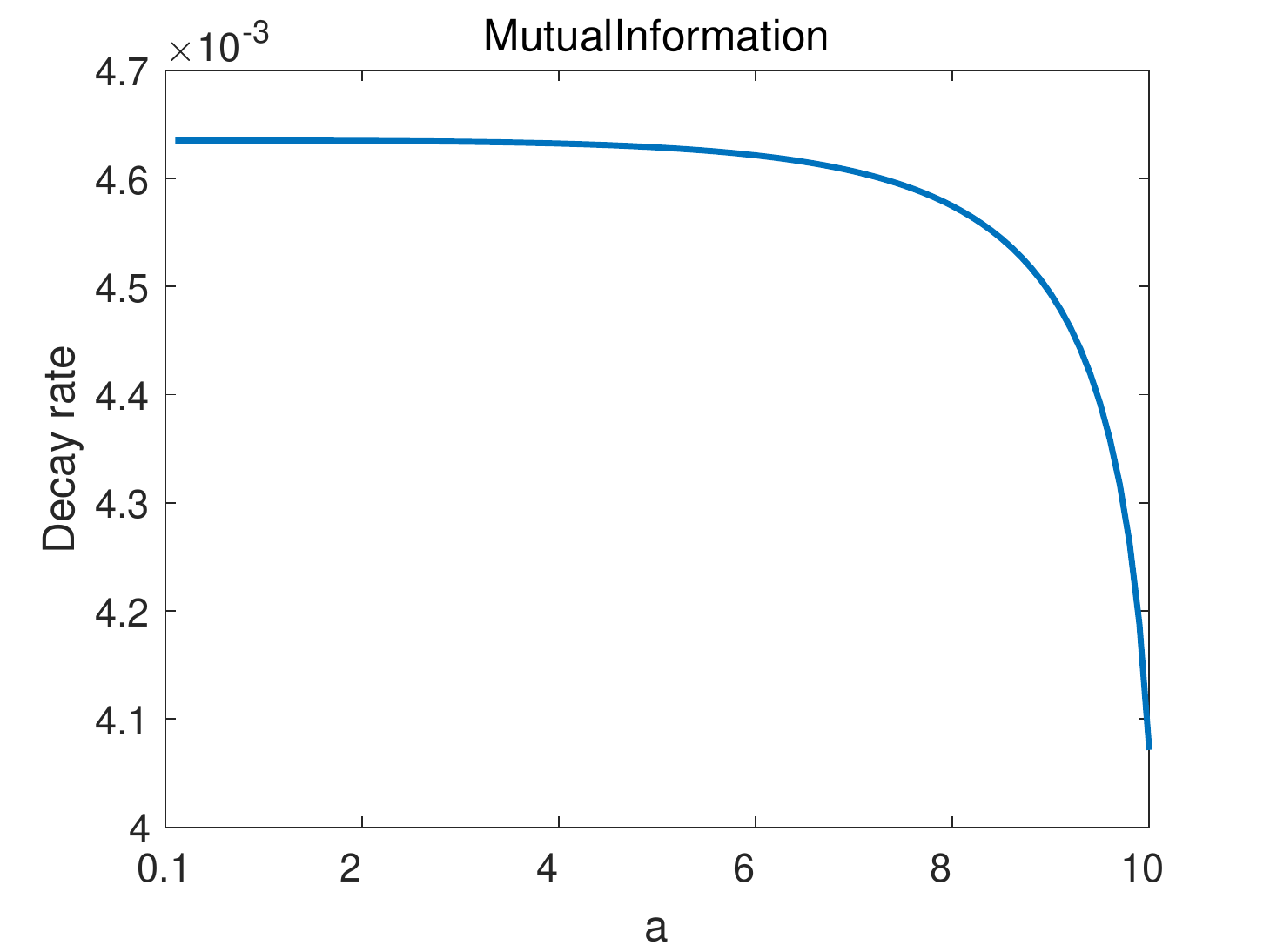}
\end{minipage}
}
\subfigure[]{
\begin{minipage}{7cm}\centering
\includegraphics[scale=0.5]{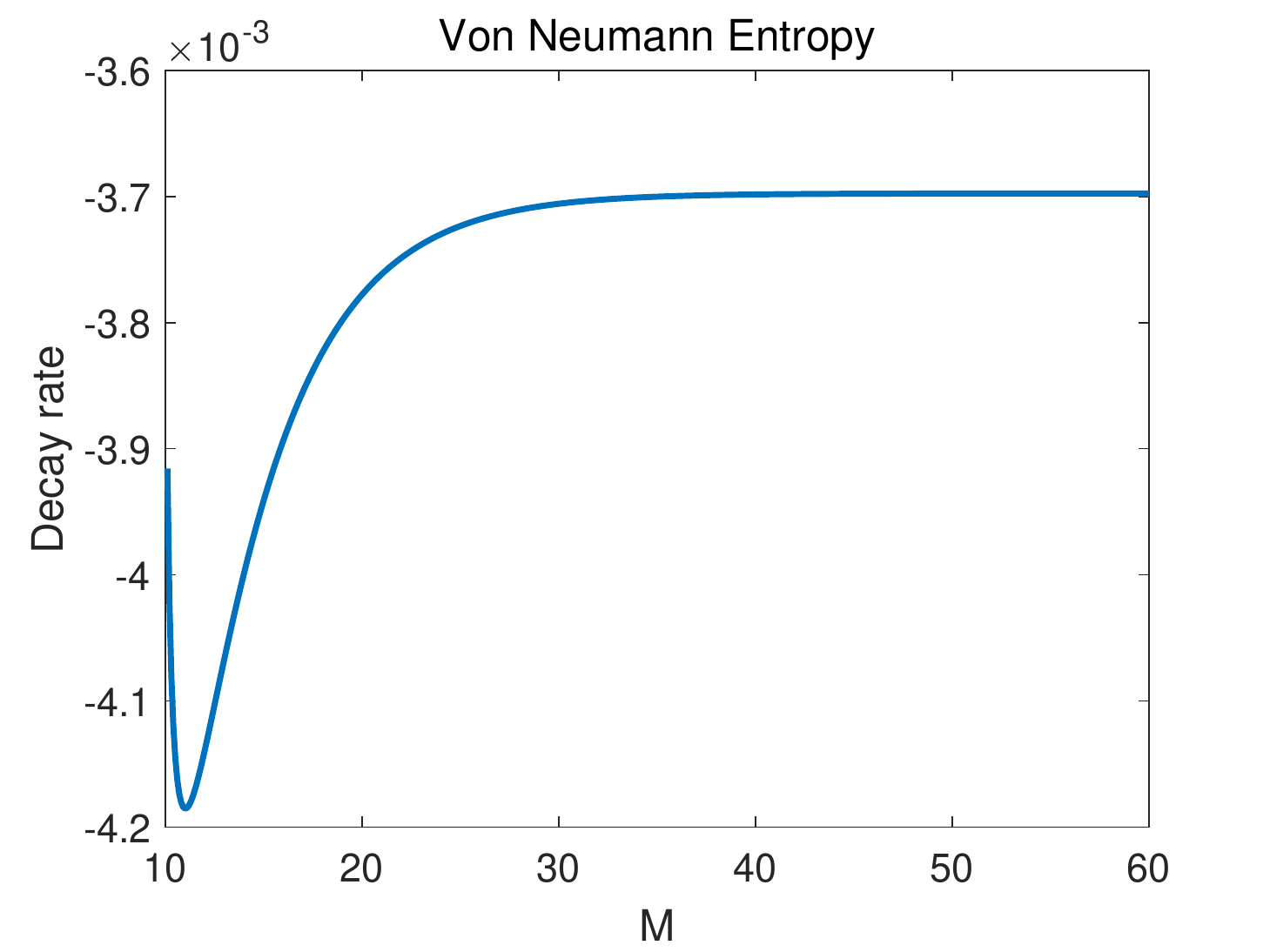}
\end{minipage}
}\subfigure[]{
\begin{minipage}{7cm}\centering
\includegraphics[scale=0.5]{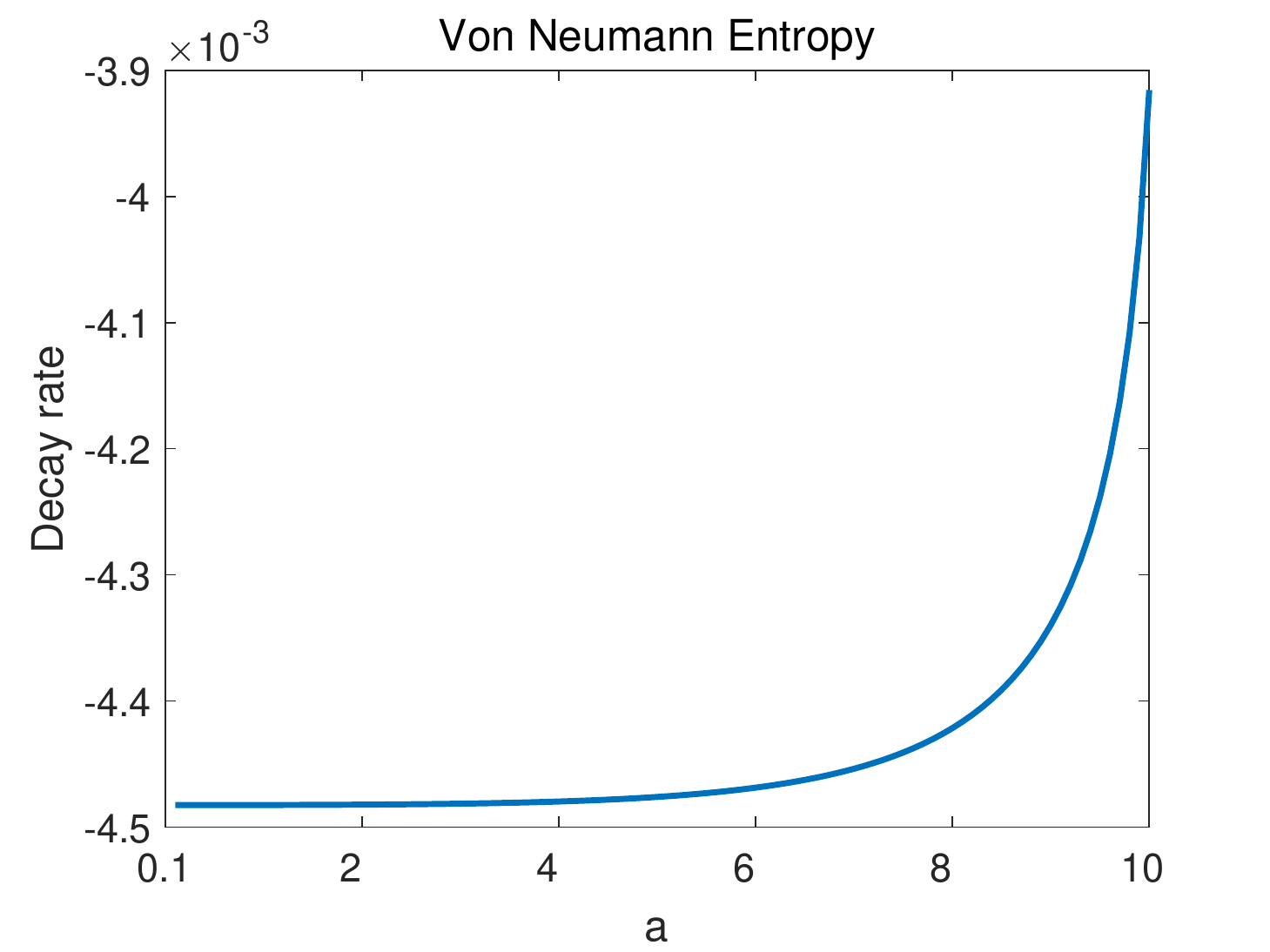}
\end{minipage}
}
\subfigure[]{
\begin{minipage}{7cm}\centering
\includegraphics[scale=0.5]{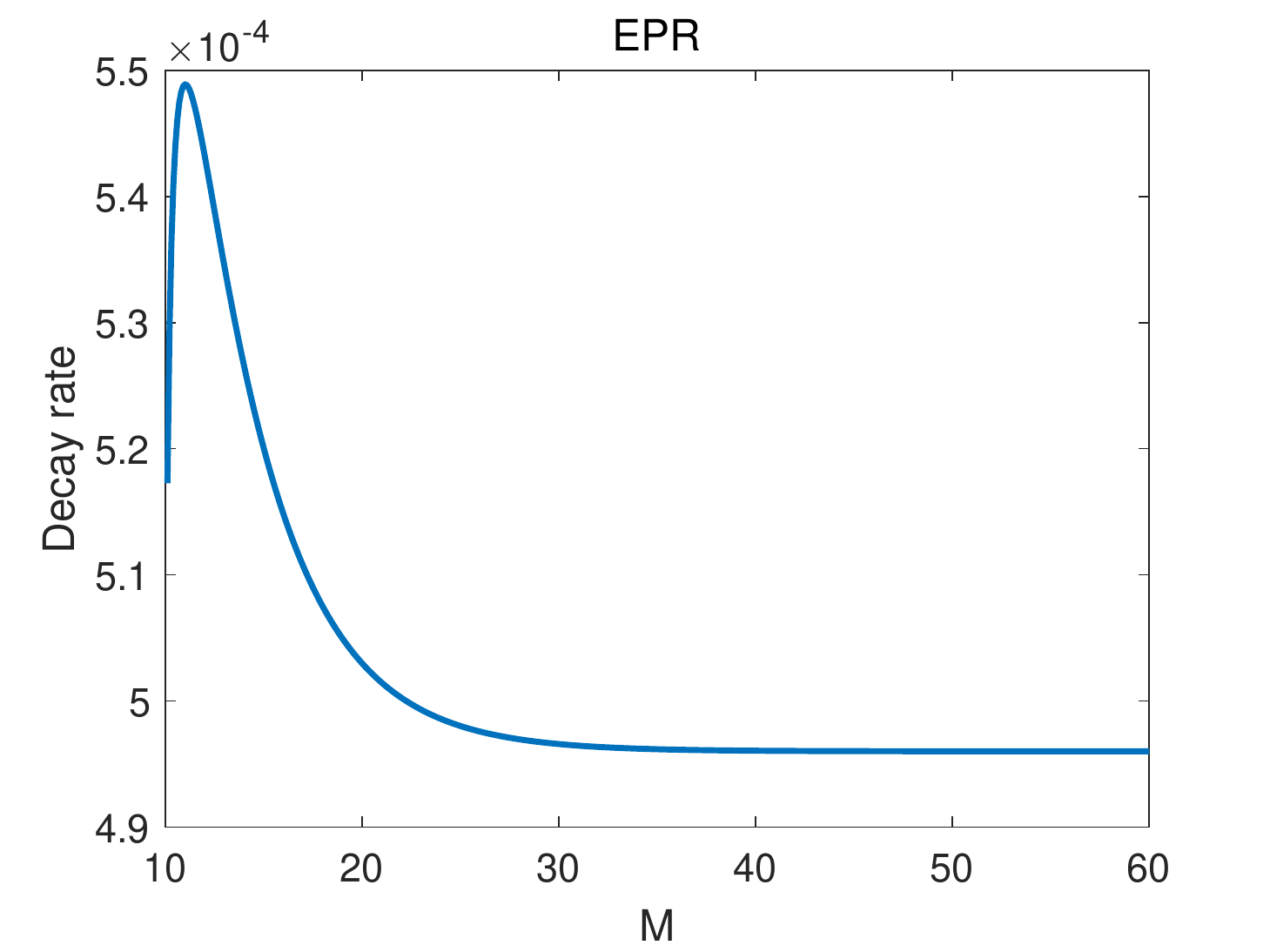}
\end{minipage}
}\subfigure[]{
\begin{minipage}{7cm}\centering
\includegraphics[scale=0.5]{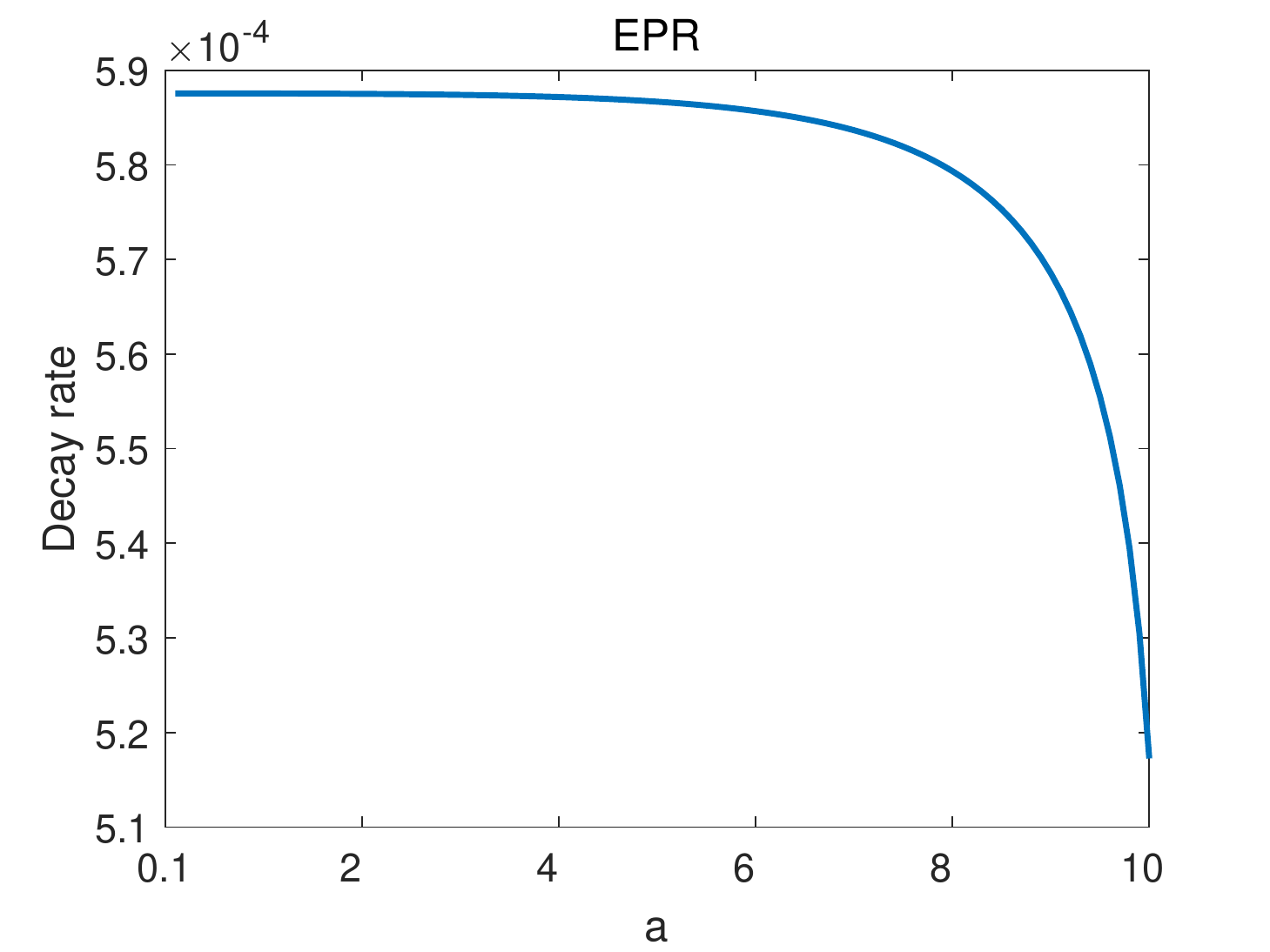}
\end{minipage}
}
\caption{The decay rates of the quantum correlations with respect to the mass when $a=10$, $t=\frac{100}{\mu^{2}}$, or w.r.t. the angular momentum per mass when $m=10$, $t=\frac{100}{\mu^{2}}$. Other parameters are the same as  those of the Fig.\ref{fig:1}}
\label{fig:8.2}
\end{figure}

\begin{figure}[htbp]
\centering
\subfigure[]{
\begin{minipage}{7cm}\centering
\includegraphics[scale=0.5]{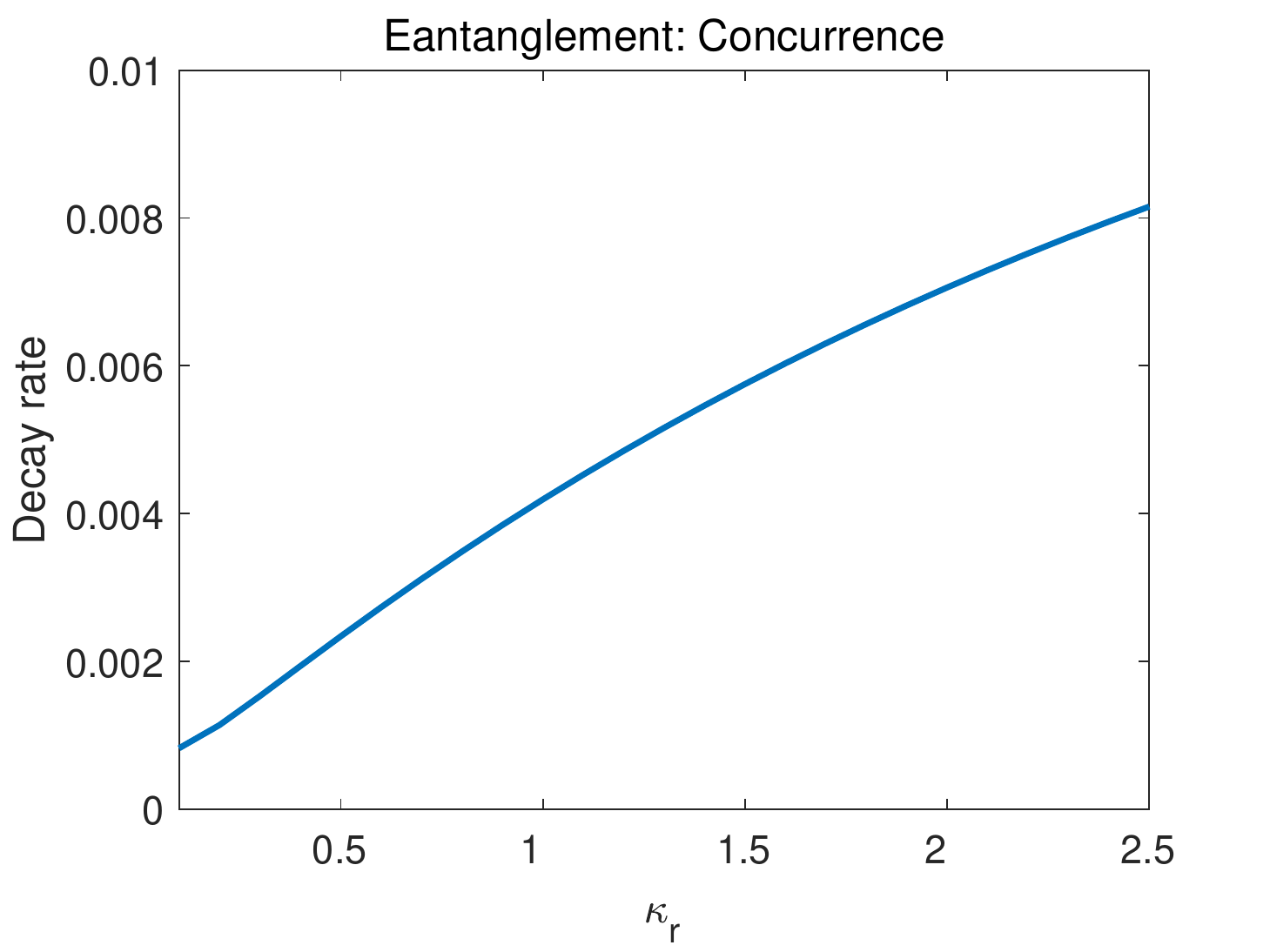}
\end{minipage}
}\subfigure[]{
\begin{minipage}{7cm}\centering
\includegraphics[scale=0.5]{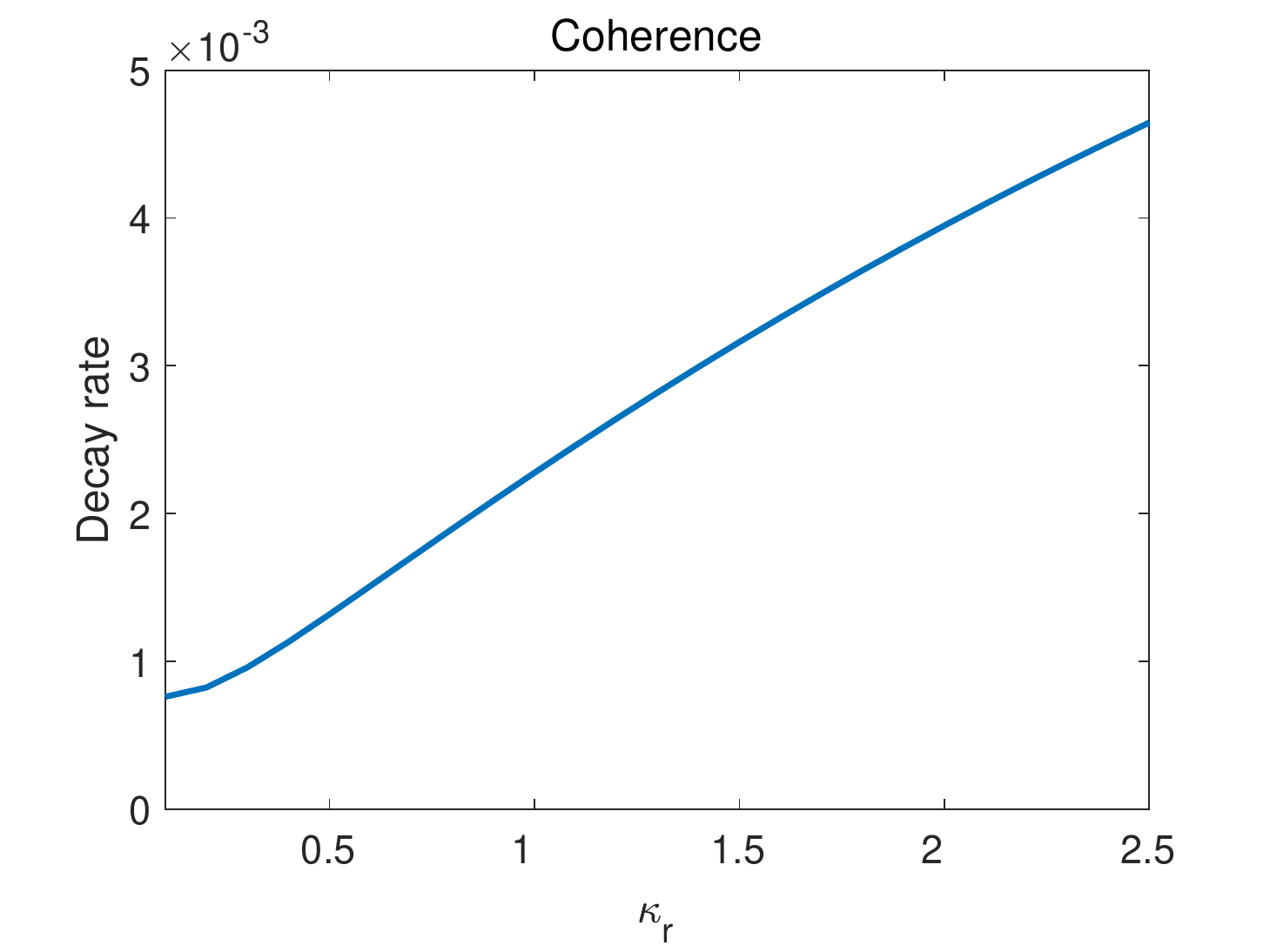}
\end{minipage}
}
\subfigure[]{
\begin{minipage}{7cm}\centering
\includegraphics[scale=0.5]{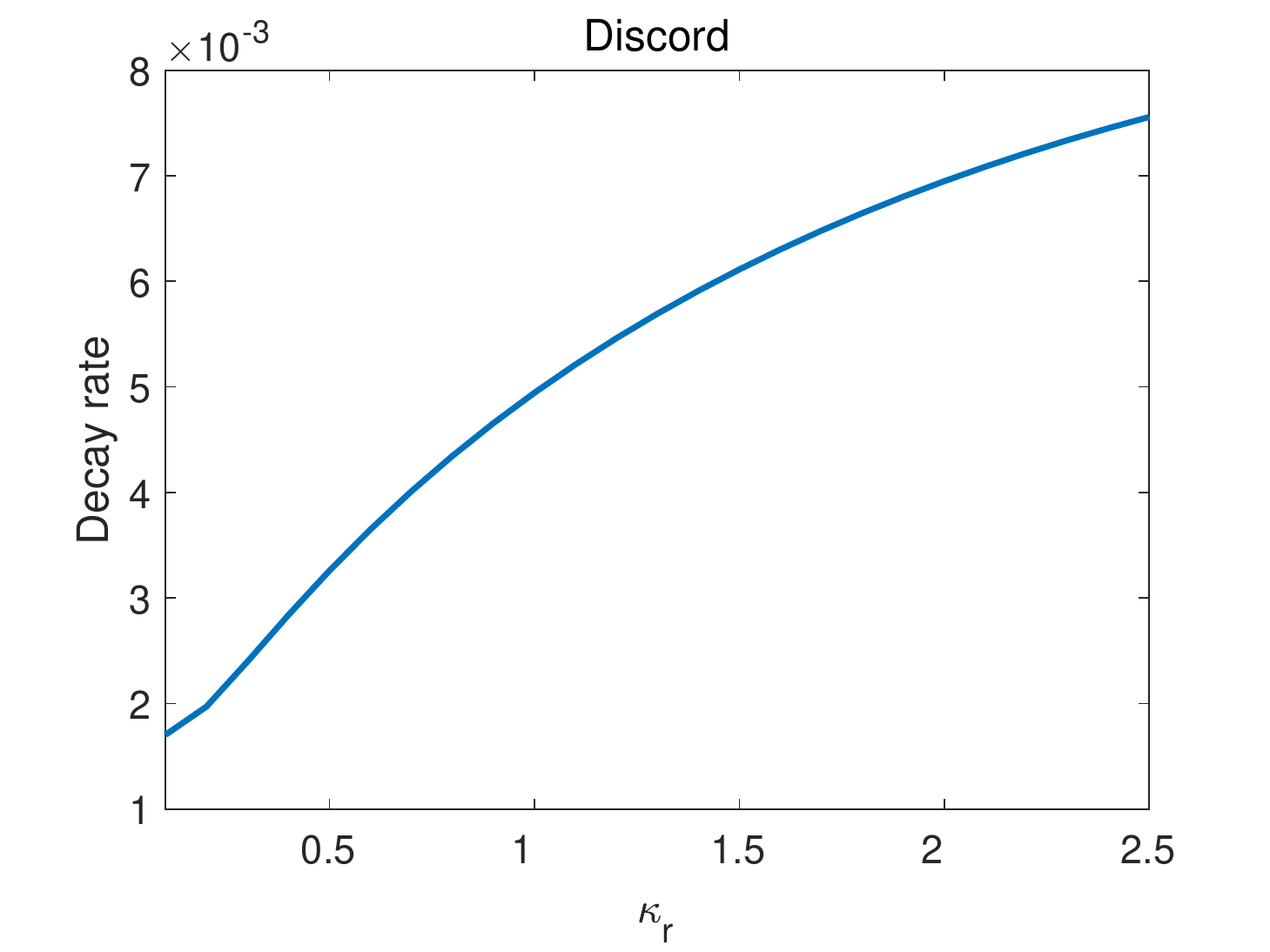}
\end{minipage}
}\subfigure[]{
\begin{minipage}{7cm}\centering
\includegraphics[scale=0.5]{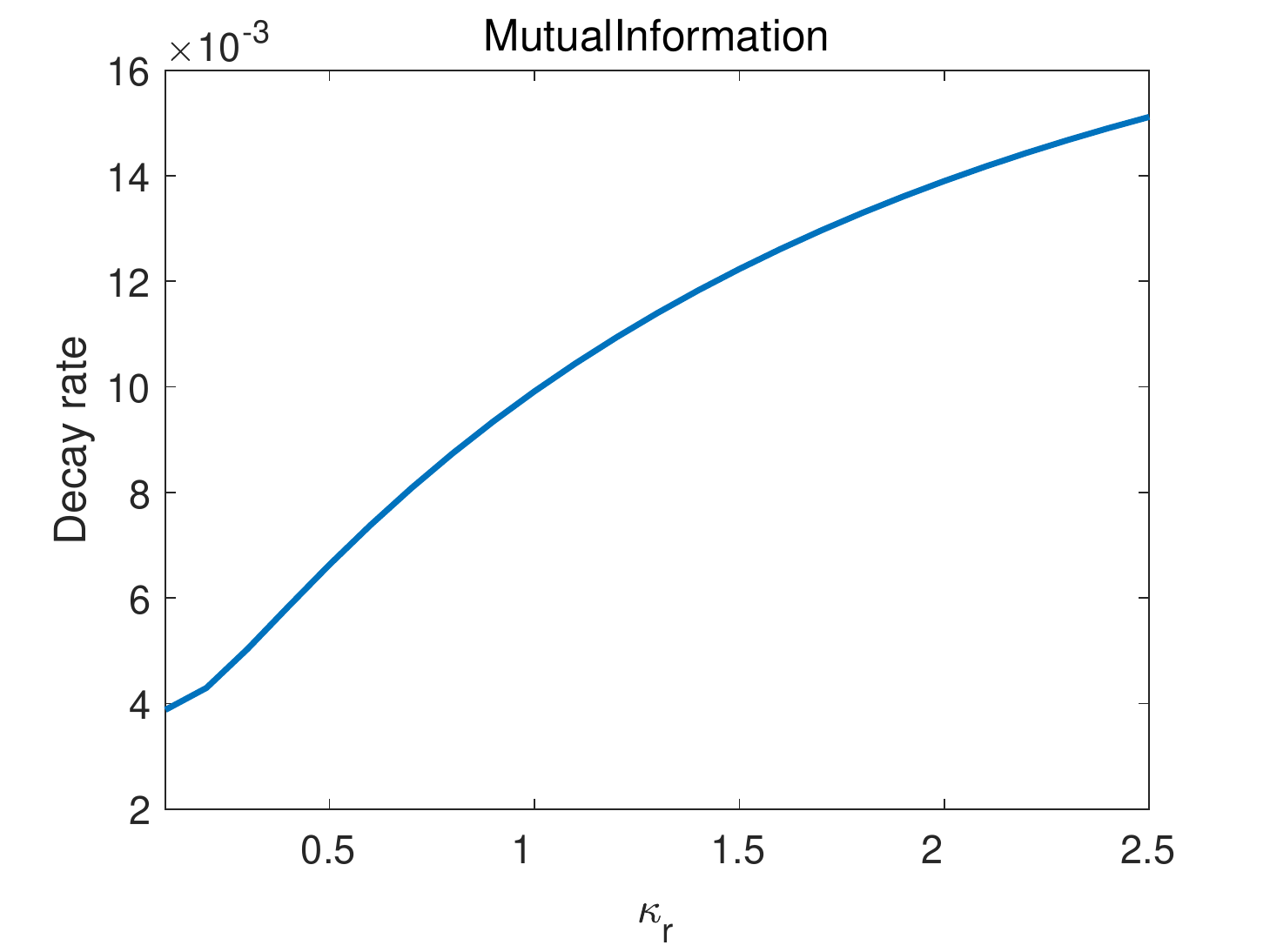}
\end{minipage}
}
\subfigure[]{
\begin{minipage}{7cm}\centering
\includegraphics[scale=0.5]{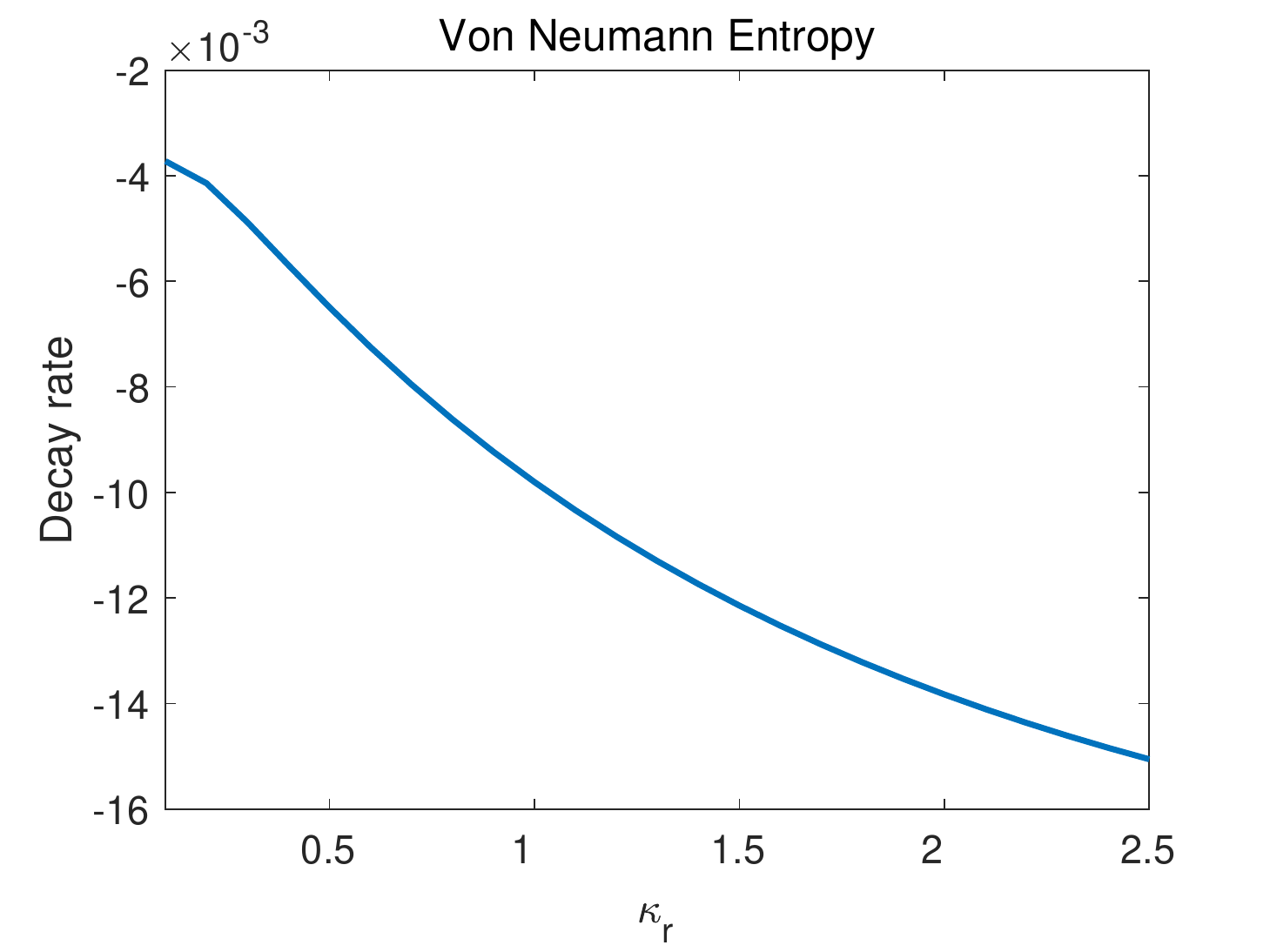}
\end{minipage}
}\subfigure[]{
\begin{minipage}{7cm}\centering
\includegraphics[scale=0.5]{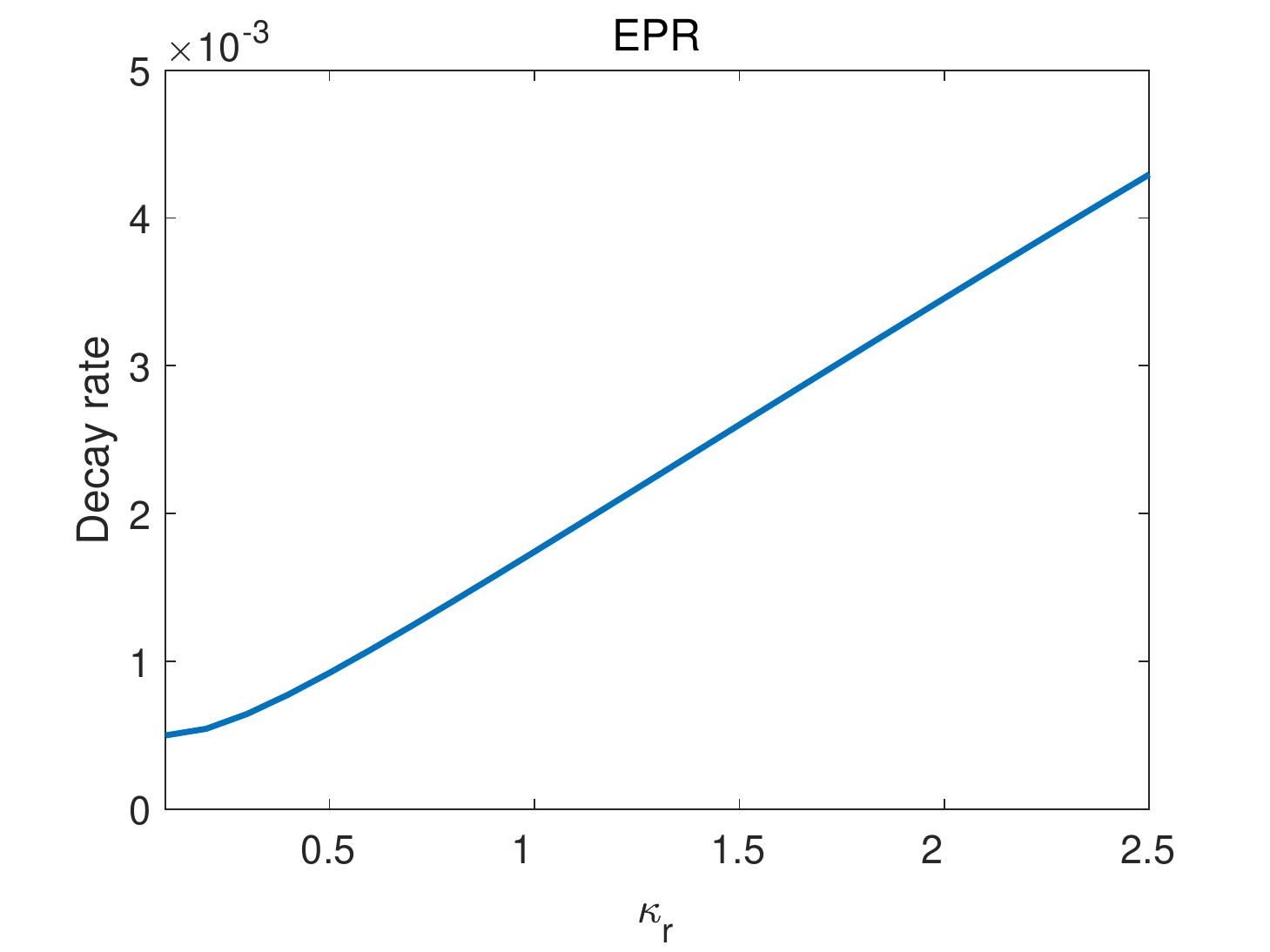}
\end{minipage}
}
\caption{The decay rates of the quantum correlations with respect to the local acceleration or curvature $\kappa_{r}$ when $t=\frac{100}{\mu^{2}}$. Other parameters are the same as those of the Fig.\ref{fig:1}}
\label{fig:8.3}
\end{figure}

The decay rates are defined as: $\frac{QC(0)-QC(t)}{t}$, where $QC$ is quantum correlation. The decay rates of correlations with respect to the mass or the angular momentum per mass are plotted in Fig.\ref{fig:8.1} \ref{fig:8.2}. The decay rates of the quantum correlations behave similarly with respect to the mass or the angular momentum per mass except the Von Neumann entropy. The decay rates of the quantum correlations increase at first and then decrease with respect to the mass. And the decay rates of the quantum correlations decrease with respect to the angular momentum per mass. However, the decay rate of the Von Neumann entropy which is negative growth decreases at first and then increases with respect to the mass, and increases with respect to the angular momentum per mass. The dependence of the decay rates on the mass and the angular momentum can be understood through the dependence of the decay rate on the $\kappa_{r}$ of the quantum correlations and the dependence of the $\kappa_{r}$ on the mass and the angular momentum. The decay rates of correlations with respect to the $\kappa_{r}$ are plotted in Fig.\ref{fig:8.3}. The decay rates of the entanglement, coherence, discord, mutual information and EPR increases as $\kappa_{r}$ increases but the decay rate of the Von Neumann entropy is negative and decreases. This shows that the thermal effect of the $\kappa_{r}$, the higher effective temperature is, the more significant dissipative effect of the environment is. Once again, we see the influence of the space time structure on the quantum correlations of the system.

\section{Nonequilibrium steady quantum correlations}
\label{sec:Nonequilibrium steady quantum correlations}

To generalize the above case to the intrinsic nonequilibrium case where the detailed balanced is not peerserved, we introduce another massless scalar field $\psi$, and assume that there is no interaction between two fields. Furthermore, we introduce the interaction between the two qubits. The hamiltonian is generalized as
\begin{equation}\begin{split}
\label{eq:54}
H_{total}&=H_{0}+H_{1}=H_{sys}+H_{field}+H_{I}\\
H_{sys}&=\frac{\omega_{1}}{2}\sigma_{z1}+\frac{\omega_{2}}{2}\sigma_{z2}+K(\sigma_{1+}\sigma_{2-}+\sigma_{1-}\sigma_{2+})\\
H_{I}&=\mu_{1}(\sigma_{1+}+\sigma_{1-})\phi(x_{1}(\tau_{1}))+\mu_{2}(\sigma_{2+}+\sigma_{2-})\psi(x_{2}(\tau_{1}))
\end{split}
\end{equation}

where $K$ is the coupling of the inter-qubits and $\mu_{1}=\mu_{2}=0.01$. The eigen energy and the eigenstates of two qubits system are: $E_{1}=-\frac{\omega_{1}+\omega_{2}}{2}$, $|\lambda_{1}\rangle=|0,0\rangle$; $E_{2}=\frac{\omega_{1}+\omega_{2}}{2}$, $|\lambda_{2}\rangle=|1,1\rangle$; $E_{3}=\kappa$, $|\lambda_{3}\rangle=\cos(\theta/2)|1,0\rangle+\sin(\theta/2)|0,1\rangle$; $E_{4}=-\kappa$, $|\lambda_{4}\rangle=-\sin(\theta/2)|1,0\rangle+\cos(\theta/2)|0,1\rangle$. We define $\kappa=\sqrt{K^{2}+(\omega_{1}-\omega_{2})^{2}/4}$ and $\theta=\arctan(2K/(\omega_{1}-\omega_{2}))$. In this paper, we only consider symmetric case which means $\omega_{1}=\omega_{2}$. In the eigen energy basis, we can define two groups of transitions operators:

\begin{equation}\begin{split}
\label{eq:55}
V_{1,1}&=\cos(\theta/2)(|\lambda_{1}\rangle\langle\lambda_{1}|+|\lambda_{4}\rangle\langle\lambda_{2}|)\\
V_{1,2}&=\sin(\theta/2)(|\lambda_{3}\rangle\langle\lambda_{2}|-|\lambda_{1}\rangle\langle\lambda_{4}|)\\
V_{2,1}&=\sin(\theta/2)(|\lambda_{1}\rangle\langle\lambda_{3}|-|\lambda_{4}\rangle\langle\lambda_{2}|)\\
V_{2,2}&=\cos(\theta/2)(|\lambda_{3}\rangle\langle\lambda_{2}|+|\lambda_{1}\rangle\langle\lambda_{4}|)
\end{split}
\end{equation}

with transition frequency $\Omega_{1}=E_{2}-E_{3}$ and $\Omega_{2}=E_{2}+E_{3}$.

A stationary detector near the horizon will experience a thermal bath with effective temperature $\frac{k_{r}}{2\pi}$. In his viewpoint, any two different points have different temperature. Now if we separate the two-qubit with a finite distance along the radius, the system is equivalently the case being connected to two independent bath. Therefore, one can derive a nonequilibrium master equation for two separated atoms in the observer's frame. Using the observer's proper time, the master equation can be derived (neglect the contribution of the principal value which can only modify the energy level) ~\cite{Dynamics of nonequilibrium thermal entanglement}~\cite{Quantum thermalization of two coupled two-level systems in eigenstate and bare-state representations}~\cite{Steady-state entanglement and coherence of two coupled qubits in equilibrium and nonequilibrium environments}

\begin{equation}\begin{split}
\label{eq:56}
\frac{d\rho_{sys}(\tau)}{d\tau}&=-i[H_{sys},\rho_{sys}(\tau)]+\mathscr{L}_{1}[\rho_{sys}]+\mathscr{L}_{2}[\rho_{sys}]\\
\mathscr{L}_{j}[\rho_{sys}]&=\sum_{\mu=1}^{2}G^{i}(-\omega_{\mu})(2V_{j,\mu}\rho_{sys}V_{j,\mu}^{\dagger}-{\rho_{sys},V_{j,\mu}^{\dagger}V_{j,\mu}})\\
&  +G^{i}(\omega_{\mu})(2V_{j,\mu}^{\dagger}\rho_{sys}V_{j,\mu}-{\rho_{sys},V_{j,\mu}V_{j,\mu}^{\dagger}})
\end{split}
\end{equation}

$G^{i}$ corresponds to the Fourier transform of Green function for different fields. From Eqn.\eqref{eq:56}, it is sufficient to obtain the steady state, and the concrete expression is in Appendix~\ref{sec:Steady state expression}. The frequency gap of the two identical atoms is different when the two atoms keep different separation distances from a stationary detector's viewpoint due to the red-shift effect. However, we are not interested in the red-shift effect here and always set the frequency gap of two atoms is the same.

%Under these consideration, we now investigate the quantum correlations of the nonequilibrium steady state in the bare basis. The $\Delta r$ measures nonequilibriumness. It is shown that the finial state contains no information about the initial state. It is possible to harvest the quantum correlation from the vacuum as we seen. The concurrence and the mutual information in general display nonmonotonic behaviors with respect to the $\Delta r$ as shown in Fig.\ref{fig:9}(a)(c). This nonmonotonic behavior can be understood as a competition between populations and coherence.~\cite{Steady-state entanglement and coherence of two coupled qubits in equilibrium and nonequilibrium environments} The discord, fidelity and Bell-CHSH increase monotonically with the nonequilibrium condition characterized by the $\Delta r$ to a constant as shown in Fig.\ref{fig:9}(b)(d). The coherence and the Von Neumann entropy decrease monotonically to a constant.

\begin{figure}[htbp]
\centering
\subfigure[]{
\begin{minipage}{7cm}\centering
\includegraphics[scale=0.5]{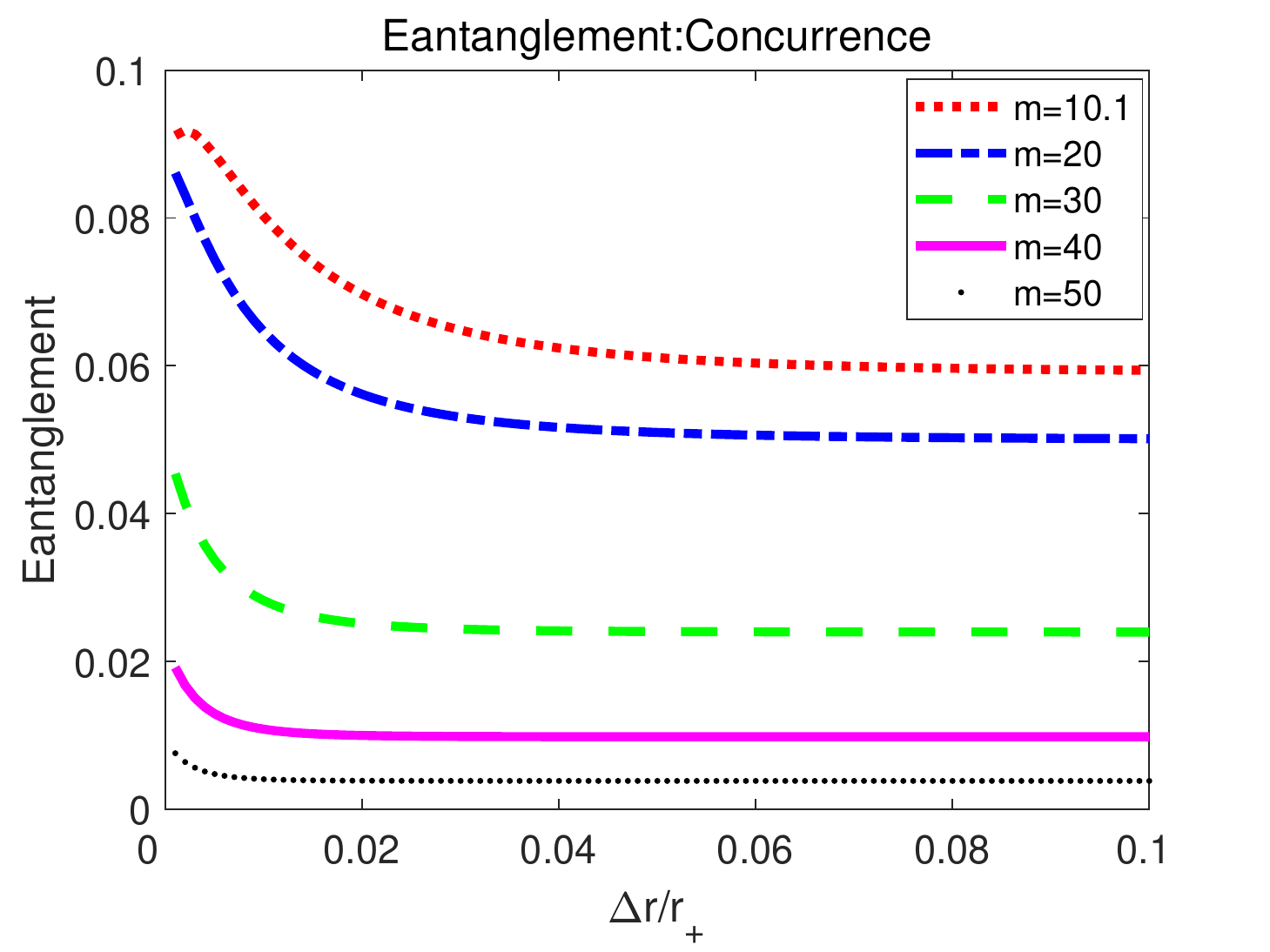}
\end{minipage}
}
\subfigure[]{
\begin{minipage}{7cm}\centering
\includegraphics[scale=0.5]{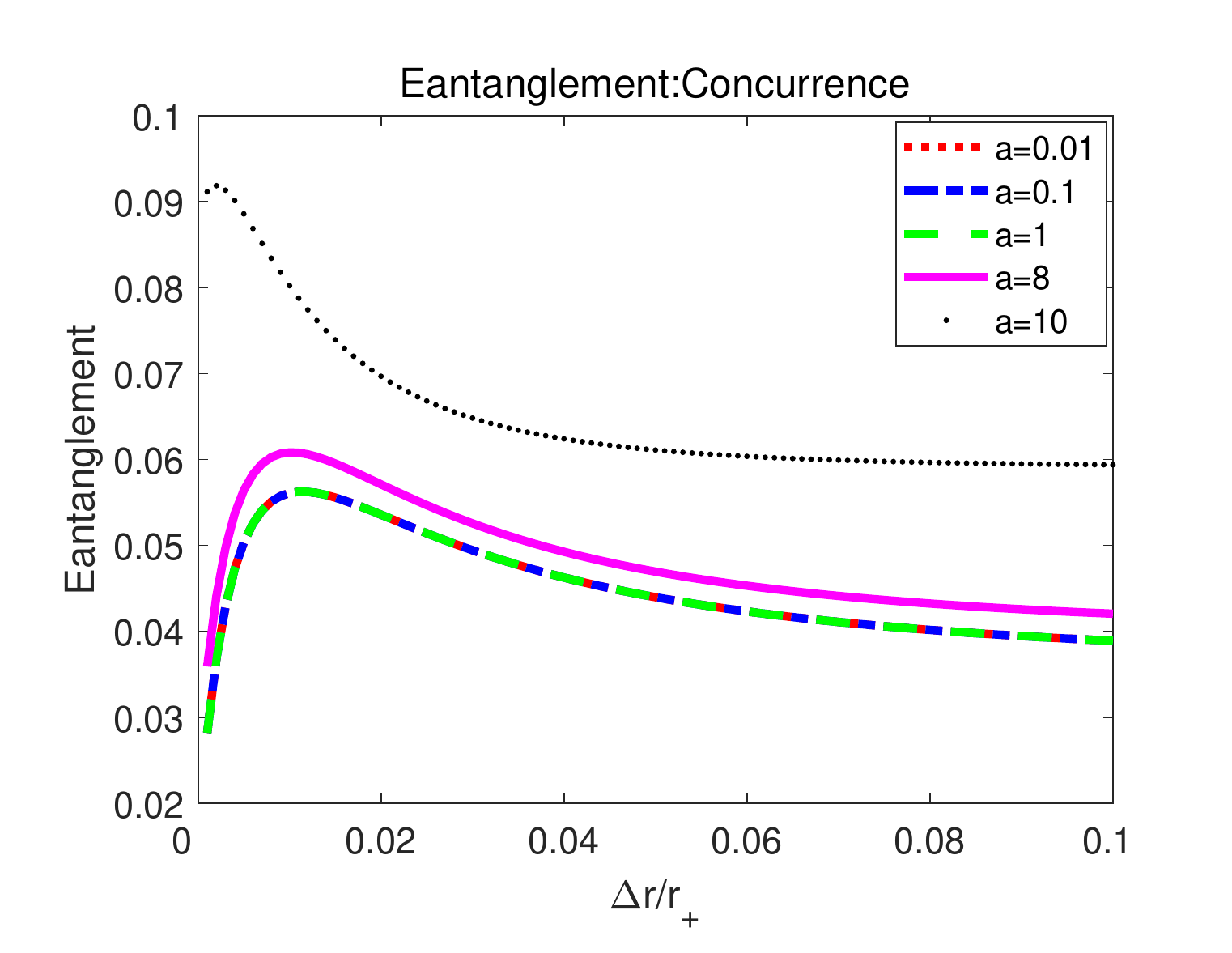}
\end{minipage}
}
\subfigure[]{
\begin{minipage}{7cm}\centering
\includegraphics[scale=0.5]{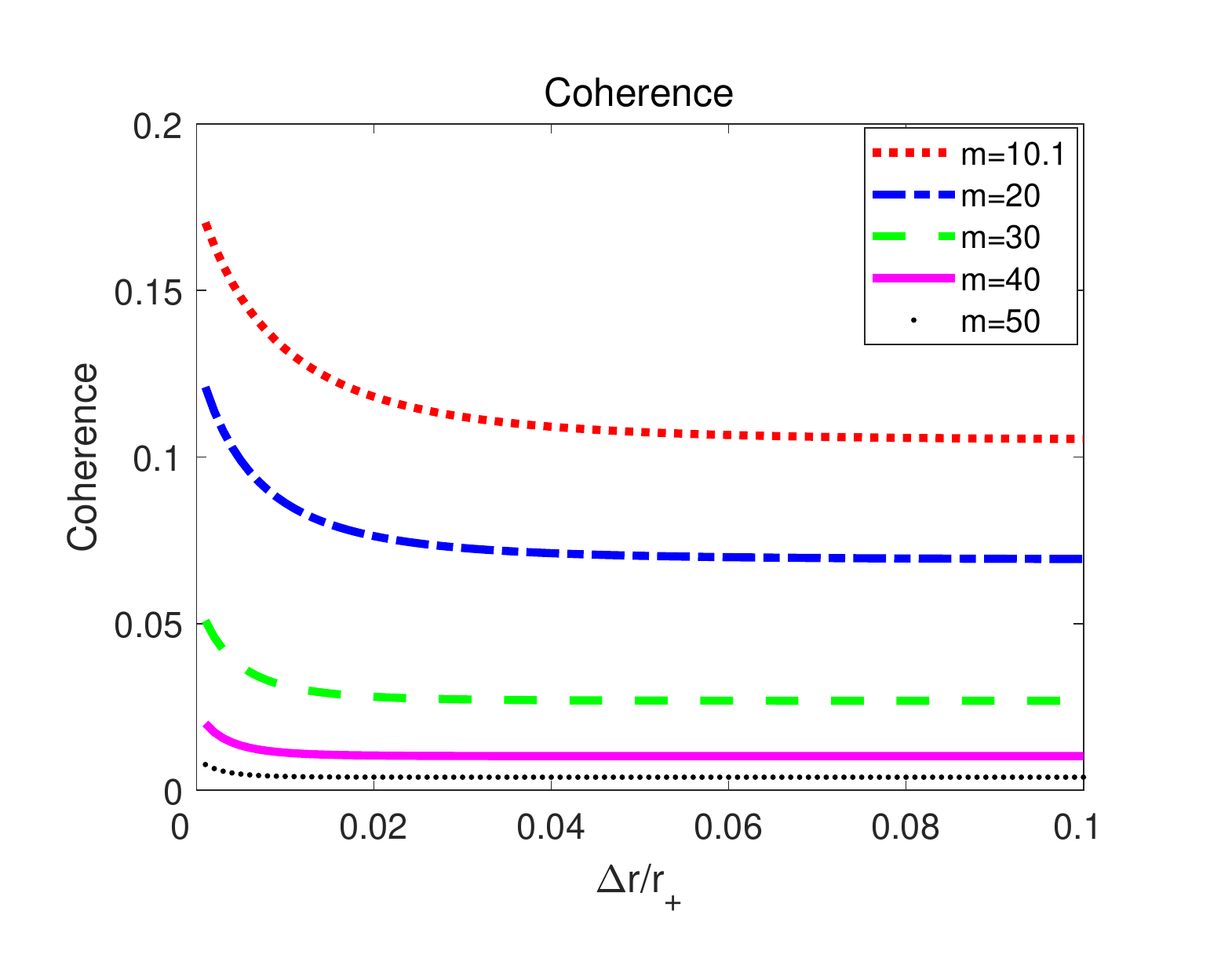}
\end{minipage}
}
\subfigure[]{
\begin{minipage}{7cm}\centering
\includegraphics[scale=0.5]{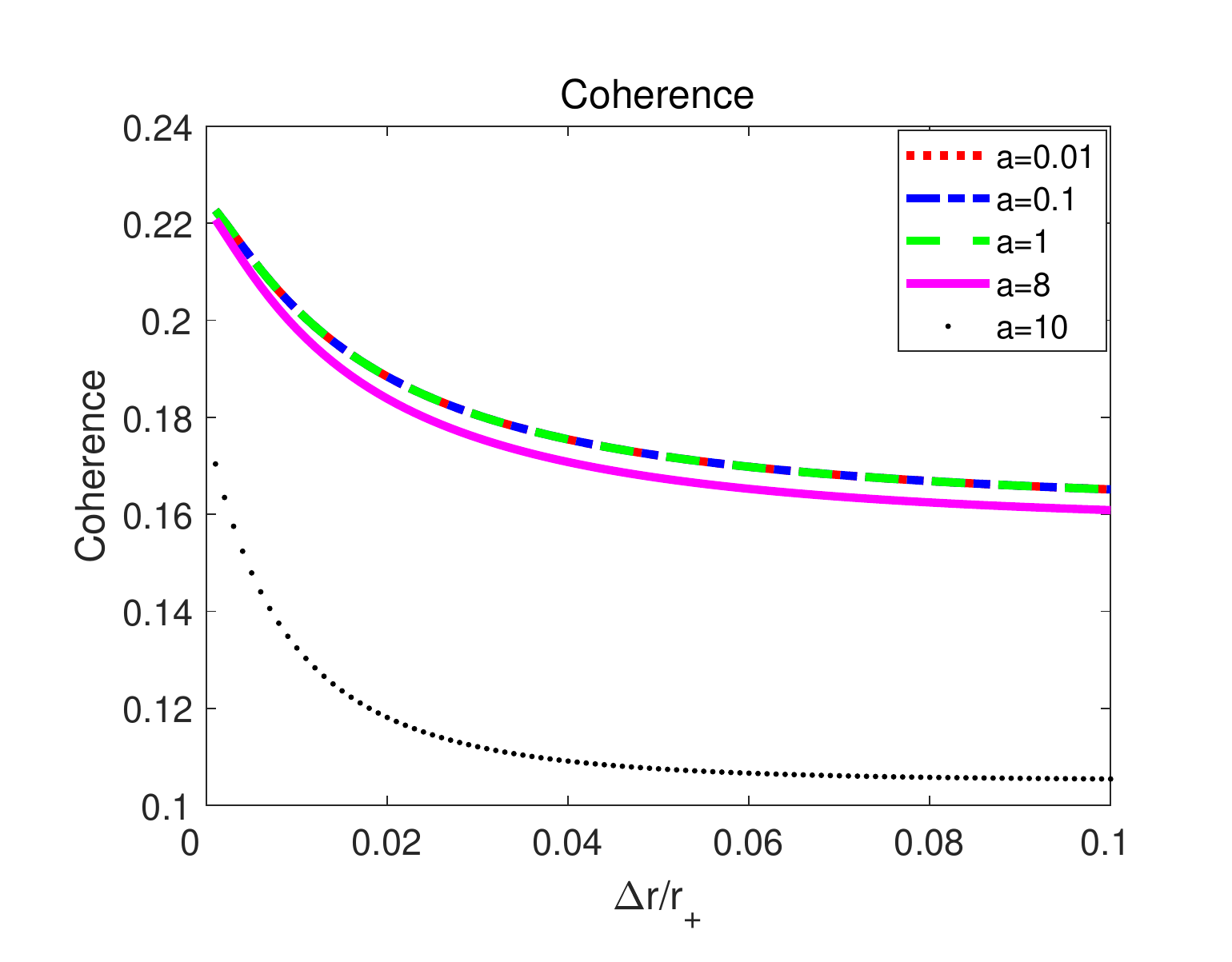}
\end{minipage}
}
\subfigure[]{
\begin{minipage}{7cm}\centering
\includegraphics[scale=0.5]{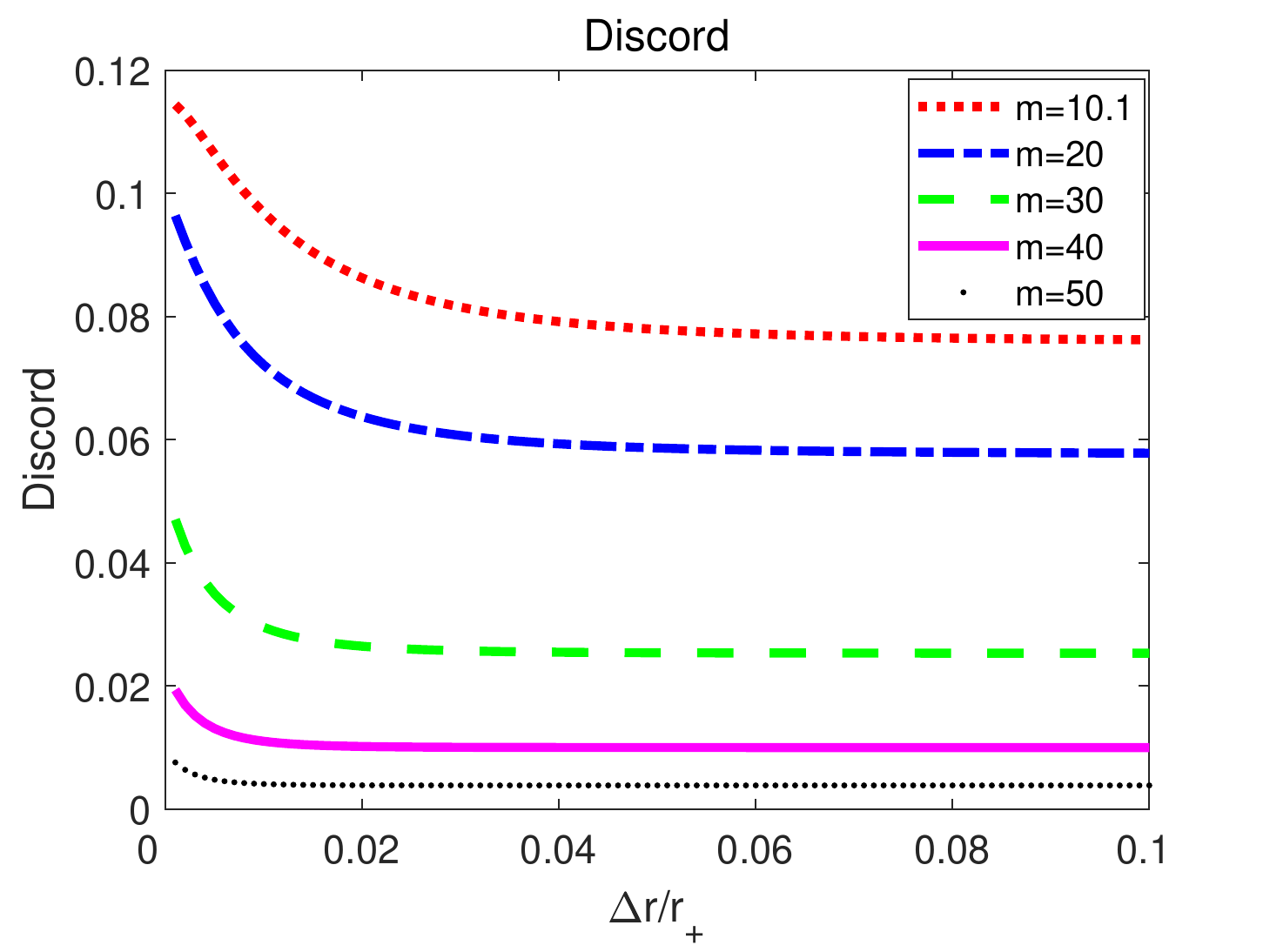}
\end{minipage}
}
\subfigure[]{
\begin{minipage}{7cm}\centering
\includegraphics[scale=0.5]{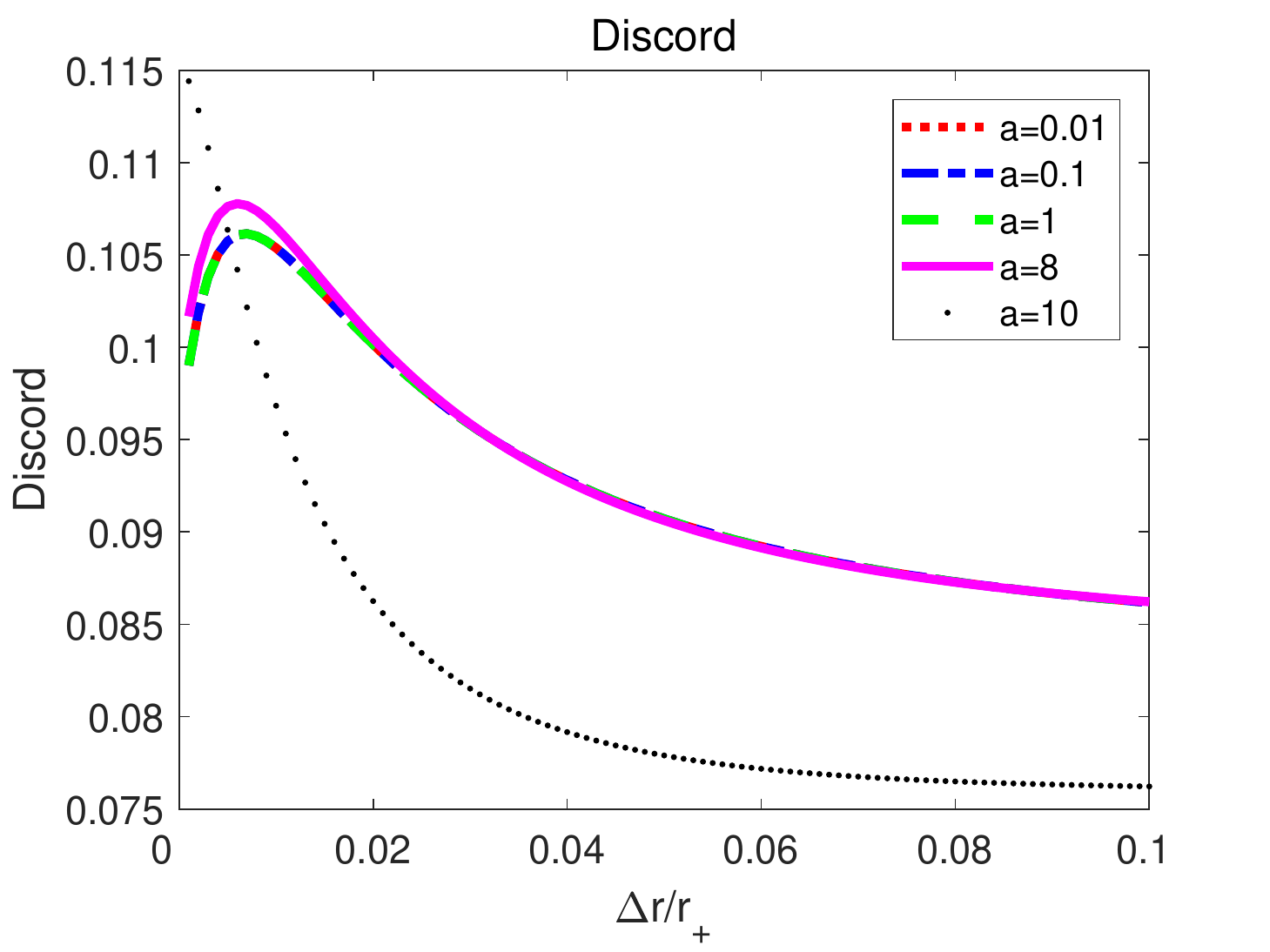}
\end{minipage}
}
\caption{We consider a nonequilibrium scenario: one of the two atoms is stationary at $1.006r_{+}$ while another is at $(1.006+\Delta r)r_{+}$. Quantum correlations at nonequilibrium steady state varying with the separation distance at angular momentum $a=10$ or the mass $M=10$.}
\label{fig:11.1}
\end{figure}

\begin{figure}[htbp]
\centering
\subfigure[]{
\begin{minipage}{7cm}\centering
\includegraphics[scale=0.5]{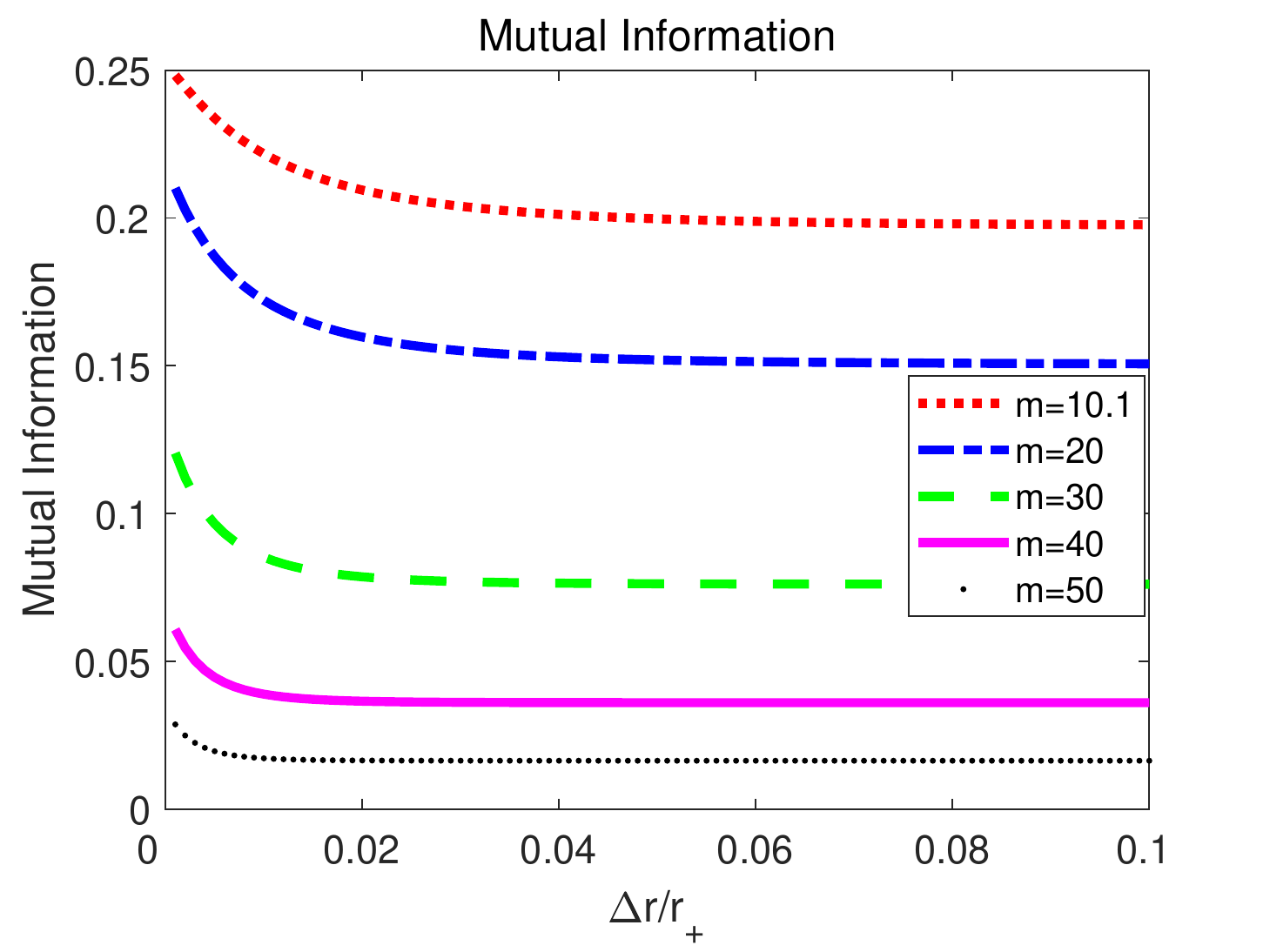}
\end{minipage}
}
\subfigure[]{
\begin{minipage}{7cm}\centering
\includegraphics[scale=0.5]{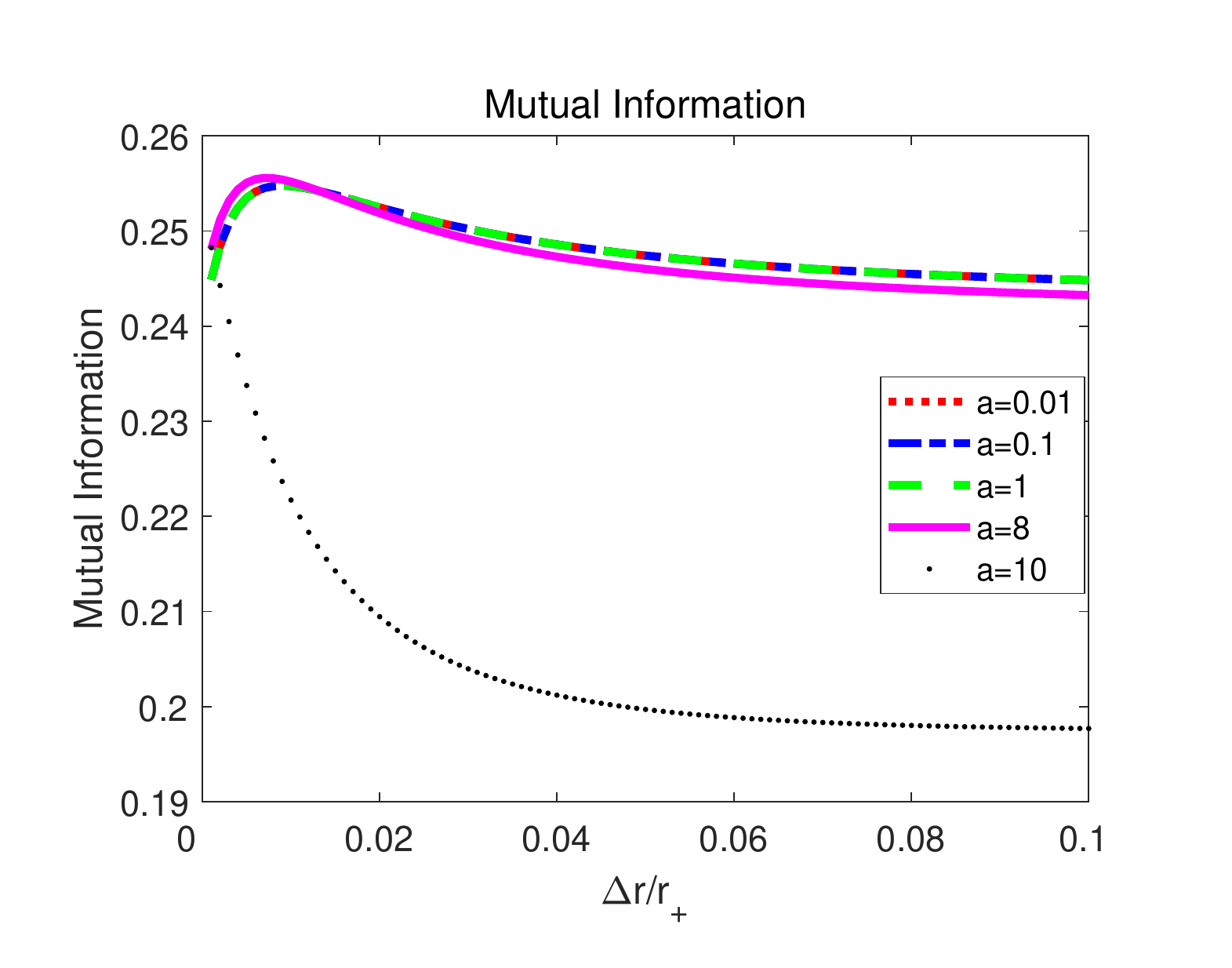}
\end{minipage}
}
\subfigure[]{
\begin{minipage}{7cm}\centering
\includegraphics[scale=0.5]{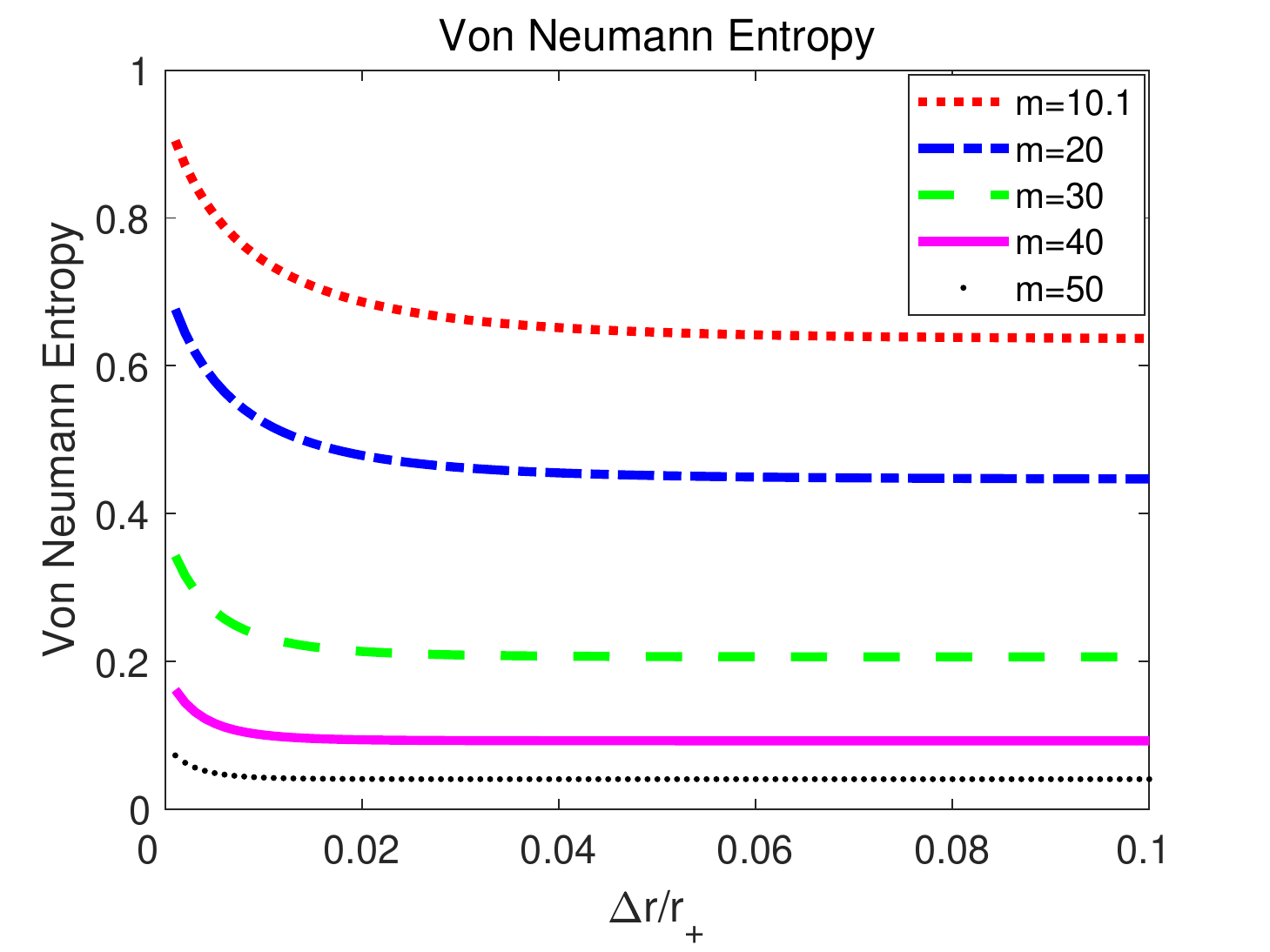}
\end{minipage}
}
\subfigure[]{
\begin{minipage}{7cm}\centering
\includegraphics[scale=0.5]{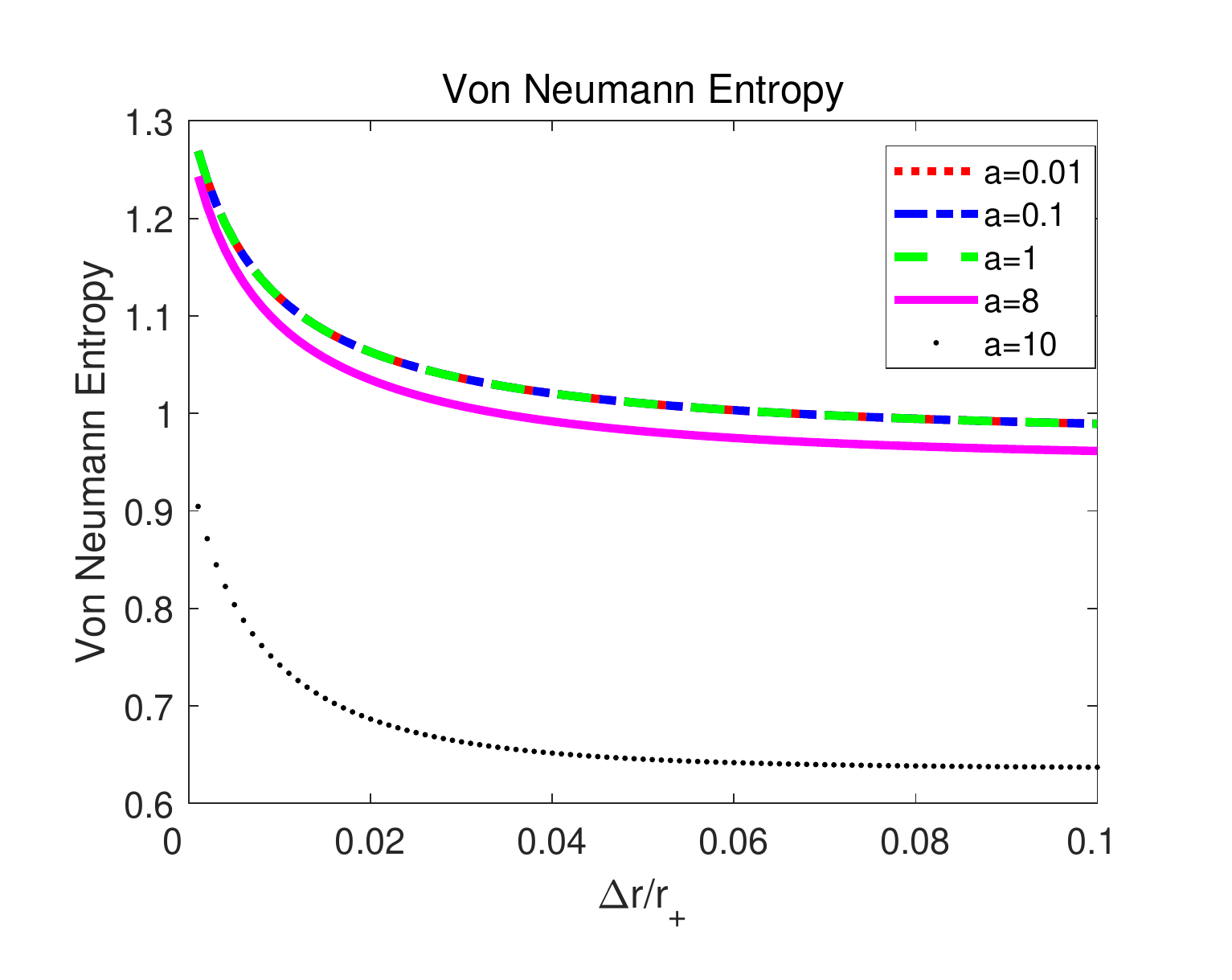}
\end{minipage}
}
\caption{We consider a nonequilibrium scenario: one of the two atoms is stationary at $1.006r_{+}$ while another is at $(1.006+\Delta r)r_{+}$. Quantum correlations at nonequilibrium steady state varying with separation distance at angular momentum $a=10$ or the mass $M=10$.}
\label{fig:11.2}
\end{figure}

\begin{figure}[htbp]
\centering
\subfigure[]{
\begin{minipage}{7cm}\centering
\includegraphics[scale=0.5]{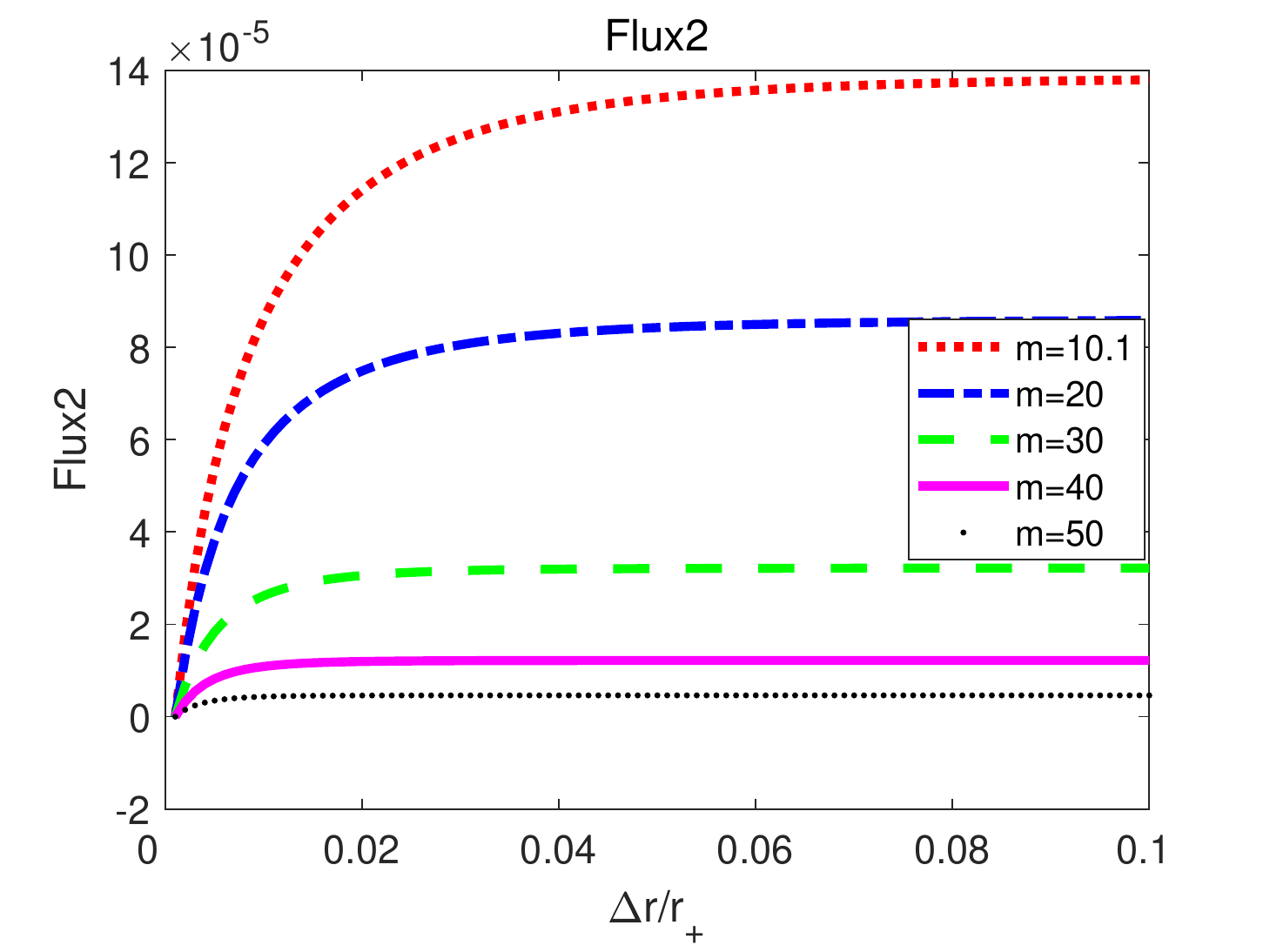}
\end{minipage}
}\subfigure[]{
\begin{minipage}{7cm}\centering
\includegraphics[scale=0.5]{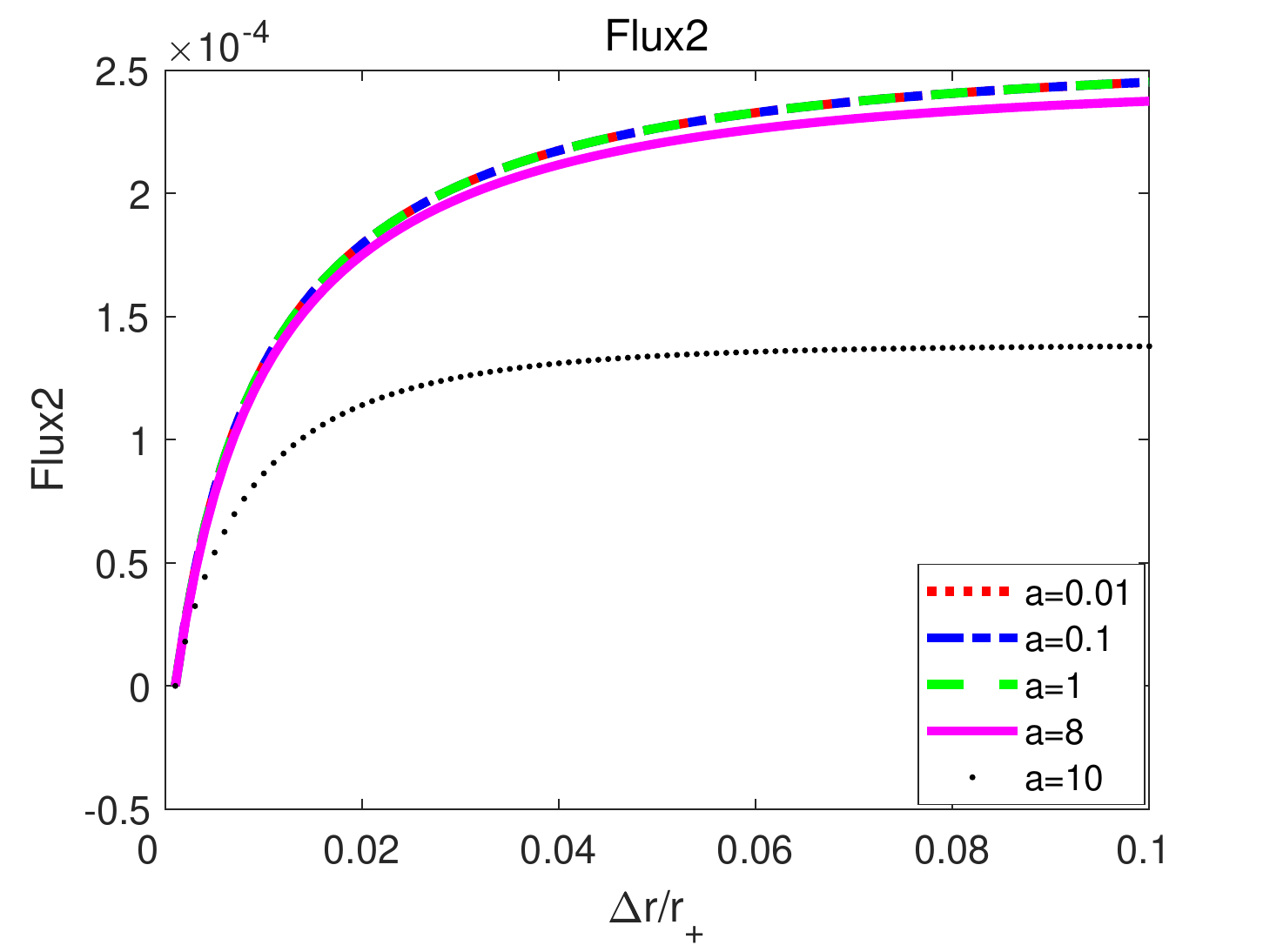}
\end{minipage}
}
\subfigure[]{
\begin{minipage}{7cm}\centering
\includegraphics[scale=0.5]{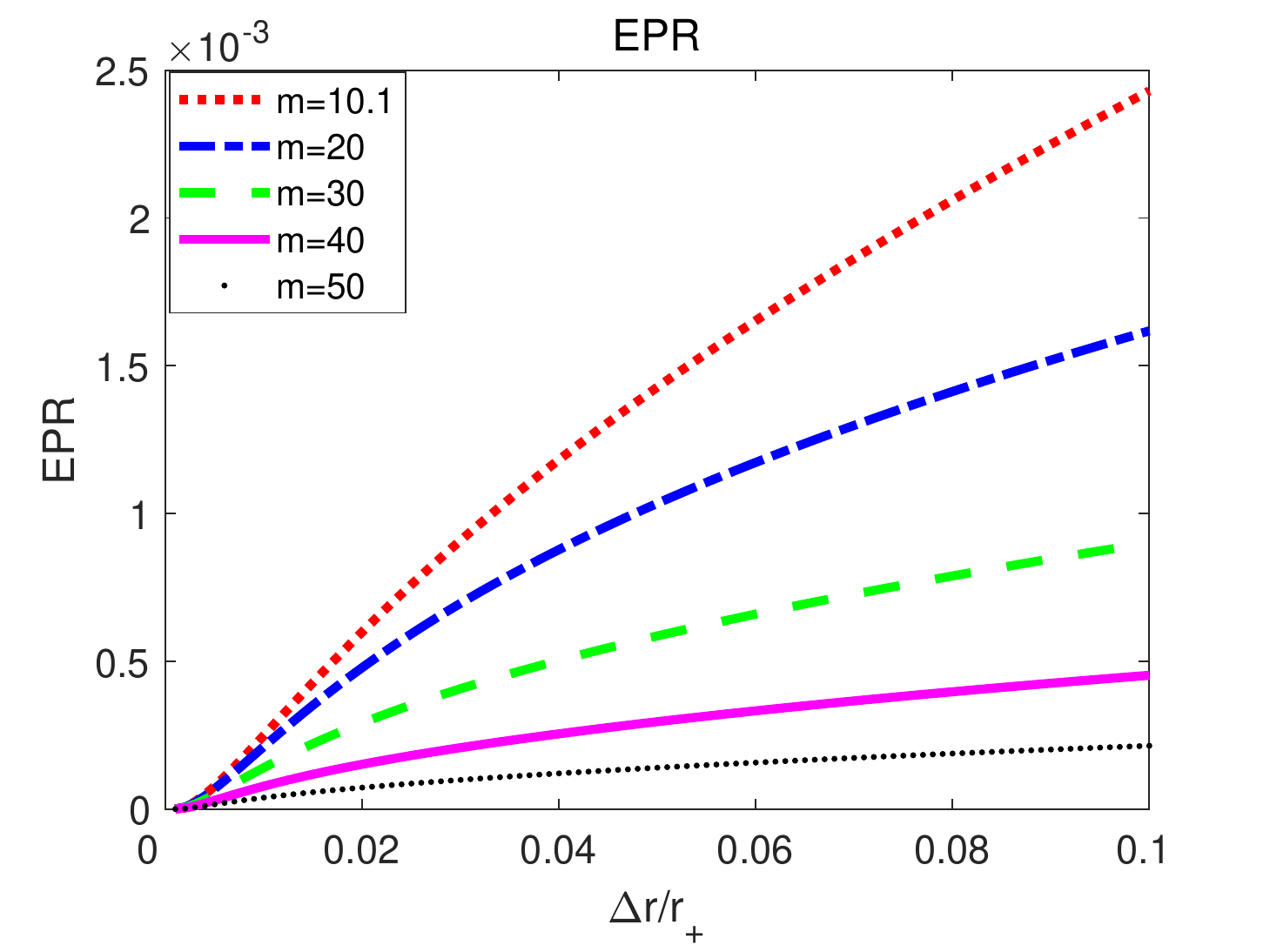}
\end{minipage}
}\subfigure[]{
\begin{minipage}{7cm}\centering
\includegraphics[scale=0.5]{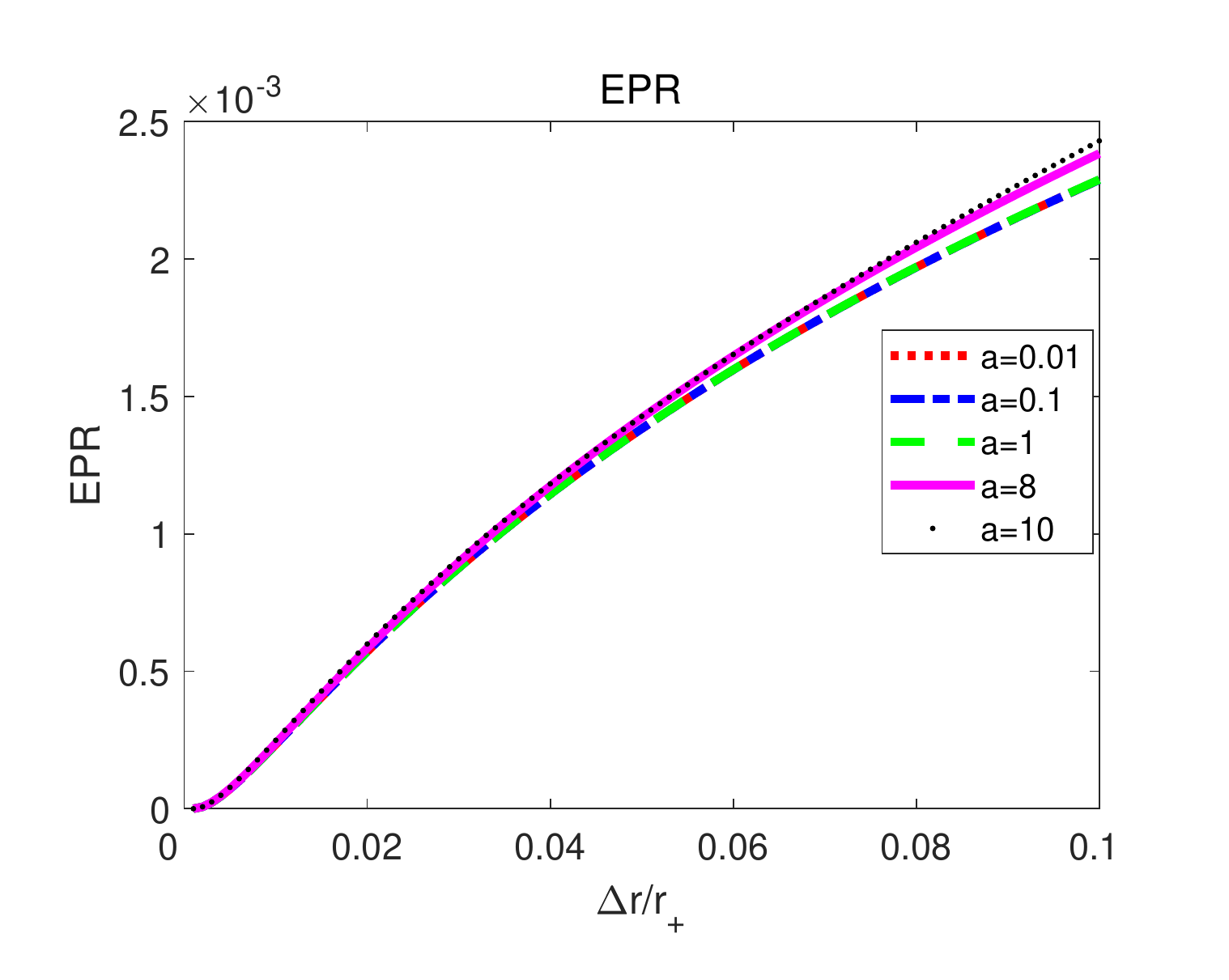}
\end{minipage}
}
\subfigure[]{
\begin{minipage}{7cm}\centering
\includegraphics[scale=0.5]{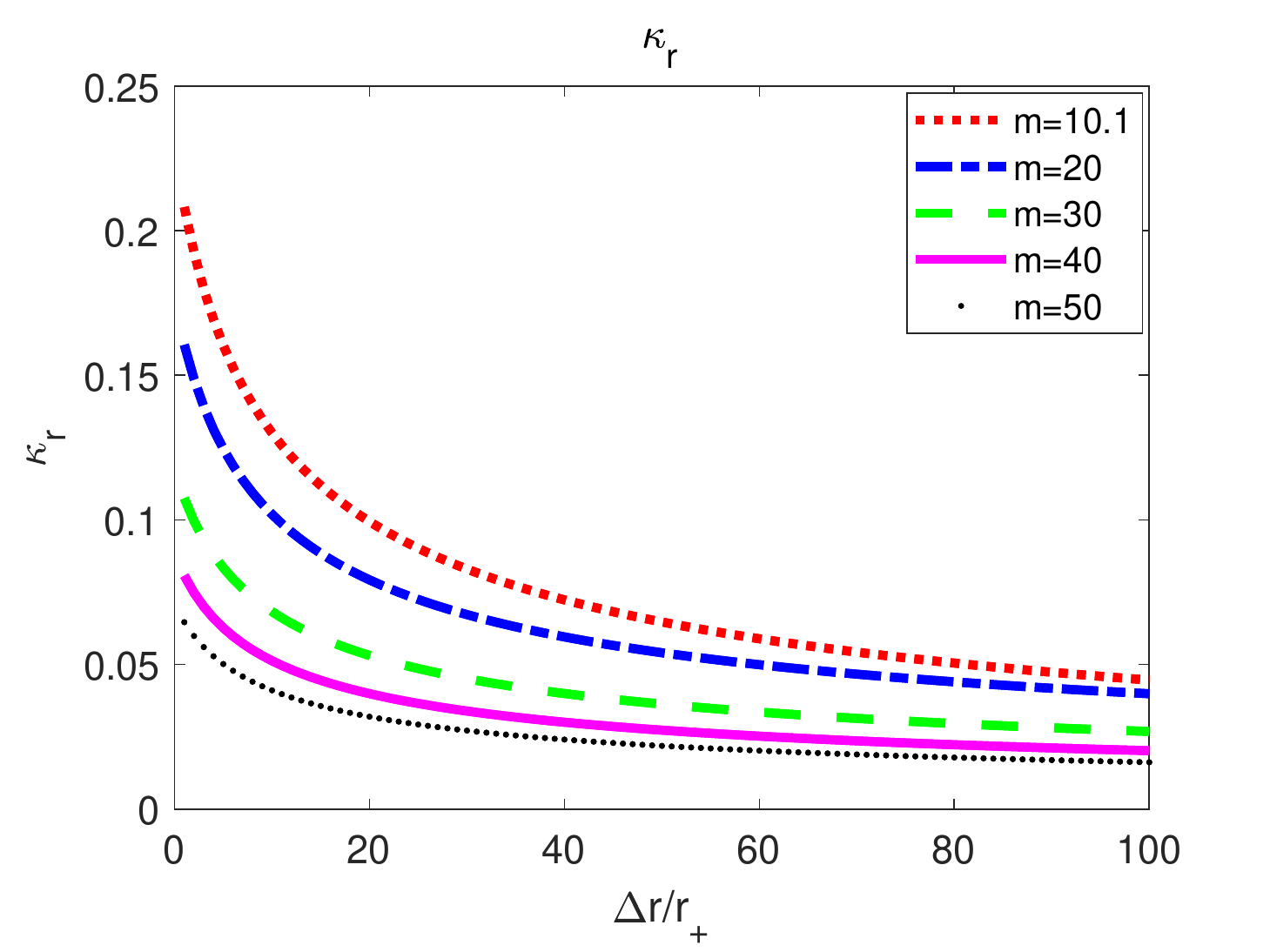}
\end{minipage}
}\subfigure[]{
\begin{minipage}{7cm}\centering
\includegraphics[scale=0.5]{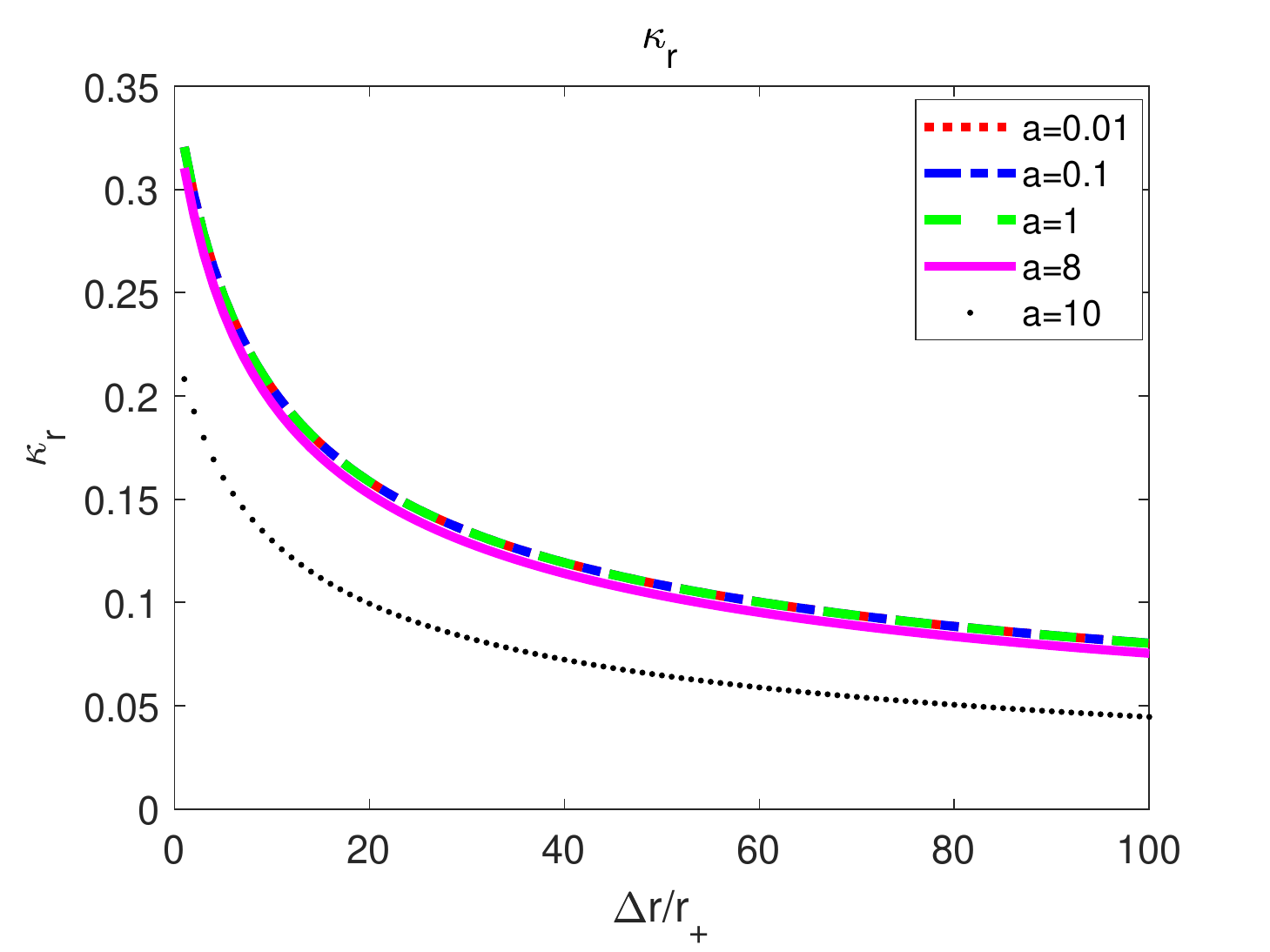}
\end{minipage}
}
\caption{The a) b) flux and c) d) EPR at nonequilibrium steady state varying with separation distance at angular momentum $a=10$ or the mass $M=10$. e) f) the $\kappa_{r}$ varying with separation distance at angular momentum $a=10$ or the mass $M=10$. One of the two atoms is stationary at $1.006r_{+}$ while another is at $(1.006+\Delta r)r_{+}$.}
\label{fig:11.3}
\end{figure}

Under these considerations, we now investigate the quantum correlations of the nonequilibrium steady state in the bare basis. The $\Delta r$ measures nonequilibrium. This is because the difference in radius reflects the difference in the local acceleration or the local space time curvature. This shows difference in temperatures through Unruh-Hawking effect. Thus the case is similar to the one of the two qubits couple to individual bath separately with the different temperature. Therefore the system is in nonoequilibrium. However, there is an interesting question? Are the quantum correlations in curved space time back ground the same or different from the case where the system is coupled to two corresponding bathes? Addressing this issue can help us to understand whether the effects of curved space time is equivalent to the temperature on the global correlation level. If there is only one field and only qubit-field interaction as the case in ~\ref{sec:Equilibrium quantum correlations in curved space time}. It has shown that the space time curvature, the acceleration and the temperature influence differently on the quantum correlations.~\cite{Cosmological quantum entanglement}~\cite{Entanglement in curved spacetimes and cosmology} Now let's go back and look at this case. Strictly speaking, the coupling of the inter-qubits $K$ relies on the distance and the space time curvature. This directly reflects that the space time curvature, the acceleration and the temperature influence differently on the system. However, we set $K$ is a constant for simplicity. The quantum correlations of the system are established by the interaction between the qubits rather than the fields since they do not correlate. The effects of the space time curvature between the two point-like qubits on the quantum correlations vanish. Even so, we can still perceive the different effects of the space time curvature and the temperature on the quantum correlations in our setting up. We found the separation distance directly determines the property of the system due to the space time curvature. However, the system coupled to two corresponding bathes is not related to the separation distance between the two qubits. Moreover, the redshift effect caused by the local curvature can modify the energy levels of the system, and further influence the quantum correlations of the system. These show the difference between the effects of the curved space time and the temperature on the global correlation level. Besides that, our system is very similar to the system coupled to two corresponding bathes. As mentioned before, for the purpose of only considering nonequilibrium effect induced by different locations, we omit the redshift effect.

It is shown that the finial state contains no information about the initial state. We plot the correlations varying with $\Delta r$ under different mass or different angular momentum. On the whole, the correlations arrive at a steady value when $\Delta r$ is large. For the entanglement, the discord and the mutual information in Fig.\ref{fig:11.1}(a)(b)(f), Fig.\ref{fig:11.2}(b), they vary non-monotonously with $\Delta r$. More importantly, they can be amplified by the nonequilibrium. The coherence and the Von Neumann entropy monotonously decrease to a constant with $\Delta r$ for both different masses and the angular momentum. The nonequilibrium appears to reduce the coherence and generates the correlation between the system and the environment.

%The behavior of discord can be understood as a competition between the decoherence effect among the systems due to the environment (field) and the system-environment (detectors-field) entanglement that such decoherence also tends to generate. ~\cite{Thermal amplification of field-correlation harvesting}~\cite{Equilibrium and Nonequilibrium Quantum Correlations Between Two Accelerated Detectors}
%When we fix the angular momentum, the mutual information with the different masses behave similarly, they decrease to a constant. We fix the mass, the mutual information with the different masses behave differently: when the angular momentum per mass is far less than the mass, the mutual information varies non-monotonously with $\Delta r$; when the angular momentum per mass is close to the mass, the mutual information decreases monotonously. The von nuemann entropy decreases to a constant with $\Delta r$.

The flux measures the energy exchange between the system and the environment. The energy flux from the $ith$ field to the system at the steady state is given by $I_{i}=Tr[\mathscr{L}_{i}(\rho_{sys})H_{sys}]$.(we can check $I_{1}+I_{2}=0$ which satisfies flux conserved.) The flux increases to a constant in Fig.\ref{fig:11.3}(a)(b). This means that the energy exchange capacity of the system has an upper bound and is limited to the environment. We can also define an effective EPR: $I(\frac{1}{T_{1}}-\frac{1}{T_{2}})$, the temperature is related to the local curvature or the acceleration $k_r$. The EPR increases when the $\Delta r$ increases as we expect in Fig.\ref{fig:11.3}(c)(d).

\begin{figure}[htbp]
\centering
\subfigure[]{
\begin{minipage}{7cm}\centering
\includegraphics[scale=0.5]{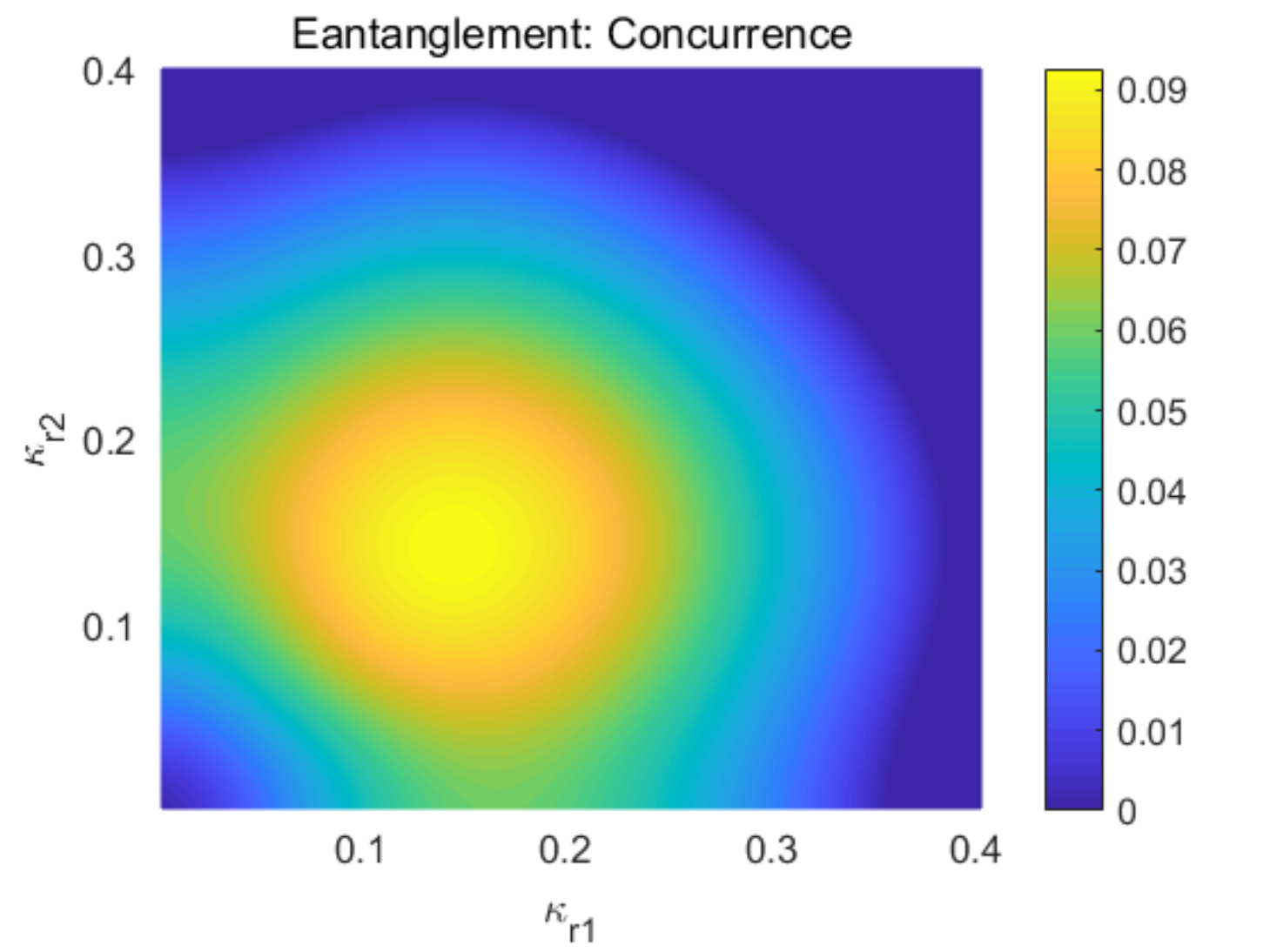}
\end{minipage}
}
\subfigure[]{
\begin{minipage}{7cm}\centering
\includegraphics[scale=0.5]{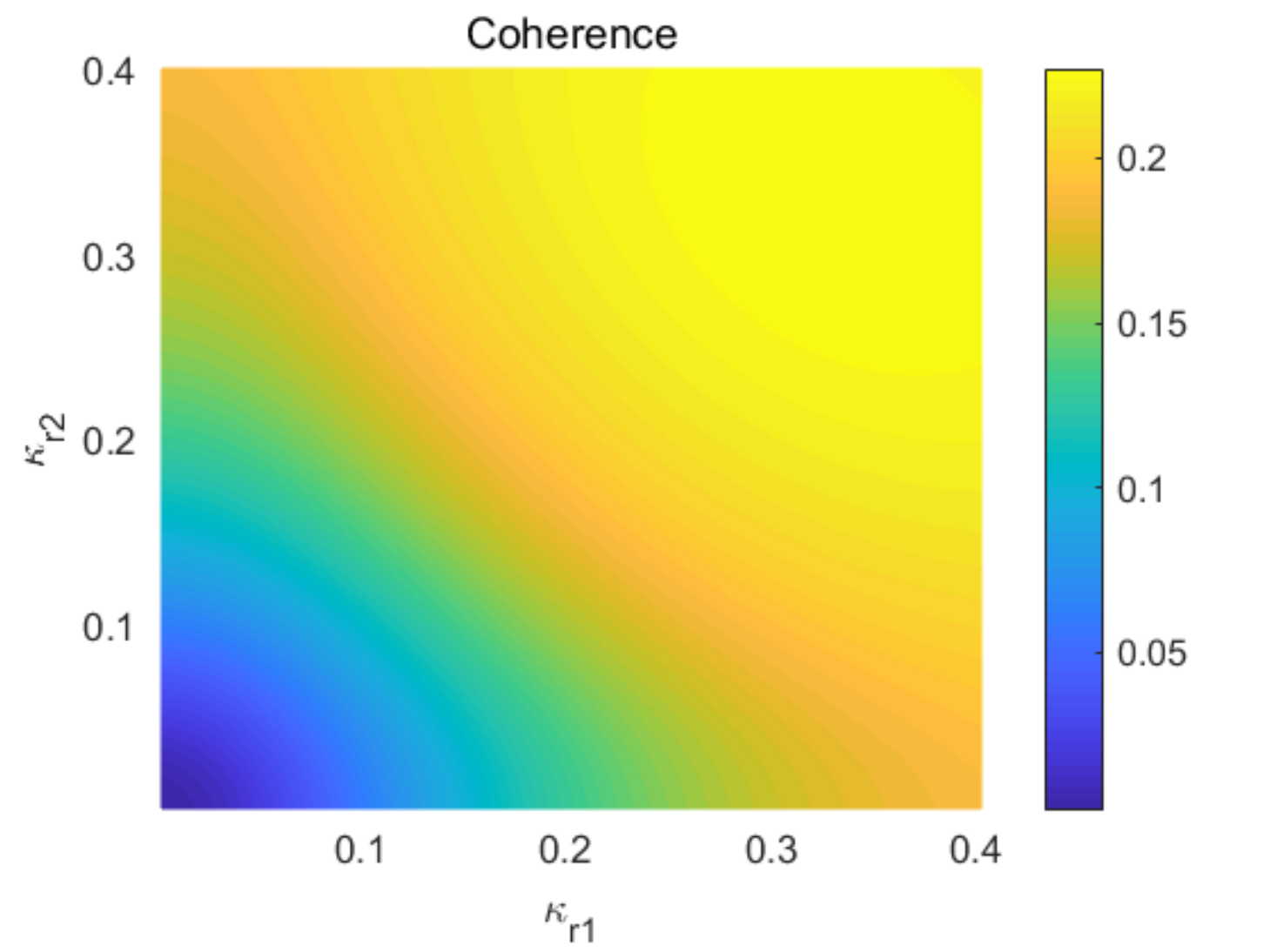}
\end{minipage}
}
\subfigure[]{
\begin{minipage}{7cm}\centering
\includegraphics[scale=0.5]{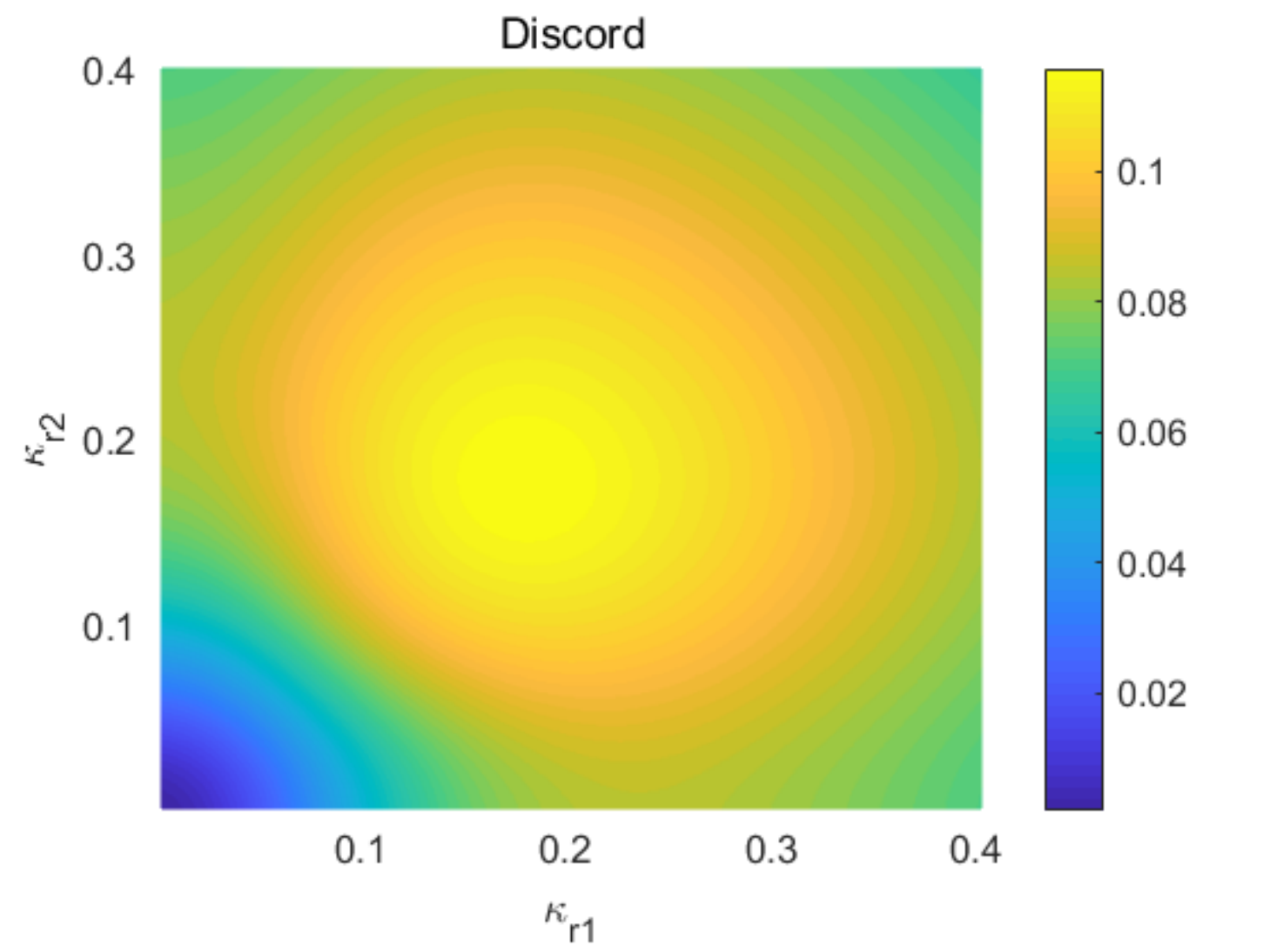}
\end{minipage}
}
\subfigure[]{
\begin{minipage}{7cm}\centering
\includegraphics[scale=0.5]{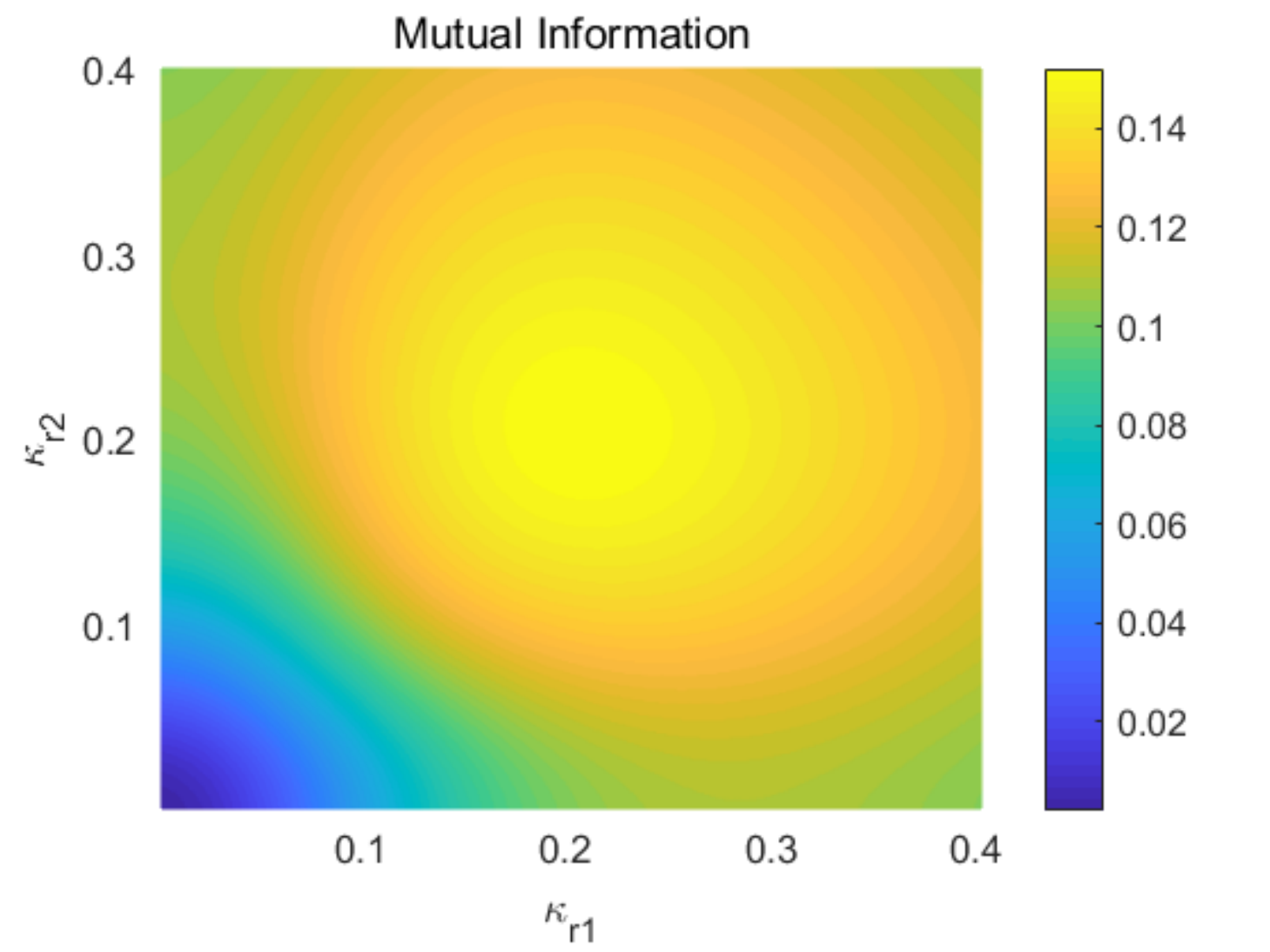}
\end{minipage}
}
\subfigure[]{
\begin{minipage}{7cm}\centering
\includegraphics[scale=0.5]{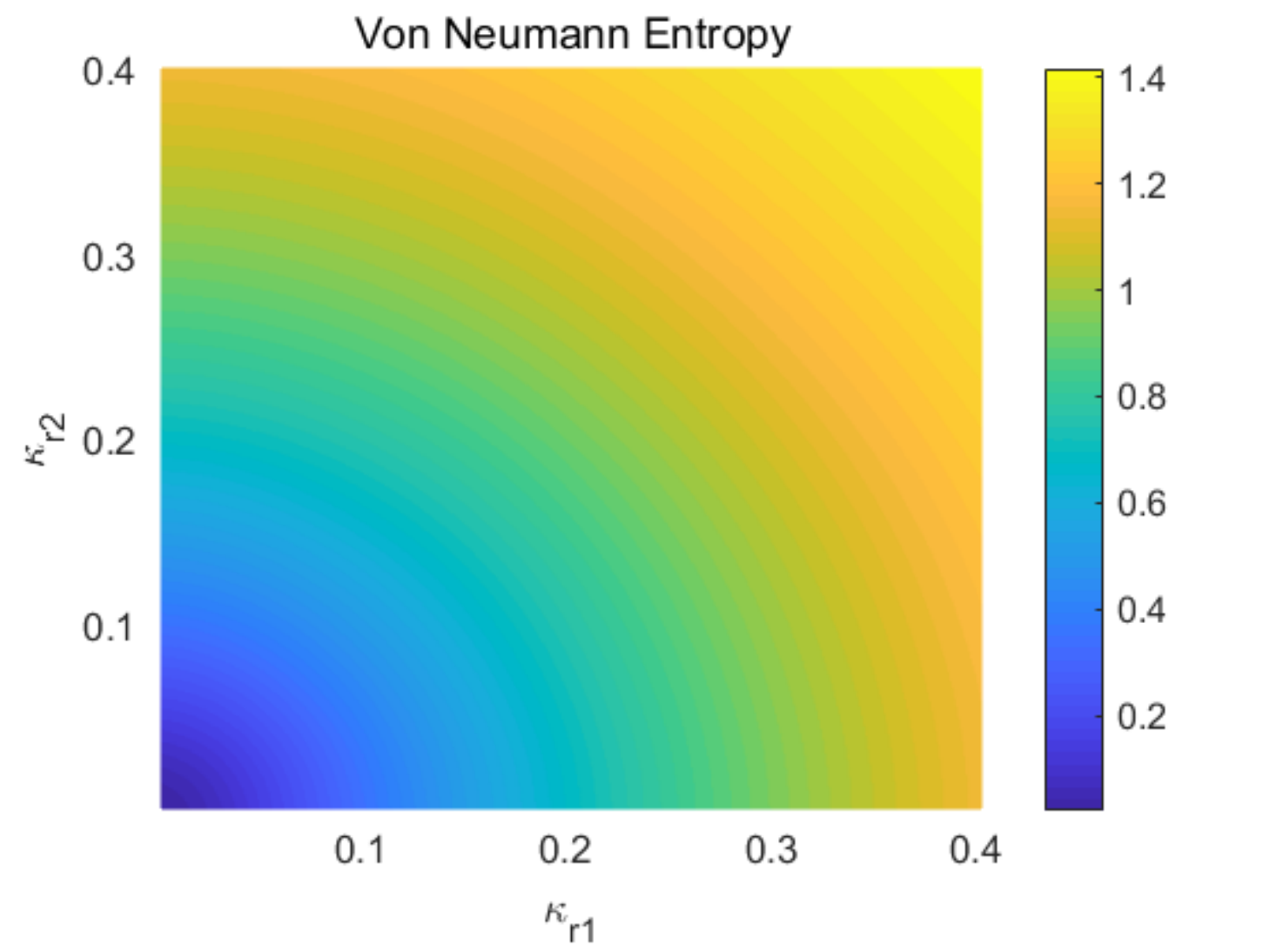}
\end{minipage}
}\subfigure[]{
\begin{minipage}{7cm}\centering
\includegraphics[scale=0.5]{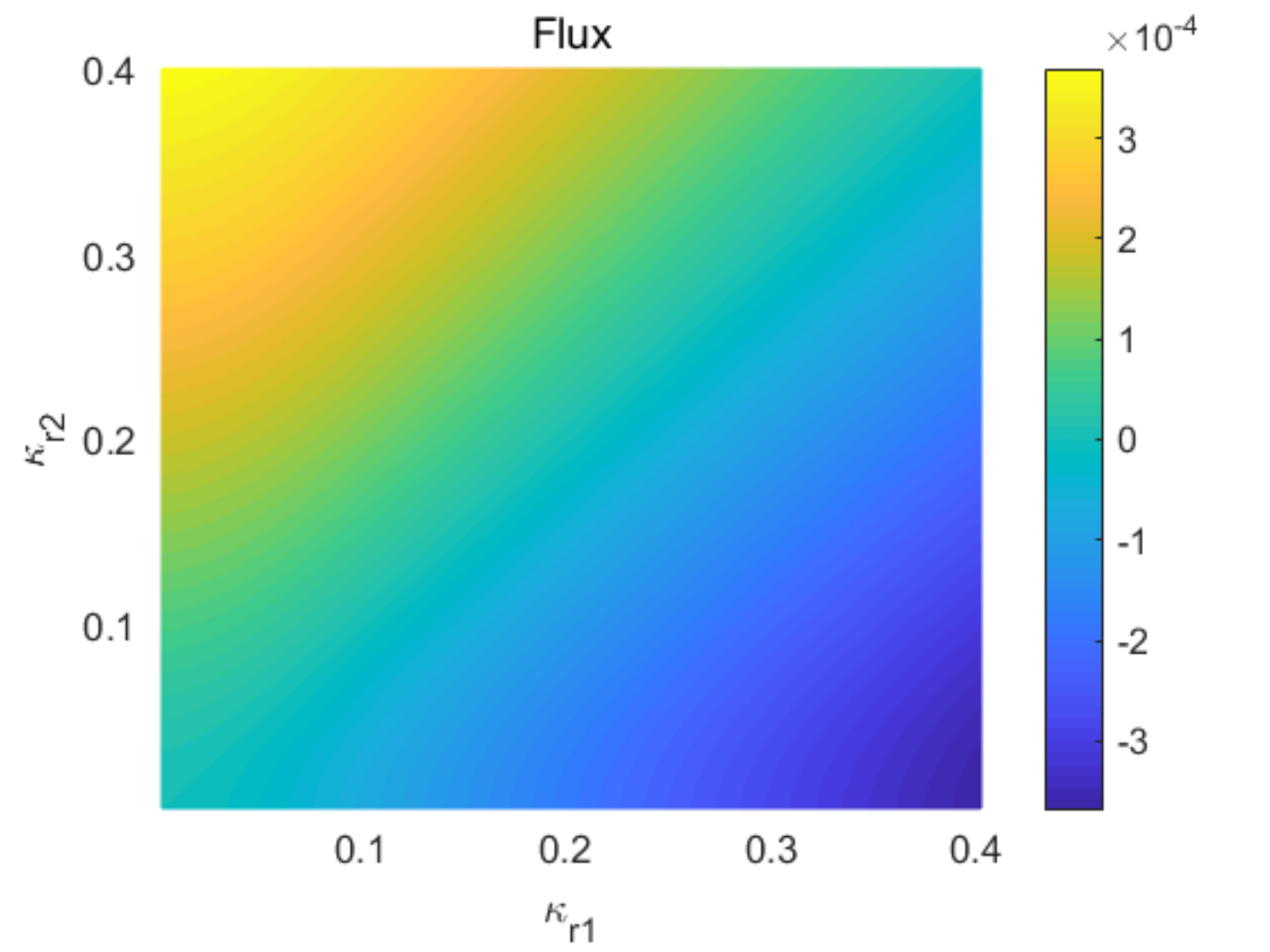}
\end{minipage}
}
\subfigure[]{
\begin{minipage}{7cm}
\includegraphics[scale=0.5]{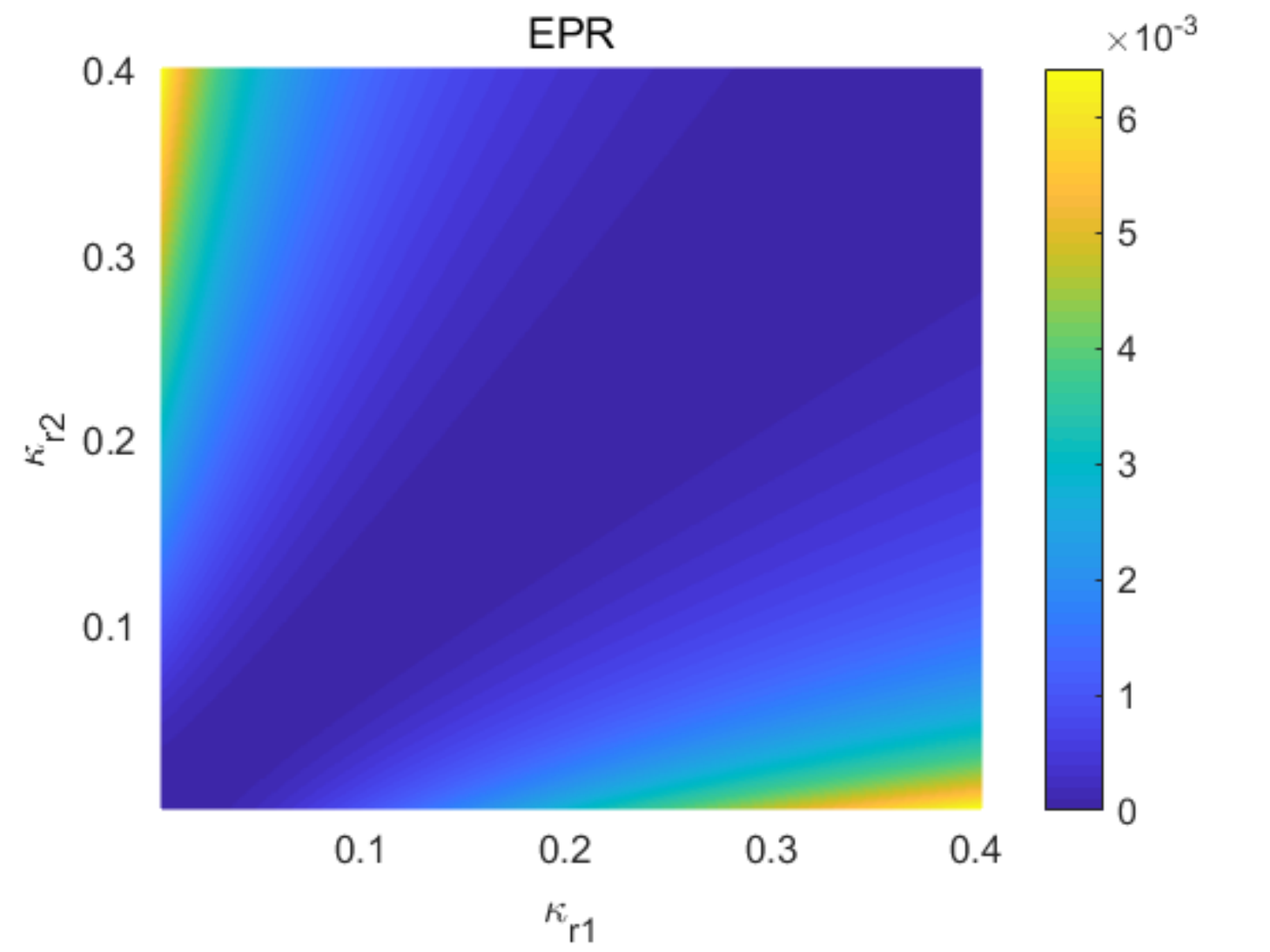}
\end{minipage}
}
\caption{Quantum correlations at nonequilibrium steady state varying with $\kappa_{r1}$ and $\kappa_{r2}$.}
\label{fig:11.4}
\end{figure}

The above non-trivial phenomena can also be understood from the dependence of the local curvature $\kappa_{r}$ on the mass and the angular momentum. The difference is that the system now is determined by not only $\kappa_{r1}$ but also $\kappa_{r2}$. This leads to the different effect compared to the previous nonequilibrium model. We plot the quantum correlations, the flux and EPR varying with both $\kappa_{r1}$ and $\kappa_{r2}$ in Fig.\ref{fig:11.4}. These figures are symmetric along the line $\kappa_{r1}=\kappa_{r2}$. We can see that the entanglement, the discord and the mutual information in Fig.\ref{fig:11.4}(a)(c)(d) show the non-monotonic behaviors as the results before, while the coherence, the Von Neumann entropy, the flux and the EPR show the monotonic behaviors. The above figures explain the Fig.\ref{fig:11.1}, \ref{fig:11.2}, \ref{fig:11.3} well according to the behaviors of $\kappa_{r}$ in Fig.\ref{fig:11.3}(e)(f). The non-monotonic behaviors of the entanglement, the discord and the mutual information can also be understood as a competition between the populations and the coherence.~\cite{Steady-state entanglement and coherence of two coupled qubits in equilibrium and nonequilibrium environments}. The concurrence in this model can be formulated as $\mathscr{C}=Max(0,\mathscr{C}_{l_{1}}-\sqrt{\rho_{11}\rho_{22}})$ where $\mathscr{C}_{l_{1}}$ is the coherence. Thus the concurrence is directly dependent on the coherence and the population. We see that both the coherence and the population vary non-monotonically. Although the discord and mutual information can not be derived with the similar formula, we believe that this competition perspective still holds for the discord and the mutual information because they measure the quantum correlations with certain similar parts in some sense. The Von Neumann entropy is amplified by the $\kappa_{r}$, the reason is clear: the higher temperature leads to the strengthening of the interaction between the system and the field. Therefore this makes more easily to produce the correlation. The coherence is representation-dependent, it vanishes in the eigen-energy basis but is non-vanishing in the bare basis. The non-vanishing coherence in the bare basis is induced by the interaction of inter-qubits and proportional to $|\rho_{33}-\rho_{44}|$. One can also understand the behavior of the coherence from a competition relationship. On the one hand, when the temperature is low, only the ground state $|\lambda_{1}\rangle$ and the first excited state $|\lambda_{4}\rangle$ are significantly occupied, then the coherence is proportional to $\rho_{44}$ and increases with temperature. As the temperature increases, the second excited state $|\lambda_{3}\rangle$ also starts to be occupied. The coherence then is proportional to $|\rho_{33}-\rho_{44}|$ which shows a competition relationship. As long as the temperature is high enough, the coherence decreases and vanishes at the infinite temperature.

Here we see that the information of the two separate qubits is encoded in space time structure again. Unlike previous nonequilibrium case, the qubits system here involves two different locations so that the intrinsic nonequilibrium emergents where detailed balanced is explicitly broken. As a result, the energy flux and associated dissipative cost emerge. They are used to support sustaining and survival of the quantum correlations for long time (at the steady state). Therefore nonequilibrium can contribute to form the quantum correlation of the system.

\section{Conclusion}
\label{sec:Conclusion}

In this paper, we focus on the quantum correlations of curved space time near the horizon of $Kerr$ black hole by using the dimensional reduction and the $Born-Markov$ master equation. We quantify the quantum correlations of the two-qubit system, and the entanglement between the system and the environment. In the equilibrium model, we can obtain a steady state which contains the initial partial information. It is possible to harvest the quantum correlations from the $Unruh$ vacuum. We investigate how the quantum correlations vary with the mass and the angular momentum. We found that the quantum correlations in the system decrease at first and then increase to a constant as the mass increases from a value close to the angular momentum per mass. The increase of angular momentum can amplify the quantum correlations. The entanglement between the system and environment behaves oppositely to the correlations in the system. Importantly, we found the increase of the local space time curvature can reduce the correlations in the system due to the thermal Unruh effect but enhance the entanglement between the system and environment.

In the second nonequilibrium transient model, we found that the information scrambles inevitably to environment. The angular momentum weakens the scrambling but the mass relates to the correlations non-monotonously seen from the decay rates of the correlations. The entanglement between the system and environment also behaves oppositely to the correlations in the system. At the fixed time, the quantum correlations are very similar to the equilibrium case: the quantum correlations in two-qubit system vary non-monotonically with the mass of the black hole and are amplified by the angular momentum, the increase of the space time curvature will suppress the quantum correlations in the system. We quantify the EPR (entropy production rate) of the system and found that it decreases in time. The EPR decreases at first and then increases to a constant with respect to the increase of mass and increases when the angular momentum increases. Besides, the space time curvature suppresses the information scrambling. We also found the local curvature can enhance the decay rates of quantum correlations and EPR, but reduce the decay rate of the Von Neumann entropy which is negative growth. The $\kappa_{r}$ not only stands for the local curvature of space time but also the thermal nature of black hole. The similar behaviors of the quantum correlations in the above two scenarios are due to the fact that the system state is determined by $\kappa_{r}$. More profoundly speaking, the features and characteristics of the system information is encoded in the space time structure.

In the third nonequilibrium model, we investigate the quantum correlations of the nonequilibrium steady state. On the whole, the quantum correlations survive and sustain at a steady when $\Delta r$ (which measures the nonequilibrium) is large. The entanglement, discord and the mutual information behave non-monotonically under certain parameters. This means the quantum correlations can be amplified by the nonequilibrium. The coherence monotonically decreases to a constant. The Von Nuemann entropy decreases to a constant which means that the nonequilibrium reduces the correlation between the system and the environment. The flux which measures the degree of the detailed balance breaking increases to a constant. The EPR as a nonequilibrium thermodynamics dissipative cost increases when the $\Delta r$ increases as we expect. We can qualitatively understand the above non-trivial behaviors by checking the dependence of the system on both the local curvatures or accelerations $\kappa_{r1}$ and $\kappa_{r2}$. In this model the information of the quantum correlations of two separate qubits system are not only encoded in the space time structure, but also from the nonequilibrium contribution.

\appendix

\section{Derivation of the Entropy production}
\label{sec:Derivation of the Entropy production}

The unitary transformation preserves the $von-Neumann$ entropy, therefore

\begin{equation}\begin{split}
\label{eq:A1}
\emph{I}_{AE:B}^{f}&=\emph{S}_{AE}^{f}+\emph{S}_{B}^{f}-\emph{S}_{ABE}^{f}\\
&=\emph{S}_{AE}^{i}+\emph{S}_{B}^{i}-\emph{S}_{ABE}^{i}\\
&=\emph{S}_{A}^{i}+\emph{S}_{E}^{i}+\emph{S}_{B}^{i}-\emph{S}_{AB}^{i}-\emph{S}_{E}^{i}\\
&=\emph{I}_{A:B}^{i}
\end{split}
\end{equation}

In the meanwhile

\begin{equation}\begin{split}
\label{eq:A2}
\emph{I}_{AE:B}^{f}&=\emph{S}_{AE}^{f}+\emph{S}_{B}^{f}-\emph{S}_{ABE}^{f}\\
&=\emph{S}_{A}^{f}+\emph{S}_{E}^{f}-\emph{I}_{A:E}^{f}+\emph{S}_{B}^{f}-\emph{S}_{ABE}^{f}\\
&=\emph{S}_{A}^{f}+\emph{S}_{E}^{f}-\emph{I}_{A:E}^{f}+\emph{S}_{B}^{f}-\emph{S}_{AB}^{f}-\emph{S}_{E}^{f}+\emph{I}_{AB:E}^{f}\\
&=\emph{S}_{A}^{f}+\emph{S}_{E}^{f}-\emph{I}_{A:E}^{f}+\emph{S}_{B}^{f}-\emph{S}_{A}^{f}-\emph{S}_{B}^{f}+\emph{I}_{A:B}^{f}-\emph{S}_{E}^{f}+\emph{I}_{AB:E}^{f}\\
&=\emph{I}_{A:B}^{f}+\emph{I}_{AB:E}^{f}-\emph{I}_{A:E}^{f}
\end{split}
\end{equation}

Combine with Eqn.\eqref{eq:A1}, we get $\Delta\emph{I}_{A:B}(t_{i}:t_{f})=\emph{I}_{AB:E}^{f}-\emph{I}_{A:E}^{f}$. In the meanwhile, the above equation implies $\Delta\emph{I}_{A:B}(t_{i}:t_{f})>0$, which is from the fact that the correlation between $AB$ and $E$ should be larger than the correlation between $A$ and $E$. Inserting Eqn.\eqref{eq:49} into Eqn.\eqref{eq:50}

\begin{equation}\begin{split}
\label{eq:A3}
\Sigma_{AB}(t_{i}:t_{f})&=\emph{I}_{AB:E}(t_{f})-\emph{I}_{A:E}(t_{f})+\Sigma_{A}(t_{i}:t_{f})\\
&=\Delta\emph{I}_{A:B}(t_{i}:t_{f})+\Sigma_{A}(t_{i}:t_{f})
\end{split}
\end{equation}

\section{Steady state expression}
\label{sec:Steady state expression}

We give a concrete expression of the steady state matrix. The computation method was given in ~\cite{Quantum thermalization of two coupled two-level systems in eigenstate and bare-state representations}~\cite{Steady-state entanglement and coherence of two coupled qubits in equilibrium and nonequilibrium environments}~\cite{Coherence enhanced quantum metrology in a nonequilibrium optical molecule}. For the steady state matrix, the off-diagonal elements vanish and the diagonal elements are

\begin{equation}\begin{split}
\label{eq:A4}
\rho_{11}&=\frac{X_{1}^{+}Y_{2}^{+}}{X_{1}Y_{2}}\\
\rho_{22}&=\frac{X_{1}^{-}Y_{2}^{-}}{X_{1}Y_{2}}\\
\rho_{33}&=\frac{X_{1}^{-}Y_{2}^{+}}{X_{1}Y_{2}}\\
\rho_{44}&=\frac{X_{1}^{+}Y_{2}^{-}}{X_{1}Y_{2}}\\
\end{split}
\end{equation}

where we define
\begin{equation}\begin{split}
\label{eq:A5}
X_{i}&=X_{i}^{+}+X_{i}^{-}\\
Y_{i}&=Y_{i}^{+}+Y_{i}^{-}\\
X_{i}^{\mp}&=2\cos^{2}(\theta/2)G^{1}(\pm\omega_{i})+2\sin^{2}(\theta/2)G^{2}(\pm\omega_{i})\\
Y_{i}^{\mp}&=2\sin^{2}(\theta/2)G^{1}(\pm\omega_{i})+2\cos^{2}(\theta/2)G^{2}(\pm\omega_{i})
\end{split}
\end{equation}

\acknowledgments

He Wang thanks to Wei Wu, Xuanhua Wang, Kun Zhang and Hong Wang for helpful discussions.

\paragraph{Note added.} This is also a good position for notes added
after the paper has been written.

% The bibliography will probably be heavily edited during typesetting.
% We'll parse it and, using the arxiv number or the journal data, will
% query inspire, trying to verify the data (this will probalby spot
% eventual typos) and retrive the document DOI and eventual errata.
% We however suggest to always provide author, title and journal data:
% in short all the informations that clearly identify a document.


\begin{thebibliography}{99}

\bibitem{Quantifying Coherence}
Baumgratz, T. , M. Cramer , and M. B. Plenio, \emph{Quantifying Coherence}, \emph{Phys. Rev. Lett.} {\bf 113.14} (2014) 140401.

\bibitem{Quantum entanglement}
Horodecki R., Horodecki P., Horodecki M., et al., \emph{Quantum entanglement}, \emph{Rev. Mod. Phys.} {\bf 81} (2009) 865

\bibitem{Entanglement of a Pair of Quantum Bits}
S. Hill and W.K. Wootters, \emph{Entanglement of a Pair of Quantum Bits}, \emph{Phys. Rev. Lett.} {\bf 78} (1997) 5022.

\bibitem{Quantum Discord}
Harold Ollivier and Wojciech H. Zurek, \emph{Quantum Discord: A Measure of the Quantumness of Correlations}, \emph{Phys. Rev. Lett.} {\bf 88} (2001) 017901.

\bibitem{Quantum information transfer and models for black hole mechanics}
Steven B. Giddings and Yinbo Shi, \emph{Quantum information transfer and models for black hole mechanics}, \emph{Phys. Rev. D} {\bf 87} (2012) 064031.

\bibitem{Black holes as mirrors}
Patrick Hayden and John Preskill, \emph{Black holes as mirrors: quantum information in random subsystems}, \emph{J. High Energy Phys} {\bf 09} (2007) 887-891.

\bibitem{Cosmological quantum entanglement}
E. Martin Martinez, Menicucci N. C., \emph{Cosmological quantum entanglement}, \emph{Class. Quantum Grav.} {\bf 29(22)} (2012) 224003.

\bibitem{Entanglement in curved spacetimes and cosmology}
E. Martin Martinez, Menicucci N. C., \emph{Entanglement in curved spacetimes and cosmology}, \emph{Class. Quantum Grav.} {\bf 31(21)} (2014).

\bibitem{Quantum Computation and Quantum Information}
M.A. Nielsen, I.L. Chuang,\emph{Quantum Computation and Quantum Information},
Cambridge University Press, Cambridge, England, 2000.

\bibitem{Teleporting}
Bennett, Charles H., et al., \emph{Teleporting an unknown quantum state via dual classical and Einstein-Podolsky-Rosen channels.}, \emph{Phys. Rev. Lett.} {\bf 70.13} (1993) 1895-1899.

\bibitem{Quantum Cryptography}
Ekert, Artur K., \emph{Quantum Cryptography Based on Bell's Theorem}, \emph{Phys. Rev. Lett.} {\bf 67(6)} (1991) 661-663.

\bibitem{Teleportation}
Horodecki R., Horodecki M., Horodecki P., \emph{Teleportation, Bell's Inequalities and Inseparability}, \emph{Phys. Lett. A} {\bf 222(1-2)} (1996) 21-25

%\bibitem{Bell inequality, Bell states and maximally entangled states for n qubits}
%N Gisin, H Bechmann-Pasquinucci, \emph{Bell inequality, Bell states and maximally entangled states for n qubits}, \emph{Phys. Lett. A} {\bf 246(1-2)} (1998) 1-6

%\bibitem{Bell's inequality, generalized concurrence and entanglement in qubits.}
%Po Yao Chang, Su Kuan Chu and Chen Te Ma, \emph{Bell's inequality, generalized concurrence and entanglement in qubits.}, \emph{Internat. J. Modern Phys. A} {\bf 34} (2019) 1950032

\bibitem{Bell's inequalities versus teleportation}
Popescu, Sandu, \emph{Title}, \emph{Bell's inequalities versus teleportation: What is nonlocality?}, \emph{Phys. Rev. Lett.}{\bf 72(6)} (1994) 797-799.

\bibitem{Notes on black hole evaporation}
Unruh, W. G., \emph{Notes on black hole evaporation}, \emph{Phys. Rev.D} {\bf 14(4)} (1976)870-892.

\bibitem{Controlling entanglement generation in external quantum fields}
F. Benatti and R. Floreanini, \emph{Controlling entanglement generation in external quantum fields}, \emph{J. Opt. B: Quantum Semiclass. Opt} {\bf 7} (2005) S429

\bibitem{Entanglement generation in uniformly accelerating atoms}
F. Benatti and R. Floreanini, \emph{Entanglement generation in uniformly accelerating atoms: Reexamination of the Unruh effect}, \emph{Phys. Rev. A} {\bf 70} (2004) 012112.

\bibitem{Thermal amplification of field-correlation harvesting}
Brown, Eric G., \emph{Thermal amplification of field-correlation harvesting}, \emph{Phys. Rev. A} {\bf 88} (2013) 062336.

\bibitem{Equilibrium and Nonequilibrium Quantum Correlations Between Two Accelerated Detectors}
H. Wang, J. Wang, \emph{Equilibrium and Nonequilibrium Quantum Correlations Between Two Accelerated Detectors}, arxiv:2010.08203.

\bibitem{Entanglement generation outside a Schwarzschild black hole and the Hawking effect}
Hu J., Yu H., \emph{Entanglement generation outside a Schwarzschild black hole and the Hawking effect}, \emph{Journal of High Energy Physics} {\bf  2011(8)} ( 2011) 1-13.

\bibitem{Quantum entanglement generation in de Sitter spacetime}
Hu J. and H. Yu, \emph{Quantum entanglement generation in de Sitter spacetime}, \emph{Phys. Rev. D} {\bf 88(10)} (2013) 1845-1858.

\bibitem{Black hole explosions?}
S. Hawking, \emph{Black hole explosions?}, \emph{Nature} {\bf 248} (1974) 30.

\bibitem{Understanding Hawking radiation in the framework of open quantum systems}
Yu H , Zhang J ., \emph{Understanding Hawking radiation in the framework of open quantum systems}, \emph{Phys. Rev.D} {\bf 77} (2008) 029904.

\bibitem{Researching on Hawking Effect in a Kerr Space Time via Open Quantum System Approach}
Liu X. M., Liu W. B., \emph{Researching on Hawking Effect in a Kerr Space Time via Open Quantum System Approach}, \emph{Advances in High Energy Physics} {\bf 2014} (2014) 1-8.

\bibitem{Hawking radiation from rotating black holes and gravitational anomalies}
K. Murata and J. Soda, \emph{Hawking radiation from rotating black holes and gravitational anomalies}, \emph{Phys. Rev. D} {\bf 74} (2006) 044018.

\bibitem{Anomalies}
S. Iso, H. Umetsu, and F. Wilczek, \emph{Anomalies, Hawking radiations, and regularity in rotating black holes}, \emph{Phys. Rev. D} {\bf 74} (2006) 044017.

\bibitem{The Theory of Open Quantum Systems}
Breuer, H. P. , and F. Petruccione, \emph{The Theory of Open Quantum Systems},
Oxford University Press (2006). p130-136

\bibitem{Completely positive dynamical semigroups of N-level systems}
Vittorio Gorini, Andrzej Kossakowski, and E. C. G. Sudarshan, \emph{Completely positive dynamical semigroups of N-level systems}, \emph{Journal of Mathematical Physics} {\bf 17} (1976) 821.

\bibitem{Quantum discord for two-qubit X-states}
Mazhar Ali, A. R. P. Rau, and G. Alber, \emph{Quantum discord for two-qubit X-states}, \emph{Phys. Rev. A} {\bf 81(4)} (2010) 82-82.

\bibitem{Quantum discord for two-qubit systems}
Shunlong Luo, \emph{Quantum discord for two-qubit systems}, \emph{Phys. Rev. A} {\bf 77} (2008) 042303.

\bibitem{Nonequilibrium effects on quantum correlations: Discord}
Xuanhua Wang and Jin Wang, \emph{Nonequilibrium effects on quantum correlations: Discord}, \emph{Phys. Rev. A} {\bf 100} (2019) 052331.

\bibitem{Black Holes: An Introduction}
Raine D., Thomas E., \emph{Black Holes: An Introduction},
Imperial College Press, Singapore: World Scientific Publishing (2005). p42

\bibitem{Are black holes springlike?}
Michael R.R.Good and Yen Chin Ong, \emph{Are black holes springlike?}, \emph{Phys. Rev. D} {\bf 91} (2015) 044031.
%\bibitem{A study of Quantum Correlations in Open Quantum Systems}
%Chakrabarty I. , Banerjee S. , Siddharth N., \emph{A study of Quantum Correlations in Open Quantum Systems}, \emph{Quantum Information and Computation} {\bf 11(7)} (2010) 541-562.

\bibitem{Loss of Spin Entanglement For Accelerated Electrons in Electric and Magnetic Fields}
J. Doukas and Lloyd C. L. Hollenberg, \emph{Loss of Spin Entanglement For Accelerated Electrons in Electric and Magnetic Fields}, \emph{J. Doukas and Lloyd C. L. Hollenberg} {\bf 79} (2009) 052109.

\bibitem{Dynamics and quantum entanglement of two-level atoms in de Sitter spacetime}
Tian Z., Jing J.,\emph{Dynamics and quantum entanglement of two-level atoms in de Sitter spacetime}, \emph{Annals of Physics} {\bf 350} (2014) 1-13.

%\bibitem{Quantum Fields in Curved Space}
%Birrell N. D., Davies P., \emph{Quantum Fields in Curved Space}
%Of Cambridge Monographs on Mathematical Physics, 1982, 36:349.

\bibitem{Entropy production as correlation between system and reservoir}
M. Esposito, K. Lindenberg, and C. Van den Broeck, \emph{Entropy production as correlation between system and reservoir}, \emph{New Journal of Physics} {\bf 12} (2010) 013013.

\bibitem{Kun Zhang}
Kun Zhang, Xuanhua Wang, Qian Zeng, Jin Wang, \emph{Conditional entropy production and quantum fluctuation theorem of dissipative information}, arXiv:2105.06419.

\bibitem{Dynamics of nonequilibrium thermal entanglement}
Ilya Sinaysky, Francesco Petruccione, and Daniel Burgarth, \emph{Dynamics of nonequilibrium thermal entanglement}, \emph{Phys. Rev. A} {\bf 78} (2008) 062301.

\bibitem{Quantum thermalization of two coupled two-level systems in eigenstate and bare-state representations}
Jie Qiao Liao, Jin Feng Huang, and Le Man Kuang, \emph{Quantum thermalization of two coupled two-level systems in eigenstate and bare-state representations}, \emph{Phys. Rev. A} {\bf 83} (2011) 052110.

\bibitem{Steady-state entanglement and coherence of two coupled qubits in equilibrium and nonequilibrium environments}
Z. Wang, W. Wu, and J. Wang, \emph{Steady-state entanglement and coherence of two coupled qubits in equilibrium and nonequilibrium environments}, \emph{Phys. Rev. A} {\bf 99} (2019) 042320.

\bibitem{Coherence enhanced quantum metrology in a nonequilibrium optical molecule}
Z. Wang, W. Wu, G. Cui, and J. Wang, \emph{Coherence enhanced quantum metrology in a nonequilibrium optical molecule}, \emph{New J. Phys} {\bf 20} (2018) 033034.


% Please avoid comments such as "For a review'', "For some examples",
% "and references therein" or move them in the text. In general,
% please leave only references in the bibliography and move all
% accessory text in footnotes.

% Also, please have only one work for each \bibitem.


\end{thebibliography}
\end{document}